\RequirePackage[2020-02-02]{latexrelease}
\documentclass[aps,prb,reprint,amsmath,amssymb,floatfix]{revtex4-2}

\usepackage{graphicx,xcolor,hyperref,diagbox,bm}

\usepackage[ignoreunlbld,norefs,nocites]{refcheck}

\begin{document}

\title{Low-density Phase Diagram of the Three-Dimensional Electron Gas}

\author{Sam Azadi}

\email{sam.azadi@physics.ox.ac.uk}

\affiliation{Department of Physics, Clarendon Laboratory, University
  of Oxford, Parks Road, Oxford OX1 3PU, United Kingdom}

\affiliation{Department of Physics, Imperial College London,
  Exhibition Road, London SW7 2AZ, United Kingdom}

\author{N.\ D.\ Drummond}

\affiliation{Department of Physics, Lancaster University, Lancaster
  LA1 4YB, United Kingdom}

\date{\today}

\begin{abstract}
Variational and diffusion quantum Monte Carlo methods are employed to
investigate the zero-temperature phase diagram of the
three-dimensional homogeneous electron gas at very low density. Fermi
fluid and body-centered cubic Wigner crystal ground state energies are
determined using Slater-Jastrow-backflow and Slater-Jastrow many-body
wave functions at different densities and spin polarizations in finite
simulation cells. Finite-size errors are removed using twist-averaged
boundary conditions and extrapolation of the energy per particle to
the thermodynamic limit of infinite system size. Unlike previous
studies, our results show that the electron gas undergoes a
first-order quantum phase transition directly from a paramagnetic
fluid to a body-centered cubic crystal at density parameter
$r_\text{s} = 86.6(7)$, with no region of stability for an itinerant
ferromagnetic fluid.
\end{abstract}

\maketitle

\section{Introduction}

The three-dimensional homogeneous electron gas (3D-HEG) has been of
fundamental interest in physics and chemistry since the early days of
quantum mechanics because it is the simplest realistic bulk electronic
system capable of exhibiting strong correlation effects
\cite{Bloch,Wigner,Wigner_1938, Landau, Pines, Giuliani}. The
electron-electron interaction strength, and therefore the coupling
between the electrons, is controlled by the electron density.  The
3D-HEG models the electrons in bulk metals, but more importantly it
has long provided a testbed for the development of ideas, concepts,
and methods in condensed matter physics. For example, the ground-state
energy of the 3D-HEG provides the starting point for most of the
exchange-correlation functionals that have enabled the widespread
success of density functional theory (DFT)\@.  In this work we focus
on the low-density energy and phase behavior of the 3D-HEG\@.

Theory plays a crucial role in the study of dilute 3D-HEGs due to the
lack of a material platform that supports a 3D electron system with
both very high quality (homogeneity) and low density. It is convenient
to characterize the density of the 3D-HEG by the dimensionless
parameter $r_\text{s}$ defined as the radius of the sphere that
contains one electron on average in units of the Bohr radius. In
Hartree atomic units the 3D-HEG Hamiltonian is $\hat{H}=-(1/2)\sum_i
\nabla_i^2+\sum_{i>j} 1/|\mathbf{r}_i-\mathbf{r}_j|$, where
$\mathbf{r}_i$ is the position of electron $i$ and the
electron-electron Coulomb interaction is in practice evaluated using
Ewald summation in a finite cell. The Coulomb interaction scales as
$1/r_\text{s}$, while the kinetic energy operator scales as
$1/r_\text{s}^2$. At high to intermediate density (small $r_\text{s}$)
the kinetic energy dominates, leading to the well-known Fermi fluid
behavior of the 3D-HEG at typical metallic densities.  On the other
hand, the Coulomb energy dominates the kinetic energy at low density
(large $r_\text{s}$), fundamentally altering the physics of the
3D-HEG\@. In the low-density limit, the ground-state wave function is
an antisymmetrized product of $\delta$-functions centered on
body-centered cubic (bcc) lattice sites to minimize the Coulomb
energy, as first predicted by Wigner \cite{Wigner}. Here, we calculate
the critical density parameter $r_\text{s}$ at which there is a
zero-temperature phase transition from a Fermi fluid to a Wigner
crystal.  Furthermore, Bloch suggested that the spin-unpolarized
(paramagnetic) Fermi fluid should make a spontaneous transition to a
spin polarized (ferromagnetic) Fermi fluid at large $r_\text{s}$
before crystallization \cite{Bloch}, because aligning the electron
spins causes the spatial wave function to be fully antisymmetric, so
that electrons do not approach each other and the Coulomb energy is
reduced.

Quantum Monte Carlo (QMC) methods have long been used to provide
accurate estimates of properties of 3D-HEGs
\cite{Ceperley77,Ceperley78,Ceperley80,Ortiz_1999,Zong,Spink,Shepherd,PRL2021}.
For example, ground state QMC energies of the 3D-HEG \cite{Ceperley80}
are employed in parameterizations of the local-density approximation
to the DFT exchange-correlation functional
\cite{Perdew_1981}. However, calculating the phase diagram is
challenging because of the tiny energy differences between competing
phases.  Previous QMC simulations have indicated that decreasing the
density of a 3D-HEG causes a continuous transition from a
spin-unpolarized (paramagnetic) fluid to a fully spin-polarized
(ferromagnetic) fluid at a density of about $r_\text{s} = 50(2)$
\cite{Zong}. The phase transition to a Wigner crystal was predicted to
take place at density parameter $r_\text{s} = 106(1)$
\cite{Zong,Neil04}. Because of the fundamental role of the 3D-HEG in
condensed matter physics, the determination of its zero-temperature
phase diagram and ground-state energy is a problem that should be
revisited from time-to-time using state-of-the-art computational
methods.

In this work we have used the continuum variational and diffusion
Monte Carlo (VMC \cite{Umrigar,Ceperley77} and DMC \cite{Ceperley80})
methods in real space to obtain 3D-HEG ground state energies at
different densities and spin polarizations. In the VMC method,
parameters in a trial wave function are optimized according to the
variational principle, with energy expectation values calculated by
Monte Carlo integration in the $3N$-dimensional space of electron
position vectors.  In the DMC method, the imaginary-time
Schr\"{o}dinger equation is used to evolve a statistical ensemble of
electronic configurations towards the ground state. Fermionic
antisymmetry is maintained by the fixed-phase approximation, in which
the complex phase of the wave function is constrained to equal that of
an approximate wave function optimized within VMC\@. For real wave
functions (which occur when the system has time-reversal symmetry,
e.g., under pure periodic boundary conditions), the fixed-phase
approximation reduces to constraining the nodal surface of the wave
function.  Henceforth we refer to ``fixed nodes'' rather than ``fixed
phases'' to avoid confusion with thermodynamic ``phases''; our actual
fluid calculations used complex wave functions and the fixed-phase
approximation, while our crystal calculations used real wave functions
and the fixed-node approximation.

Fixed-node DMC finds the variational lowest-energy state with the same
nodal surface as the trial wave function. Thus the topology of the
trial wave function's nodal surface selects the quantum state under
study.  The DMC energy with an antisymmetric trial wave function is an
upper bound on the fermionic ground-state energy; furthermore, the
error in the DMC energy of any quantum state approximated by the trial
wave function is second order in the error in the trial nodal surface.

In a finite cell the eigenfunctions of the 3D-HEG Hamiltonian must all
be homogeneous (i.e., must satisfy the many-body Bloch theorems
\cite{rajagopal_1994,rajagopal_1995} with an infinitesimal ``primitive
cell'') and hence eigenvalue crossings as a function of the single
parameter $r_\text{s}$ are avoided by the von Neumann-Wigner theorem.
The true ground state energy per electron $E(r_\text{s})$ of the
3D-HEG in a finite periodic cell of a given shape, electron number
$N$, and spin-polarization $\zeta=(N_\uparrow-N_\downarrow)/N$ is
therefore a smooth function of $r_\text{s}$.  The ground-state static
structure factor $S({\bf G})=\left< \hat{\rho}(-{\bf
  G})\hat{\rho}({\bf G}) \right> - \langle \hat{\rho}(-{\bf G})\rangle
\langle \hat{\rho}({\bf G})\rangle$, where $\hat{\rho}({\bf G})=\sum_i
\exp(i{\bf G}\cdot\hat{\bf r}_i)$, describes the Fourier components of
the pair density and therefore shows whether the 3D-HEG is fluid-like
or crystal-like; this too is a smooth function of $r_\text{s}$ in a
given finite cell.  In fact there is a different $E(r_\text{s})$ curve
for each system size $N$, spin polarization, cell shape, and choice of
twisted boundary conditions.  For example a bcc simulation cell with
$N$ a cubic number strongly favors crystalline behavior.  The fluid
energy per particle fluctuates quasirandomly with system size $N$,
cell shape, and twisted boundary conditions due to momentum
quantization effects.  For a sufficiently large periodic cell of a
given shape, there must be a narrowly avoided crossing of energy
levels as a function of $r_\text{s}$ near the crystallization density,
with $S({\bf G}_\text{prim})$ changing significantly near the avoided
crossing, where $\mathbf{G}_\text{prim}$ is a primitive-cell
reciprocal lattice point of the bcc Wigner crystal, resulting in a
smooth crossover from Fermi fluid to ``floating''
\cite{Bishop_1982,Lewin_2019} crystal behavior.  For the infinite
3D-HEG, however, the center-of-mass kinetic energy per electron
vanishes and hence broken-translational-symmetry crystal wave
functions are degenerate with floating crystal wave functions.  The
avoided crossing of energy levels therefore becomes a true crossing of
energy levels with different symmetry.  Furthermore, $E(r_\text{s})$
ceases to depend on the simulation cell shape and choice of twisted
boundary conditions. At $r_\text{s}\rightarrow 0$, the interaction
potential is negligible and we have a homogeneous ground-state fluid
wave function.  At $r_\text{s} \to \infty$, the kinetic energy is
negligible and we have a bcc crystal. The symmetry of the ground state
of the infinite 3D-HEG must therefore change at some finite
$r_\text{s}$, i.e., there is a zero-temperature phase transition
\cite{Wigner}. The charge density $\rho({\bf G}_\text{prim})=\langle
\hat{\rho}({\bf G}_\text{prim}) \rangle$ is an appropriate order
parameter for the fluid-to-crystal transition, being zero in the fluid
phase and nonzero in the crystal phase.  The crystallization
transition is expected to be first order, corresponding to a crossing
of crystal and fluid energy levels as functions of $r_\text{s}$ with
the order parameter being nonzero at the crossing point in the crystal
phase. The following numerical results provide some evidence
confirming that the Wigner crystal charge density is nonzero at the
crystallization density.

\section{Calculating the zero-temperature phase diagram}

\subsection{QMC methodology}

In QMC studies of the phase diagram of the 3D-HEG, we look for a
first-order phase transition by calculating the DMC energy as a
function of density parameter $r_\text{s}$ for trial wave functions
that model the ground-state fluid and the ground-state crystal. For
the fluid phases we use Slater determinants of plane-wave orbitals,
multiplied by Jastrow correlation factors that do not alter the nodal
surface \cite{Ceperley77,Ceperley78,Ceperley80}.  We evaluate the
orbitals at quasiparticle coordinates related to the actual electron
coordinates by continuous backflow (BF) transformations
\cite{Feynman56,Lee81,Kwon98,Pablo}, allowing variational optimization
of the nodal surface without changing its topology. The nodal topology
of the fluid trial wave function is therefore the same as that of a
Slater determinant of plane waves, i.e., the wave function of a free
electron gas.  This model is not exact: as directly revealed by full
configuration interaction QMC calculations, the exact ground-state
wave function of the 3D-HEG is in fact a linear combination of many
ground- and excited-state Slater determinants of plane waves
\cite{Shepherd,Ruggeri18}.  Nevertheless, the single-determinant
Slater-Jastrow-BF (SJB) model of the fluid phase is reasonably
accurate because, by construction, it always leads to a significantly
lower variational energy than the Hartree-Fock wave function (a single
determinant of plane waves), and Hartree-Fock theory itself becomes
arbitrarily accurate at high density ($r_\text{s}\to 0$), where it
provides the first two terms in the high-energy expansion of the
3D-HEG energy \cite{GellMann}.  Furthermore, Landau's Fermi liquid
theory \cite{Landau} requires that low-lying excited states of the
Fermi fluid are adiabatically connected to the corresponding excited
states of a free electron gas, implying that the relevant parts of the
nodal surface of the Fermi fluid must be qualitatively the same as
that of a single determinant of plane waves.  The release-node method
\cite{Ceperley80}, in which walkers are equilibrated in fixed-node DMC
and then briefly allowed to cross nodes and change the sign of their
weights, is able to move nodes but is unlikely to be able to change
the nodal topology; we expect this approach to give similar results to
the fixed-node SJB-DMC method.

Our model of the Wigner crystal is a Slater determinant of
single-Gaussian orbitals centered on bcc lattice sites and made
periodic by summing over images, multiplied by a Jastrow correlation
factor.  It therefore explicitly breaks translational symmetry,
leading to an $O(N^{-1})$ finite-size (FS) error due to the
center-of-mass kinetic energy \cite{Suppl}.  The Gaussian exponent was
determined using a formula that minimizes the fixed-node DMC energy in
a large 216-electron cell \cite{Neil04,Suppl}. The Slater determinant
of Gaussian orbitals describes the ground state of an Einstein model
of a vibrating electron lattice.  In principle a better model of the
ground state of a vibrating electron lattice would be provided by an
antisymmetrized product of Gaussian functions of quasiharmonic normal
coordinates \cite{carr_crystal,carr_anharmonic}.  However, our Jastrow
factor and BF function allow an approximate description of this
quasiharmonic behavior, with additional flexibility.

Our Jastrow factor and BF functions contained polynomial and
plane-wave expansions in electron-electron separation
\cite{Drummond04,Pablo}. For the Wigner crystal the Jastrow factor
also contained a plane-wave expansion in electron position.   The wave
functions were optimized by variance minimization
\cite{Umrigar_1988a,Neil05} followed by energy minimization
\cite{Toulouse}.  The \textsc{casino} package was used for all our QMC
calculations \cite{casino}.

Monte Carlo-sampled canonical twist-averaged boundary conditions
(TABC) were used to reduce quasirandom single-particle FS errors in
Fermi fluid energies due to momentum quantization effects
\cite{Lin,SamPrb19}. Twist averaging
is unnecessary for Wigner crystals, which have localized orbitals and
do not have Fermi surfaces.  Systematic FS errors due to the use of
the Ewald interaction rather than $1/r$ to evaluate the interaction
between each electron and its exchange-correlation hole and the
neglect of long-range two-body correlations were removed by fitting
$E(N)=E(\infty)+b/N$ to the twist-averaged DMC energy data at
different system sizes \cite{Chiesa_2006}. This also removed the FS
bias in Wigner crystal energies due to the center-of-mass kinetic
energy.  We examine the performance of analytic expressions
\cite{Chiesa_2006,Drummond_2008}.

We studied the fluid phase at $r_\text{s}=30$, 40, 50, 60, 70, 80, and
100. For each density, QMC calculations were performed for simulation
cells with $130 \leq N \leq 274$. The energies were calculated for
spin polarizations $\zeta=0$, 0.25, 0.5, 0.75, and
1. Antiferromagnetic and ferromagnetic bcc crystalline phases were
investigated at $r_\text{s}=80$, 90, 100, and 125 for simulation cells
with $N=64$, 216, and 512 electrons.  Our DMC energies were
extrapolated linearly or quadratically to zero time step, with the
target walker population being varied in inverse proportion to the
time step. The energies and variances calculated using Slater-Jastrow
(SJ) and SJB wave functions for different system sizes are reported in
the Supplemental Material \cite{Suppl}.
The computational and technical details are discussed in the following
sections.

\subsection{Fluid phase wave function}\label{sec:fluid_wf}

For the fluid phase of the three-dimensional homogeneous electron gas
(3D-HEG) we used a Slater-Jastrow-backflow (SJB) trial spatial wave
function $\Psi({\bf R})=e^{J({\bf R})}S({\bf X}({\bf R}))$, where
${\bf R}=({\bf r}_1,\ldots,{\bf r}_N)$ is the $3N$-dimensional vector
of electron coordinates. The antisymmetric Slater part $S$ is a
product of determinants of single-particle orbitals for spin-up and
spin-down electrons. The single-particle orbitals in $S$ are of the
free-electron form $\psi_\mathbf{k}(\mathbf{r})=\exp(i\mathbf{k} \cdot
\mathbf{r})$, where wavevector $\mathbf{k}$ is a reciprocal lattice
vector of the simulation cell offset by twist vector
$\mathbf{k}_\text{s}$, where $\mathbf{k}_\text{s}$ lies in the
supercell Brillouin zone. The Jastrow exponent $J$, which is symmetric
under electron exchange, takes the form
\begin{equation}
 J(\mathbf{R})  = \sum_{i<j}^{N} u(r_{ij}) + \sum_{i<j}^{N}
 p(\mathbf{r}_{ij}),
\end{equation}
where $u$ is a smoothly truncated, isotropic polynomial function of
minimum-image electron-electron distance $r_{ij}$, and $p$ is a
plane-wave expansion in electron-electron separation ${\bf r}_{ij}$
\cite{Drummond04}. The $u$ term is of form
\begin{eqnarray}
 u(r) & = & {(r - L_u)}^C \Theta (L_u - r) \nonumber \\ & & {} \times
 \left( \alpha_0 + \left[\frac{\Gamma}{{(-L_u)}^C}+\frac{\alpha_0
     C}{L_u}\right]r + \sum_{l=2}^{N_u} \alpha_{l} r^l
 \right), \label{eq:u}
\end{eqnarray}
where the cutoff length $L_u$ is less than or equal to the radius of
the largest sphere that can be inscribed in the Wigner-Seitz cell of
the simulation cell, $C=3$ specifies how smooth the function is at the
cutoff length, $\Theta$ is the Heaviside step function, and
$\{\alpha_l\}$ are optimizable parameters, which differ for parallel-
and antiparallel-spin electrons. To satisfy the Kato cusp conditions
\cite{kato_pack}, $\Gamma=1/2$ for opposite-spin electrons and
$\Gamma=1/4$ for same-spin electrons. We chose $N_u=8$. The $p$ term
has the symmetry of the simulation-cell Bravais lattice and allows a
description of correlation in the ``corners'' of the simulation
cell. Its form is
\begin{equation}
 p(\mathbf{r})=\sum_A a_A \sum_{\mathbf{G} \in A^+} \cos(\mathbf{G}
 \cdot \mathbf{r}), \label{eq:p_term}
\end{equation}
where $A$ represents a star of symmetry-equivalent, nonzero,
simulation-cell reciprocal-lattice vectors ${\bf G}$, and $A^+$ is a
subset of $A$ that consists of one out of each $\pm {\bf G}$ pair.
The $\{a_A\}$ are optimizable parameters.  We used 46 stars of
${\bf G}$ vectors in $p$.

Including a backflow transformation in the trial wave function, the
Slater part of the wave function $S$ is evaluated at transformed
``quasiparticle'' coordinates ${\bf X}(\mathbf{R})=\mathbf{R}+{\bm
  \xi}(\mathbf{R})$, where
\begin{equation}
 {\bm \xi}_i(\mathbf{R})=\sum_{j\neq i}^{N} \eta(r_{ij})
 \mathbf{r}_{ij} + \sum_{j\neq i}^N {\bm \pi}({\bf r}_{ij})
\end{equation}
is the backflow displacement of electron $i$. $\eta$ is a cuspless,
smoothly truncated, isotropic polynomial function of minimum-image
electron-electron distance $r_{ij}$. The polynomial coefficients are
optimizable parameters, and are different for parallel- and
antiparallel-spin electrons \cite{Pablo}. The form of $\eta(r)$
is mathematically equivalent to that of the Jastrow $u(r)$ term
[Eq.\ (\ref{eq:u}), with $\Gamma=0$ for same-spin electrons and
  optimizable for opposite-spin electrons].  Typically we used
$N_\eta=8$ in the polynomial expansions. The ${\bm \pi}$ term has the
form of the gradient of a Jastrow $p$ term [Eq.\ (\ref{eq:p_term})]:
\begin{equation}
{\bm \pi}({\bf r}) = -\sum_A c_A \sum_{{\bf G} \in A^+} \sin({\bf G}
\cdot {\bf r}) \, {\bf G},
\end{equation}
where the $c_A$ are optimizable parameters.  As the gradient of a
scalar field, the ${\bm \pi}$ term is irrotational. We used 44 stars
of ${\bf G}$ vectors in ${\bm \pi}$. The backflow transformation
preserves the antisymmetry of the Slater wave function. The parameters
in the Jastrow factor and backflow function were optimized by variance
minimization and energy minimization.

\subsection{Crystal phase wave function}

The trial wave functions for our Wigner crystal calculations were of
later-Jastrow (SJ) form, apart from some test calculations with SJB
wave functions.  The orbitals in the Slater determinants consisted of
Gaussian functions centered on body-centered cubic (bcc) lattice sites
within the simulation cell, made periodic by summing over
simulation-cell images:
\begin{equation} \phi_{{\bf R}_\text{p}}({\bf r}) = \sum_{{\bf R}_\text{s}}
\exp\left(-C\left|{\bf r}-{\bf R}_\text{p}-{\bf
  R}_\text{s}\right|^2\right), \label{eq:crysorbs} \end{equation}
where ${\bf R}_\text{p}$ is a primitive-cell lattice point within the
supercell (which indexes the orbital) and ${\bf R}_\text{s}$ is a
simulation-cell lattice point.  In practice the sum contained only
those Gaussian functions $\exp\left(-C\left|{\bf r}-{\bf
  R}_\text{p}-{\bf R}_\text{s}\right|^2\right)$ whose value at the
closest point to ${\bf R}_\text{p}+{\bf R}_\text{s}$ in the
Wigner-Seitz simulation cell containing ${\bf r}$ was greater than
$10^{-7}$.  It was verified that truncating the sum in this manner
does not introduce statistically significant errors at the densities
considered in this work: see Table \ref{table:VMC_E_v_g} and note that
the effects of truncating the sum are reduced in larger simulation
cells; our production Wigner crystal calculations used cell sizes of
$N=64$, 216, and 512 electrons.

\begin{table}[htbp!]
\centering
\caption{Convergence of variational Monte Carlo (VMC) energy with
  respect to a parameter $g$ controlling the number of periodic images
  retained in the Wigner crystal orbitals of Eq.\ (\ref{eq:crysorbs}).
  We only retained those Gaussian functions $\exp\left(-C\left|{\bf
    r}-{\bf R}_\text{p}-{\bf R}_\text{s}\right|^2\right)$ whose value
  at the closest point of the Wigner-Seitz simulation cell containing
  ${\bf r}$ was greater than $10^{-g}$. Results are shown for a
  ferromagnetic bcc Wigner crystal at $r_\text{s}=90$ with $N=27$
  electrons in the simulation cell and Gaussian exponent $C=0.000148$
  a.u.  An SJB wave function optimized with $g=7$ was used in each
  case. \label{table:VMC_E_v_g}}
\begin{tabular}{lcc}
\hline \hline

$g$ & No.\ terms in Eq.\ (\ref{eq:crysorbs}) & SJB-VMC energy (Ha/el.)
\\

\hline

1 & 8 & $-0.0084224(3)$ \\

2 & 24 & $-0.00849361(7)$ \\

3 & 27 & $-0.00849268(7)$ \\

5 & 27 & $-0.00849263(7)$ \\

7 & 27 & $-0.00849253(7)$ \\

15 & 46 & $-0.00849254(7)$ \\

25 & 91 & $-0.00849253(6)$ \\

\hline \hline
\end{tabular}
\end{table}

\begin{figure}[htbp!]
\centering
\includegraphics[clip,height=5cm]{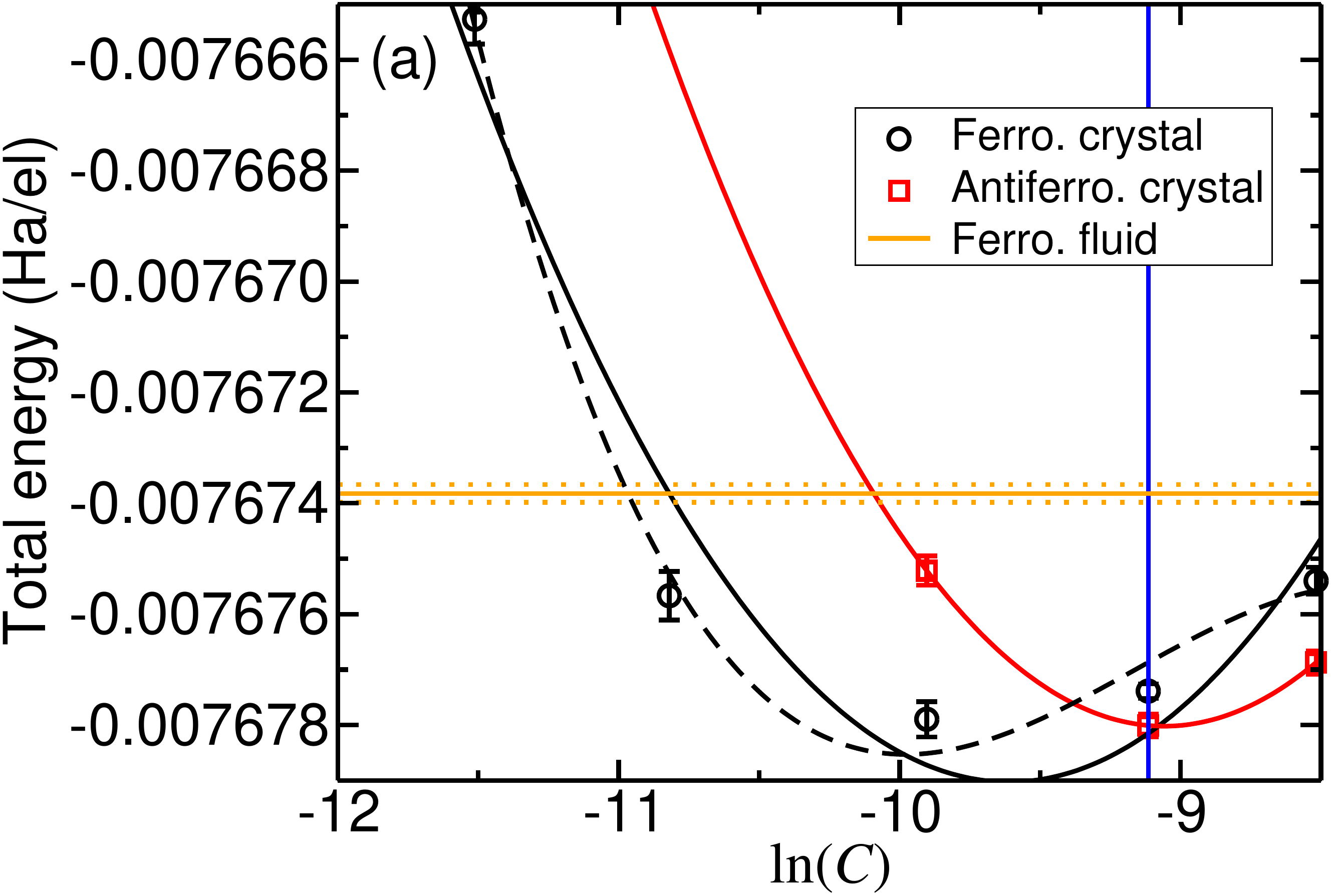}
\hfill \includegraphics[clip,height=5cm]{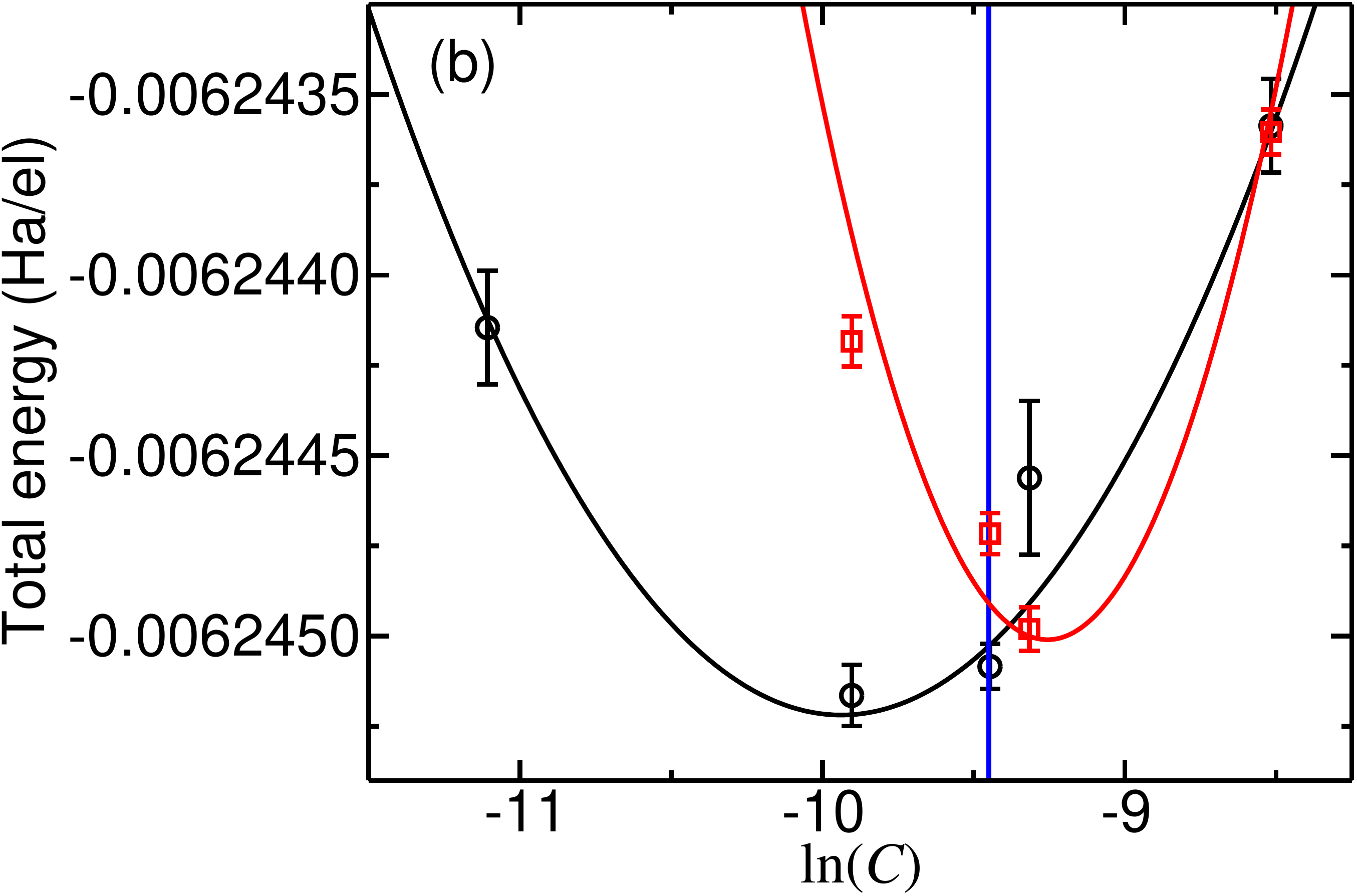}
\caption{DMC energy against logarithm of Gaussian exponent $C$ for
  ferromagnetic and antiferromagnetic bcc Wigner crystals (a) at
  $r_\text{s}=100$ and (b) at $r_\text{s}=125$, using SJ trial wave
  functions.  The energy has been extrapolated to zero time step and
  infinite system size in each case.  The solid curves show quadratic
  fits to the energies of ferromagnetic and antiferromagnetic crystals
  as functions of $\ln(C)$, while the dashed curve shows a quartic
  fit.  The horizontal line in (a) shows the twist-averaged
  paramagnetic Fermi fluid energy (with the dotted lines indicating
  error bars), while the vertical line shows the Gaussian exponent
  given by Eq.\ (\ref{eq:C_formula}). \label{fig:E_v_C_rs100}}
\end{figure}

In Fig.\ \ref{fig:E_v_C_rs100} we plot diffusion Monte Carlo (DMC)
energy against the logarithm of the Gaussian exponent $C$ for bcc
Wigner crystals at $r_\text{s}=100$ and $r_\text{s}=125$.  It is clear
that the formula \cite{Neil04}
\begin{equation} C=0.11r_\text{s}^{-3/2} \label{eq:C_formula} \end{equation}
provides near-optimal exponents, especially for antiferromagnetic
Wigner crystals.  By using Eq.\ (\ref{eq:C_formula}) we achieve
greater consistency between system sizes and densities than would
result from separately optimizing $C$ in each case.  At small system
size there is a tendency for $C$ to be underestimated relative to the
thermodynamic limit, to reduce the center-of-mass kinetic energy.  For
both ferromagnetic and antiferromagnetic crystals the error in the DMC
energy from using the formula is comparable with the statistical error
bars on the data, and is clearly much smaller than the difference
between the fluid and crystal energies, even at $r_\text{s}=100$,
close to the crystallization density.  A slightly more accurate
expression for the optimal Gaussian exponent for ferromagnetic
crystals would be $C_\text{ferro}=0.068r_\text{s}^{-3/2}$; however,
for consistency, we have used Eq.\ (\ref{eq:C_formula}) for all our
production calculations.  By contrast, within Hartree-Fock (HF) theory
the low-density Gaussian exponent is
$C_\text{HF}=1/(2r_\text{s}^{3/2})$ \cite{Neil04,Wigner_1938}.

Our Wigner crystal Jastrow factors were of the same form as the Fermi
fluid Jastrow factors described in the previous section, except
that we used 7 stars of reciprocal lattice vectors in the plane-wave
two-body term $p$, and we also used a plane-wave one-body term of the
form
\begin{equation}
 q(\mathbf{r})=\sum_B b_B \sum_{\mathbf{G} \in B^+} \cos(\mathbf{G}
 \cdot \mathbf{r}),
\end{equation}
where $B$ represents a star of symmetry-equivalent, nonzero
primitive-cell reciprocal-lattice vectors ${\bf G}$, and $B^+$ is a
subset of $B$ that consists of one out of each $\pm {\bf G}$ pair.
The $\{b_B\}$ are optimizable parameters.  We used 7 stars of ${\bf
  G}$ vectors in $q$.  The $q$ term has the symmetry of the Wigner
crystal lattice and allows a description of anisotropic warping of the
Gaussian orbitals.

For a ferromagnetic bcc Wigner crystal at $r_\text{s}=100$ with $N=64$
electrons, we performed test calculations using a backflow function of
the form described in Sec.\ \ref{sec:fluid_wf}, but with 8 stars in
the ${\bm \pi}$ term.  The VMC and DMC energies and the VMC variances
obtained with SJ and SJB wave functions, using either
Eq.\ (\ref{eq:C_formula}) or VMC energy minimization to determine the
orbital Gaussian exponent $C$, are shown in Table
\ref{table:SJ_and_SJB_ferro_rs100_N064}.  Optimizing the Gaussian
exponent and backflow function lower the VMC energy and variance.
However, we cannot assume that optimizing an overall wave function
leads to an improved nodal surface.  In fact the DMC energies obtained
using SJ and SJB wave functions and either Eq.\ (\ref{eq:C_formula})
or VMC energy minimization to determine $C$ are all in statistical
agreement with each other.  Because SJ quantum Monte Carlo (QMC)
calculations are much cheaper and allow us to explore larger system
sizes, we have used SJ wave functions in our production calculations.

\begin{table*}[htbp!]
\centering
\caption{VMC energy and variance and DMC energy for a ferromagnetic
  Wigner crystal at density parameter $r_\text{s}=100$ with $N=64$
  electrons, using SJ and SJB wave functions and using either
  Eq.\ (\ref{eq:C_formula}) or VMC energy minimization to determine
  the Gaussian orbital exponent
  $C$. \label{table:SJ_and_SJB_ferro_rs100_N064}}
\begin{tabular}{lcccc}
\hline \hline

Wave function & $C$ (a.u.) & VMC energy (Ha/el.) & VMC variance (Ha$^2$)
& DMC energy (Ha/el.) \\

\hline

SJ & $0.00011$ [Eq.\ (\ref{eq:C_formula})] & $-0.0076719(3)$ &
$0.00000329$ & $-0.00769616(8)$ \\

SJ & $0.0001553$ (opt.) & $-0.0076737(2)$ & $0.00000274$ &
$-0.0076957(2)$ \\

SJB & $0.00011$ [Eq.\ (\ref{eq:C_formula})] & $-0.0076786(2)$ &
$0.00000252$ & $-0.00769619(9)$ \\

SJB & $0.0001503$ (opt.) & $-0.0076803(2)$ & $0.00000201$ &
$-0.0076960(1)$ \\

\hline \hline
\end{tabular}
\end{table*}

\subsection{Finite-size effects in fluid phases}\label{sec:fs_fluid}

For twist averaging we used random twists $\{\mathbf{k}_\text{s}\}$
rather than a grid of twists for the following reasons. (i) Using
random twists is similar to adding three more dimensions to the
$3N$-dimensional integrals evaluated in QMC, and Monte Carlo
integration is efficient in high-dimensional spaces. (ii) A truly
systematic approach to twist averaging should use ${\bf k}_\text{s}$
in each supercell Brillouin zone (BZ) defined by the reoccupancies of
the plane-wave orbitals, and the corresponding results should be
weighted by the size of that BZ\@.  However, the complexity of the
nested BZs grows very rapidly with the number of electrons $N$
\cite{Lin}, making this approach infeasible for large system
sizes.  On the other hand, a regular grid-based approach effectively
gives a random sampling of the nested BZs. (iii) Monte Carlo sampling
of $\mathbf{k}_\text{s}$ is easily extensible: if the error is too
large, it can be reduced by including more random twists.

To twist average, we used the HF kinetic energy ($T_\text{HF}$) and
exchange energy ($X_\text{HF}$) as correlators and fit
\begin{eqnarray}
   E({\bf k}_\text{s}) & = & \langle E \rangle_\text{TA} +
   c\left[T_\text{HF}({\bf k}_\text{s}) - \left< T_\text{HF}
     \right>_\text{TA} \right] \nonumber \\ & & {} +
   d\left[X_\text{HF}({\bf k}_\text{s}) - \left< X_\text{HF}
     \right>_\text{TA} \right]
\end{eqnarray}
to the DMC energy per particle at a given system size, where ${\bf
  k}_\text{s}$ is the twist, and $\langle E \rangle_\text{TA}$, $c$,
and $d$ are fitting parameters.  The twist-averaged (TA) HF energies
$\left< T_\text{HF} \right>_\text{TA}$ and $\left< X_\text{HF}
\right>_\text{TA}$ are cheap to evaluate and were obtained using
billions of twists.

Analytical expressions have been derived for the leading-order
\cite{Chiesa_2006} and next-to-leading-order \cite{Drummond_2008}
systematic finite-size (FS) corrections to the energy per particle of
a 3D homogeneous electron gas in a finite periodic cell in which the
interaction between the particles is of Ewald form and the system is
assumed to be well-described by a SJ wave function.  The leading-order
and next-to-leading-order FS corrections to the energy per electron of
a Fermi fluid are
\begin{equation} \Delta E = \frac{\sqrt{3}}{2N r_\text{s}^{3/2}}
- \frac{C_\text{3D}}{\pi r_\text{s}^2 {(2N)}^{4/3}}
\left[{(1+\zeta)}^{2/3}+{(1-\zeta)}^{2/3}\right], \label{eq:fs_corrs}
\end{equation}
where $C_\text{3D}=5.083$ in a face-centered cubic simulation cell
(for fluid phases) and $5.086$ in a bcc simulation cell (for crystal
phases).  As explained in Ref.\ \onlinecite{Holzmann_2016}, backflow
correlations lead to additional, negative $O(r_\text{s}^{-2}N^{-1})$
FS corrections to the energy per particle, approximately given by
$\Delta E_\text{BF}=-T_\text{HF}/(3N)$, where $T_\text{HF}$ is the HF
kinetic energy per particle.  There is also a nonsystematic FS error
in the canonical ensemble TA energy per particle of a Fermi fluid due
to the incorrect shape of the TA Fermi surface; this error has an
envelope that decays as $O(r_\text{s}^{-2}N^{-4/3})$ \cite{Lin}.

It is reasonable to assume that Eq.\ (\ref{eq:fs_corrs}) holds
approximately for Wigner crystals, although the static structure
factor differs between fluids and crystals, so the corrections should
not really be exactly the same.

\begin{figure}[htbp!]
\centering
\includegraphics[clip,height=5cm]{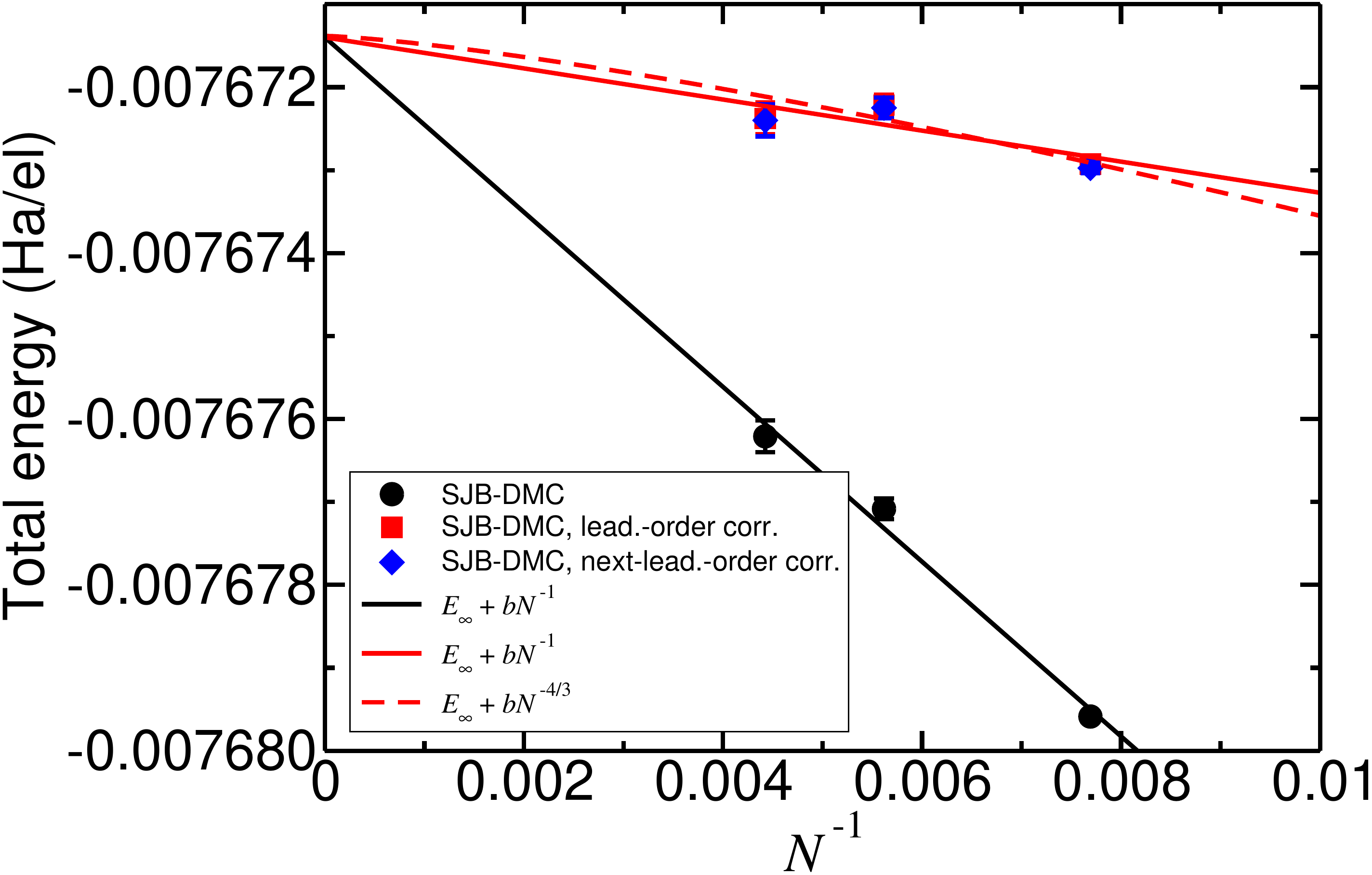}
\caption{TA SJB-DMC energy against inverse of system size for a
  paramagnetic Fermi fluid at $r_\text{s}=100$. Also shown are the
  effects of applying the first and second terms of
  Eq.\ (\ref{eq:fs_corrs}), and the effects of fitting various
  FS scaling laws to the resulting
  data. \label{fig:fs_corrs_para_fluid_rs100}}
\end{figure}

The TA SJB-DMC energy of a paramagnetic Fermi fluid at
$r_\text{s}=100$ is plotted against the reciprocal of system size in
Fig.\ \ref{fig:fs_corrs_para_fluid_rs100}.  The results of adding in
the leading-order correction of Eq.\ (\ref{eq:fs_corrs}) and the
leading-order plus next-to-leading-order FS corrections are
also shown.  At these low densities the next-to-leading-order
correction is negligible in comparison with the leading-order
correction (and also the backflow correction is relatively small).
Nevertheless, the leading-order FS correction does not remove
all the systematic FS errors.  Near the crystallization
density the single-determinant wave function form is increasingly
inappropriate, and it is possible that other FS effects may
be present in the correlations implicitly described by DMC\@.  It is
clear from Fig.\ \ref{fig:fs_corrs_para_fluid_rs100} that the
leading-order analytic corrections do not remove all FS
effects at $r_\text{s}=100$.  We therefore believe the most accurate
treatment of systematic FS effects is to extrapolate to
infinite system size using the $O(N^{-1})$ FS-error scaling
implied by the leading-order theory.

\begin{figure}[htbp!]
\centering
\includegraphics[clip,height=5cm]{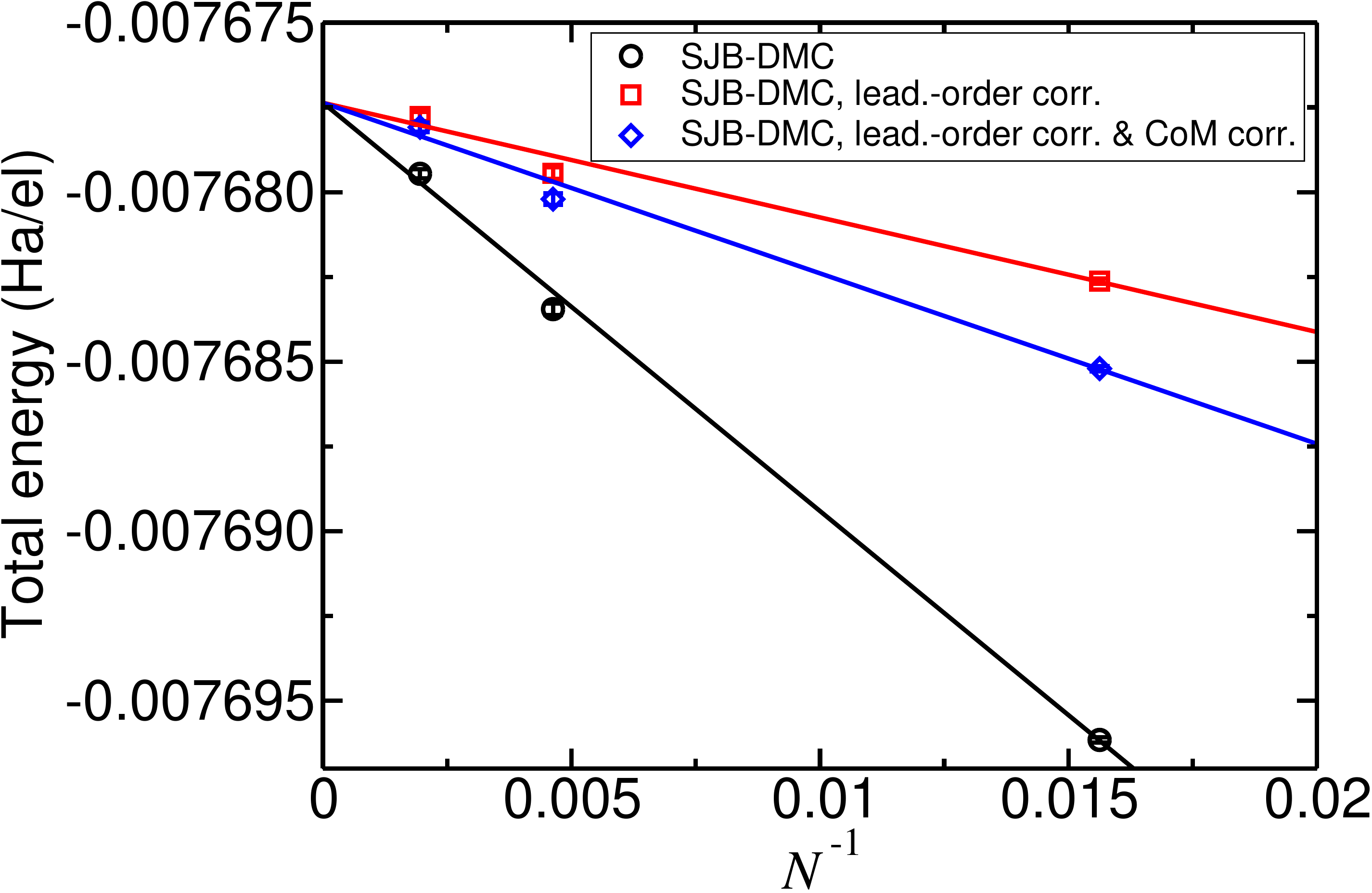}
\caption{SJ-DMC energy against inverse of system size for a
  ferromagnetic bcc Wigner crystal at $r_\text{s}=100$. Also shown are
  the effects of applying the first term of Eq.\ (\ref{eq:fs_corrs})
  and of subtracting an estimate of the center-of-mass kinetic energy,
  together with fits of $E(N)=E(\infty)+bN^{-1}$ to the resulting
  energy data. \label{fig:fs_corrs_ferro_crystal_rs100}}
\end{figure}

The SJ-DMC energy of a ferromagnetic Wigner crystal at
$r_\text{s}=100$ is plotted against system size in
Fig.\ \ref{fig:fs_corrs_ferro_crystal_rs100}.  It appears that the
leading-order $O(N^{-1})$ behavior is not completely eliminated by
either the long-range FS correction of
Eq.\ (\ref{eq:fs_corrs}) or the subtraction of the center-of-mass
kinetic energy.  Equation (\ref{eq:fs_corrs}) was derived for a fluid
phase; the static structure factor and long-range two-body Jastrow
factor are different in a crystal phase, which would lead to a
different prefactor.  Once again the best policy would appear to be to
regard the theory of FS effects as providing the appropriate
scaling law to fit to the data.

\begin{figure}[htbp!]
    \centering
    \begin{tabular}{c c}
    \includegraphics[clip,scale=0.21]{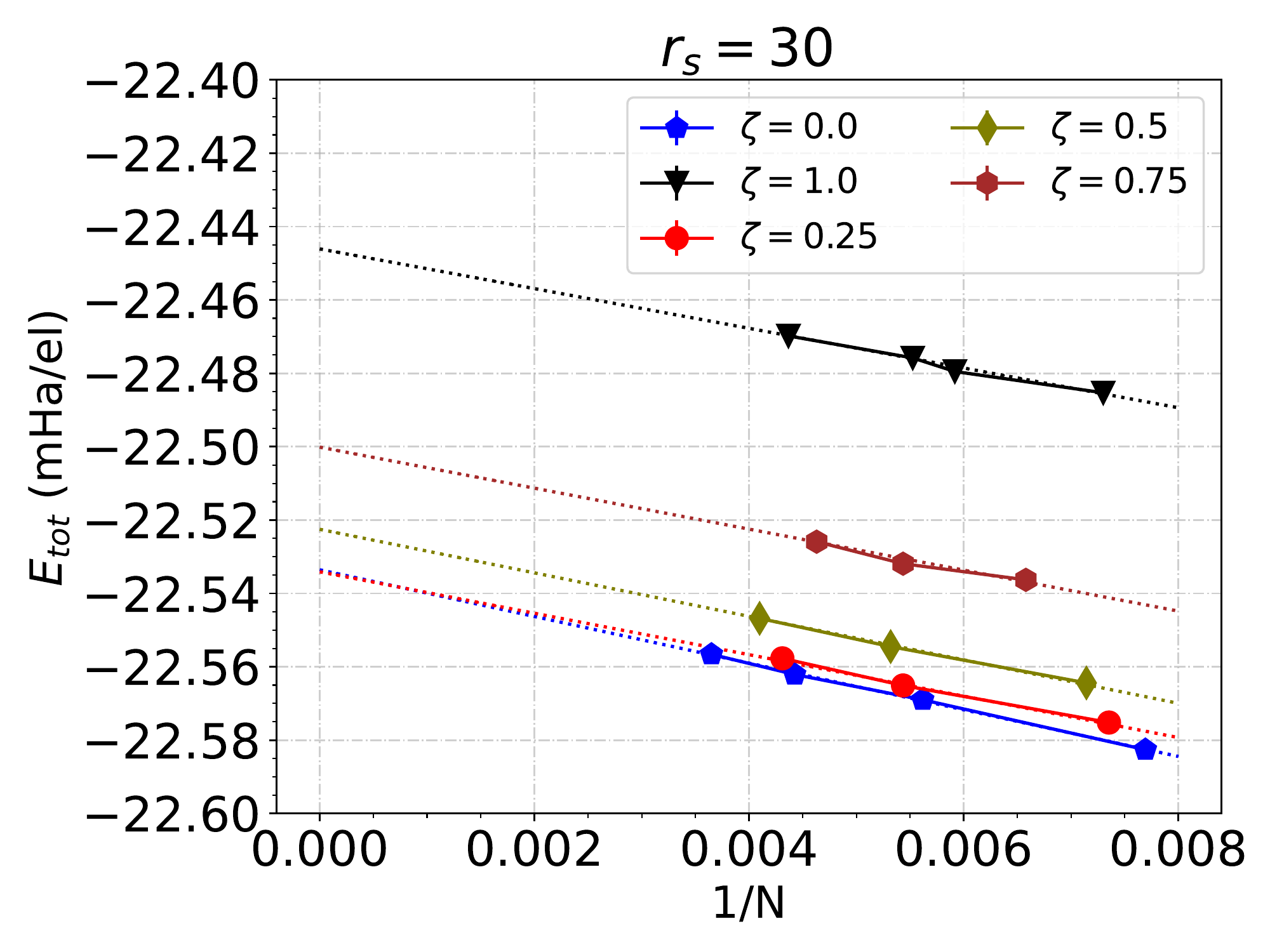}&
    \includegraphics[clip,scale=0.21]{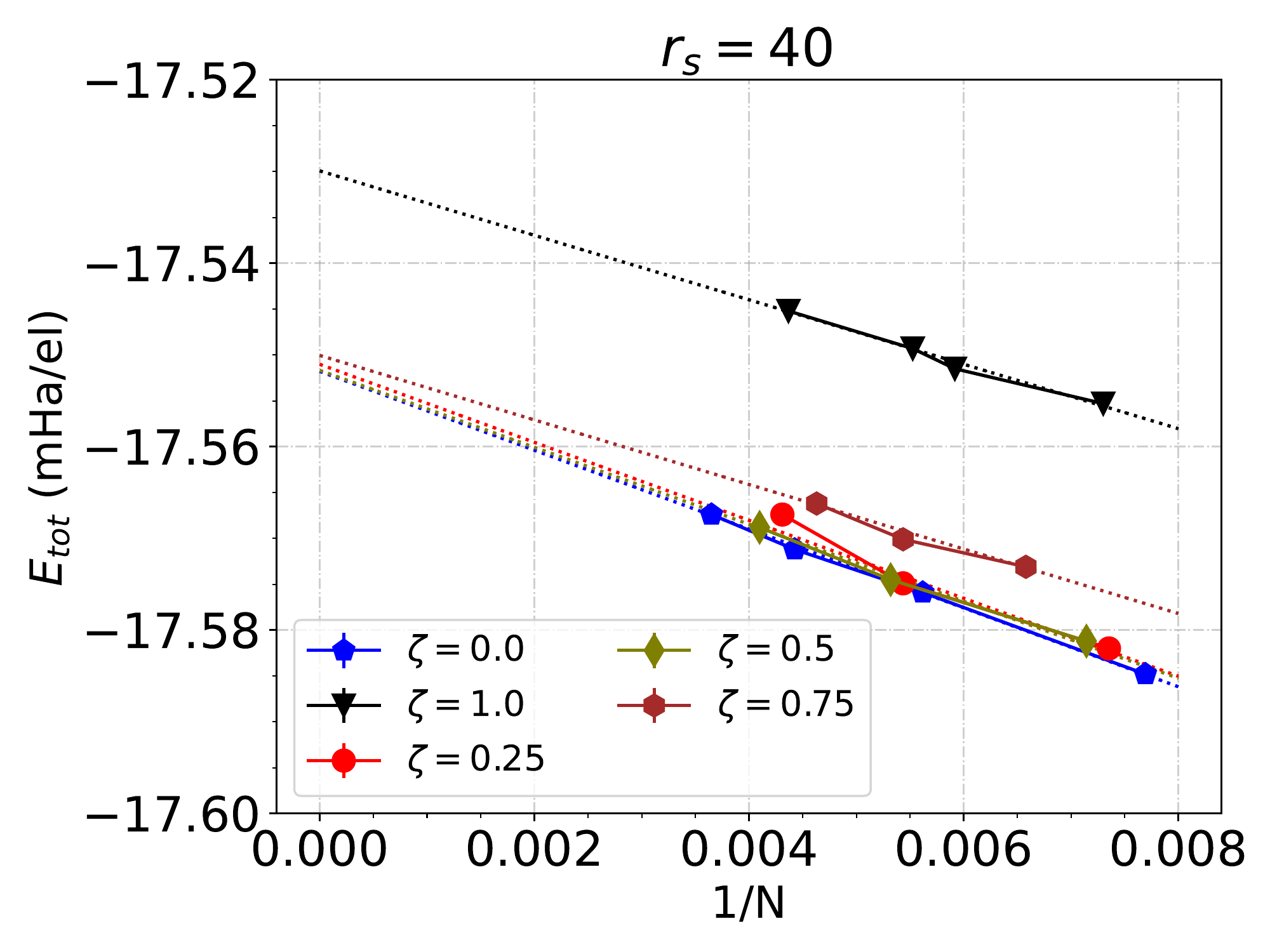}\\
    \includegraphics[clip,scale=0.21]{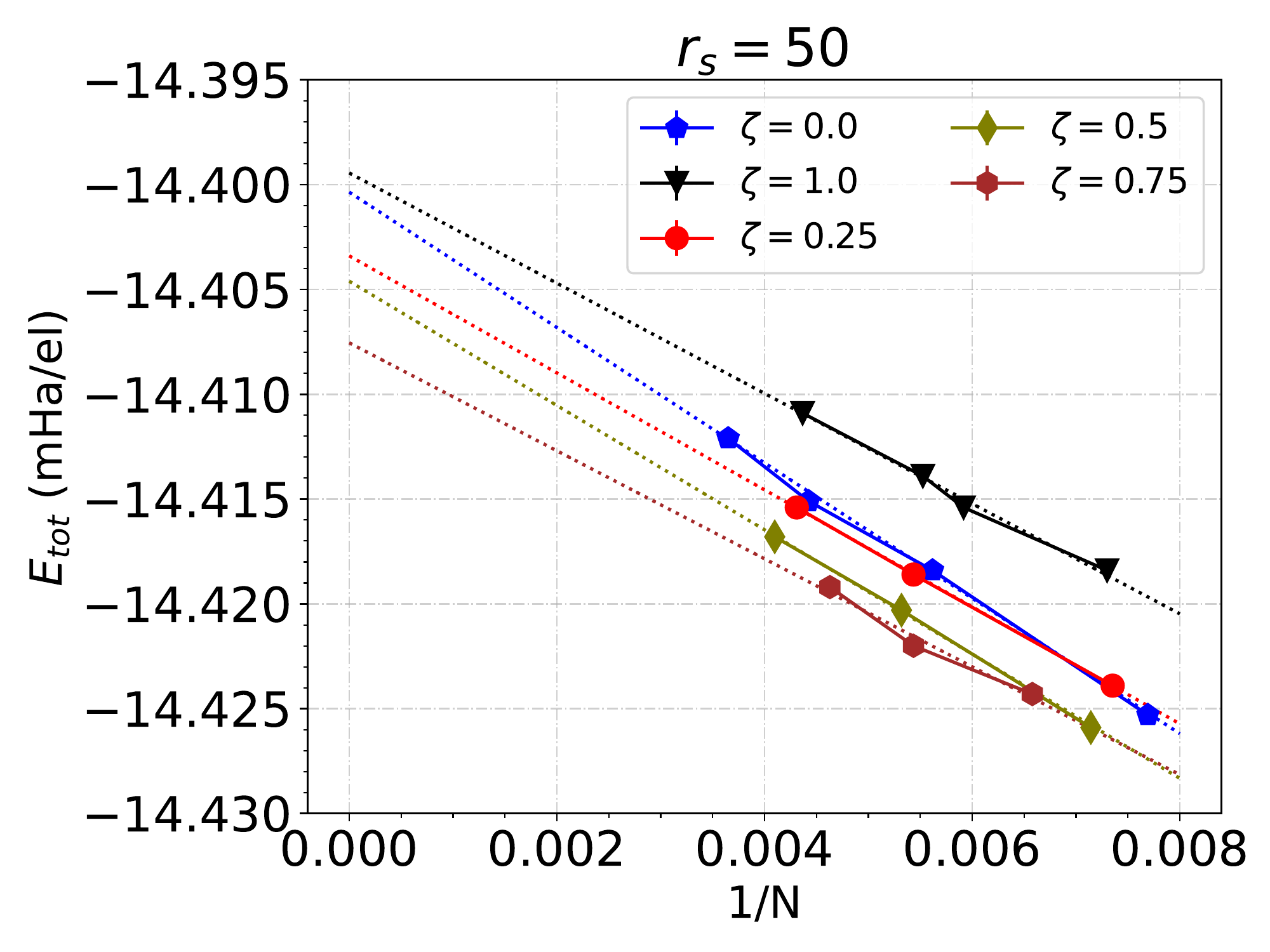}&
    \includegraphics[clip,scale=0.21]{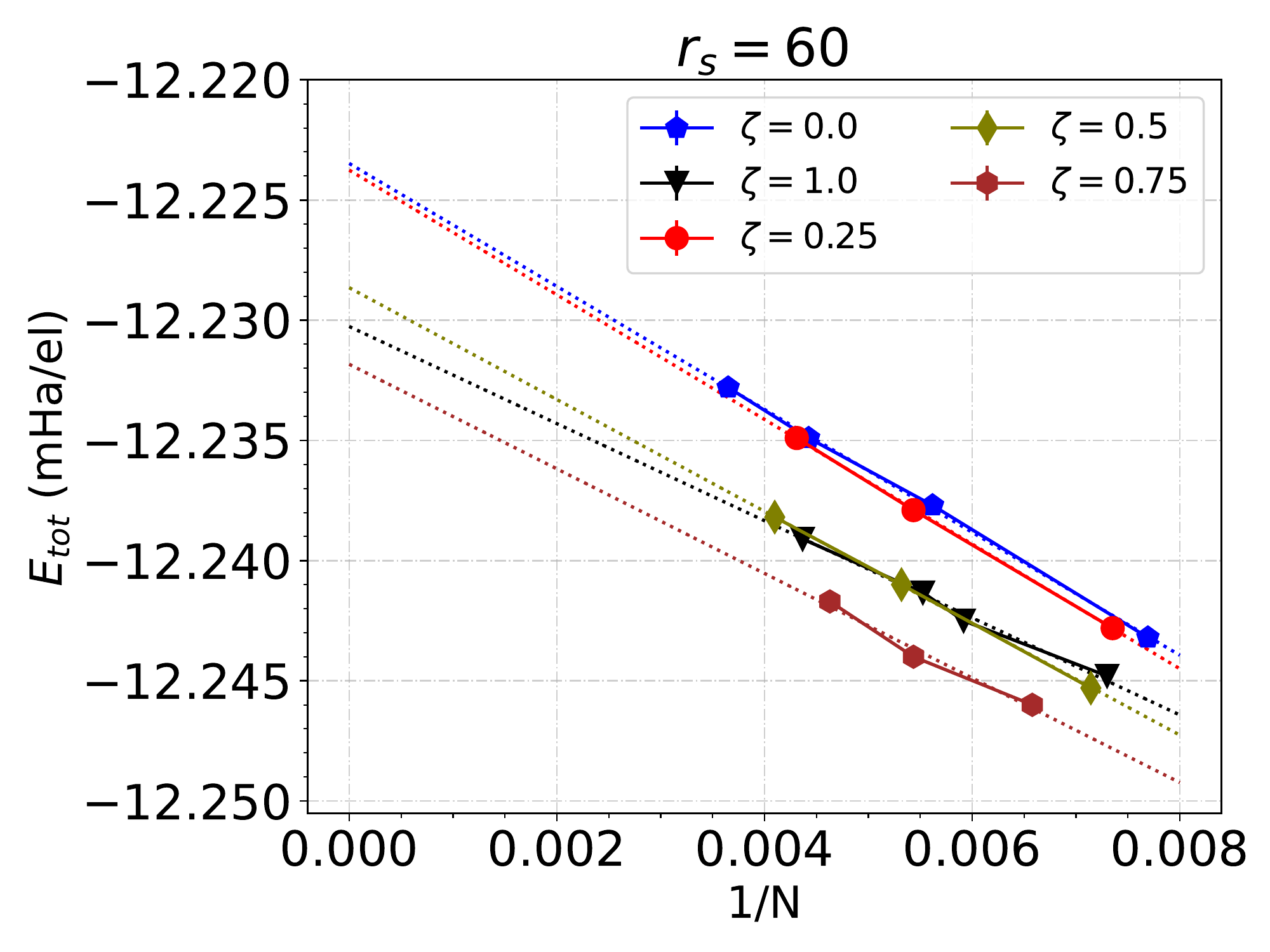}\\
    \includegraphics[clip,scale=0.21]{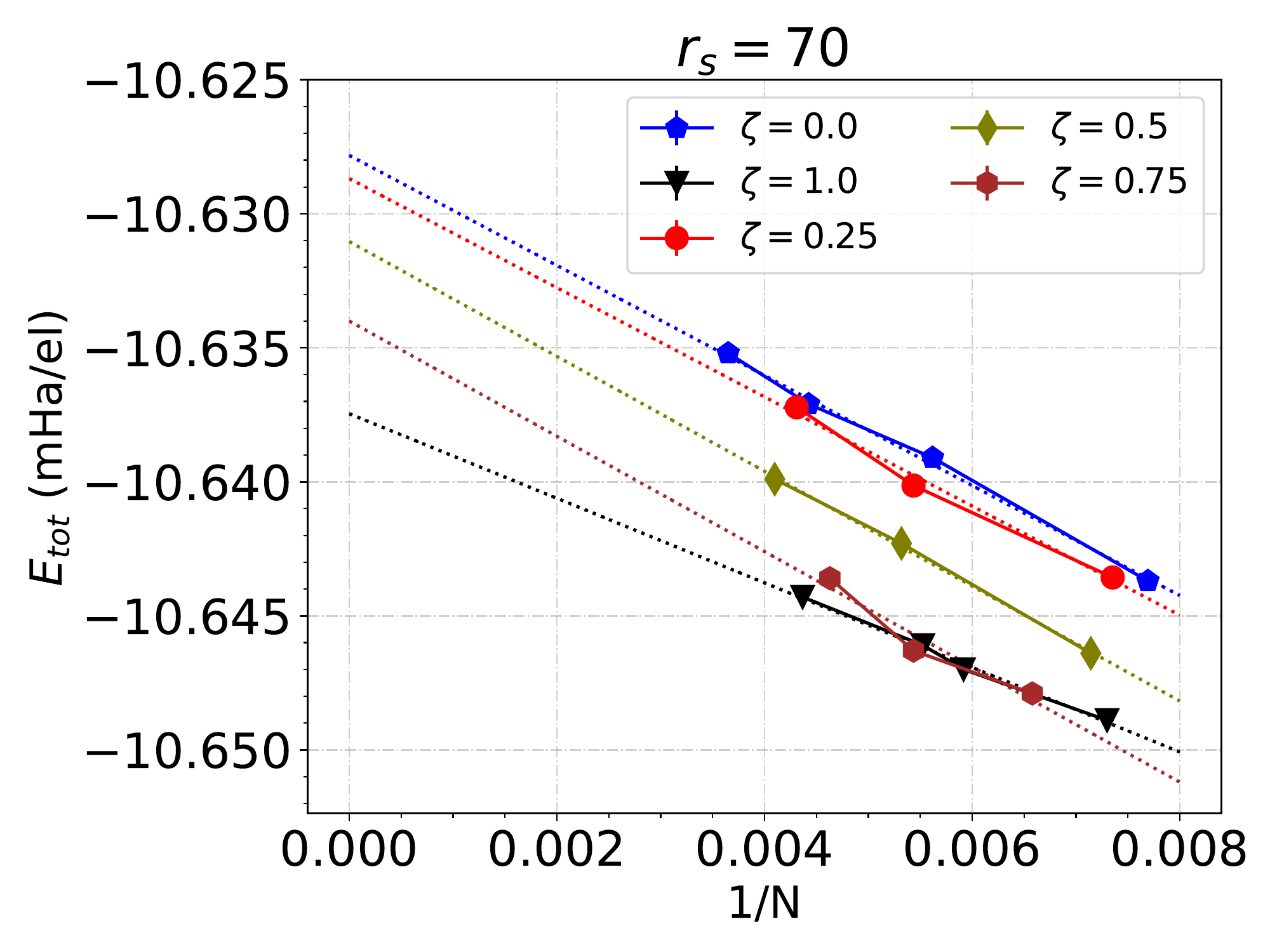}&
    \includegraphics[clip,scale=0.21]{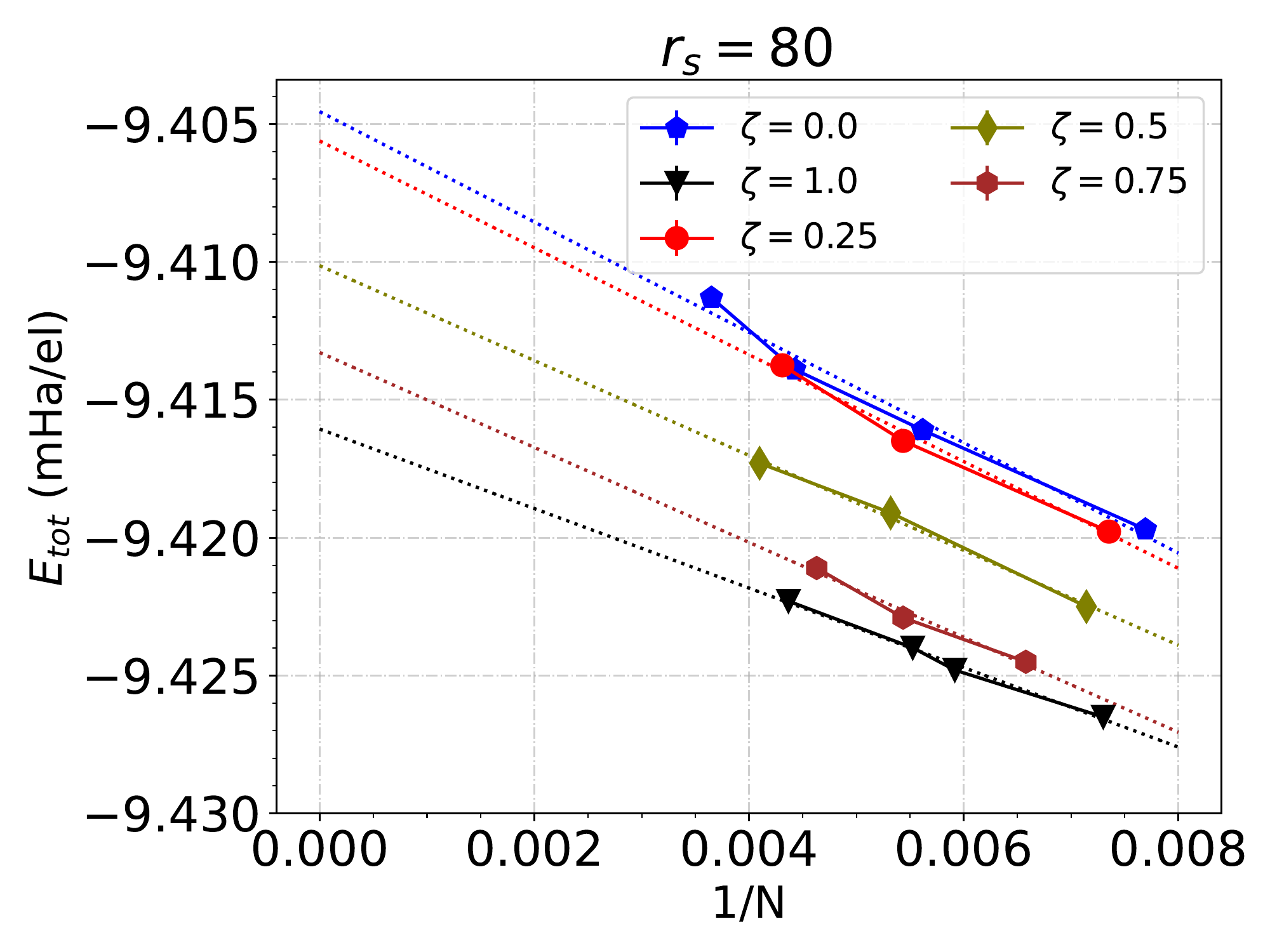}\\
    \end{tabular}
    \caption{\label{DMCN} TA Fermi fluid SJ-DMC energies as functions
      of system size $N$ at different spin polarizations $\zeta$. The
      dotted line represents the extrapolation to infinite system
      size. These SJ-DMC energies have not been extrapolated to zero
      time step.}
\end{figure}

SJ-DMC energies as functions of system size $N$ for different spin
polarizations ($\zeta=0$, 0.25, 0.5, 0.75, and 1) are illustrated in
Fig.\ \ref{DMCN}.
The SJ-DMC energies in the infinite system size limit for different
densities and polarizations are listed in Table \ref{Ezeta}.
\begin{table*}[!htbp]
\centering
\caption{\label{Ezeta} TA SJ-DMC energy of the Fermi fluid
  extrapolated from different system sizes ($130 < N < 274$) with 300
  random twists at each $N$ to the thermodynamic limit.  Energies are
  in mHa/el. The numbers in parentheses indicate statistical
  errors. The DMC energies were not extrapolated to zero time step.}
 \begin{tabular}{lccccc}
 \hline \hline

 \diagbox{$r_\text{s}$}{$\zeta$} & 0.0 & 0.25 & 0.50 & 0.75 & 1.0 \\

 \hline

 30 & $-22.5336(7)$ & $-22.534(3)$  & $-22.522(1)$  & $-22.500(4)$ &
 $-22.446(3)$ \\

 40 & $-17.5518(4)$ & $-17.551(4)$  & $-17.552(1)$  & $-17.550(2)$ &
 $-17.523(1)$ \\

 50 & $-14.4003(4)$ & $-14.4034(1)$ & $-14.4046(4)$ & $-14.407(2)$ &
 $-14.399(1)$  \\

 60 & $-12.2235(2)$ & $-12.2237(1)$ & $-12.2286(1)$ & $-12.231(2)$ &
 $-12.2303(8)$ \\

 70 & $-10.6278(5)$ & $-10.629(1)$  & $-10.6310(4)$ & $-10.634(3)$ &
 $-10.6374(5)$ \\

 80 & $-9.404(1)$   & $-9.406(1)$   & $-9.4101(6)$  & $-9.413(1)$  &
 $-9.4161(5)$  \\

\hline\hline
\end{tabular}
\end{table*}

\begin{figure}[htbp!]
    \centering
    \begin{tabular}{c c}
    \includegraphics[clip,scale=0.21]{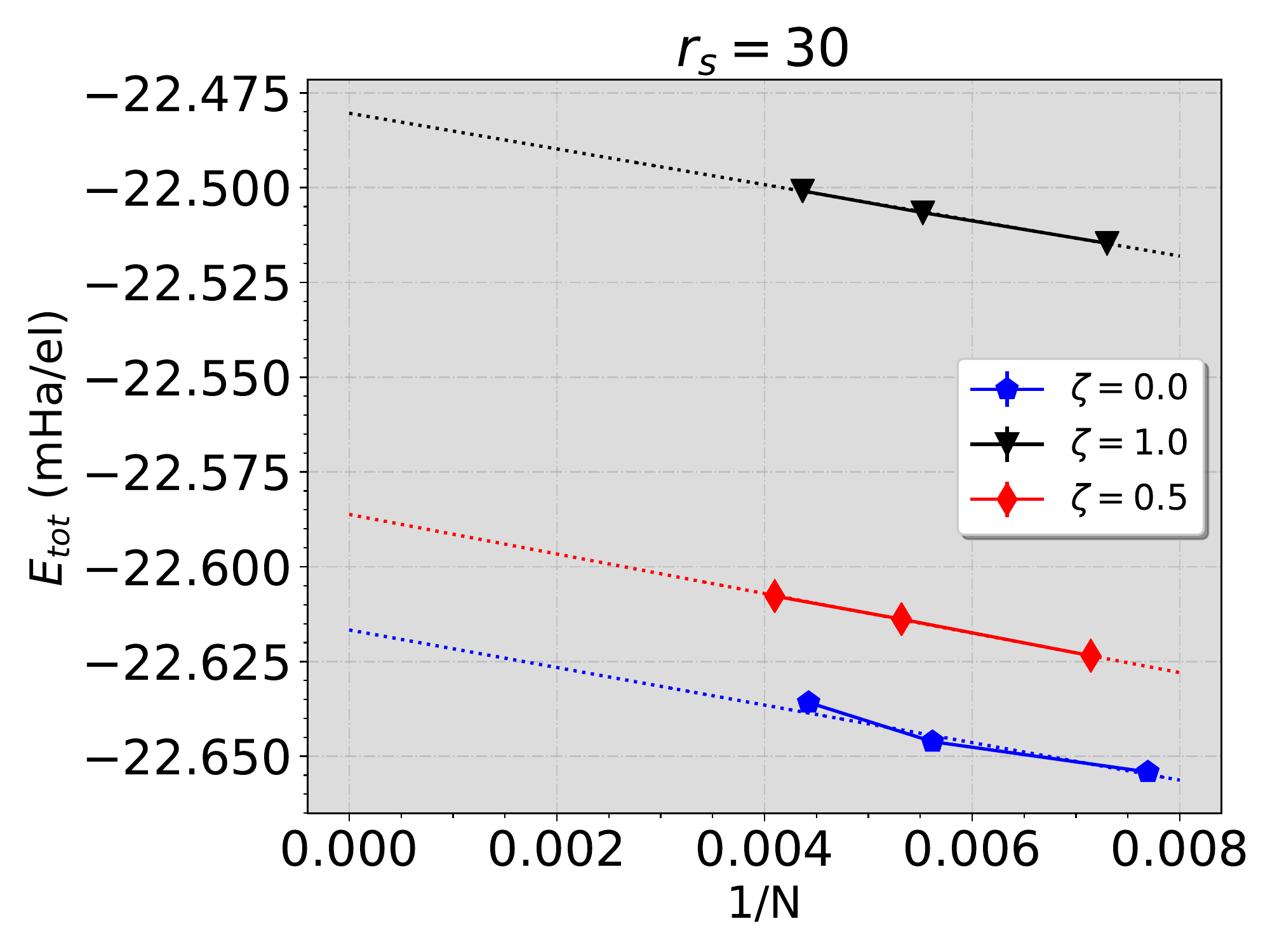}&
    \includegraphics[clip,scale=0.21]{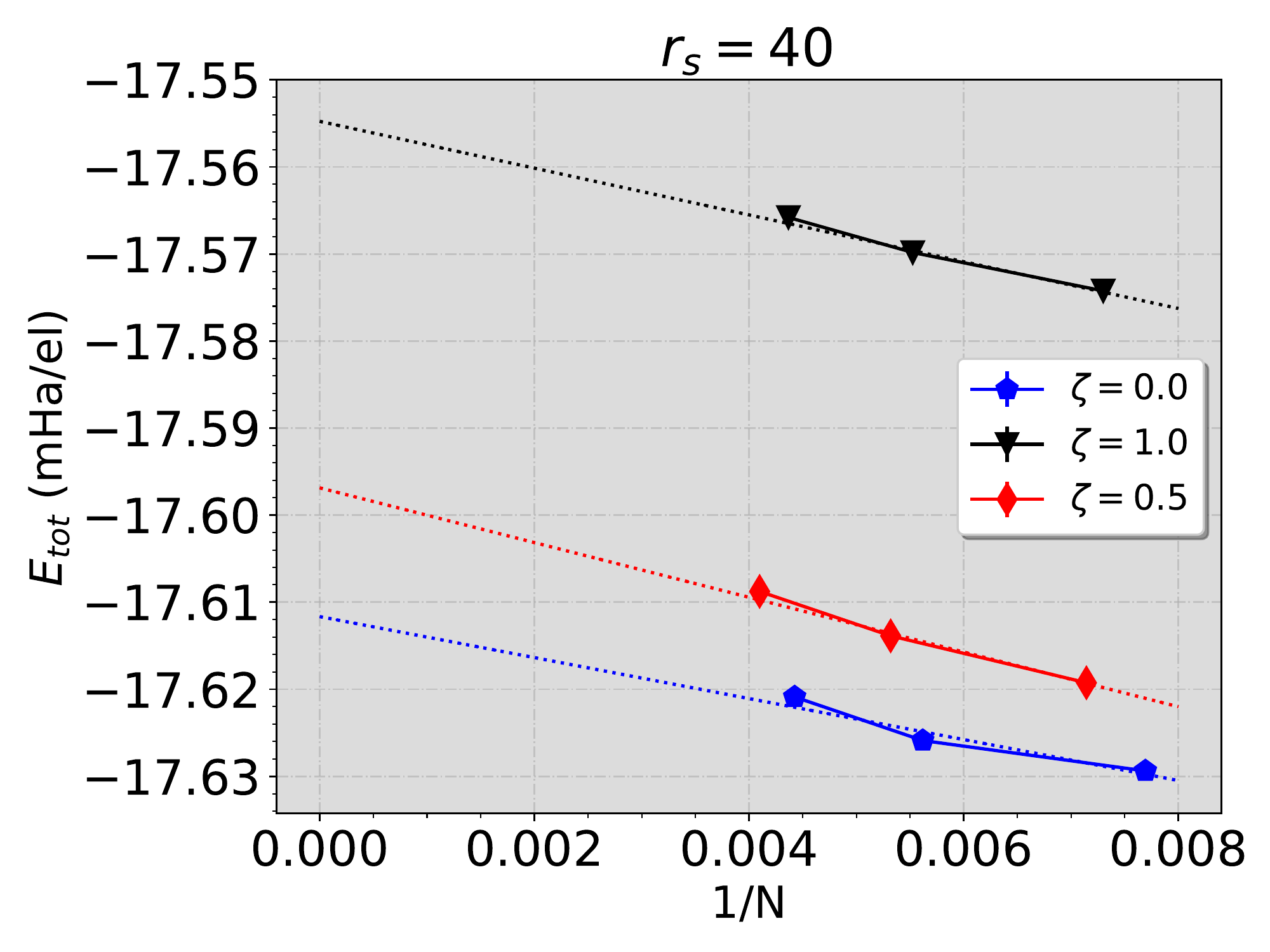}\\
    \includegraphics[clip,scale=0.21]{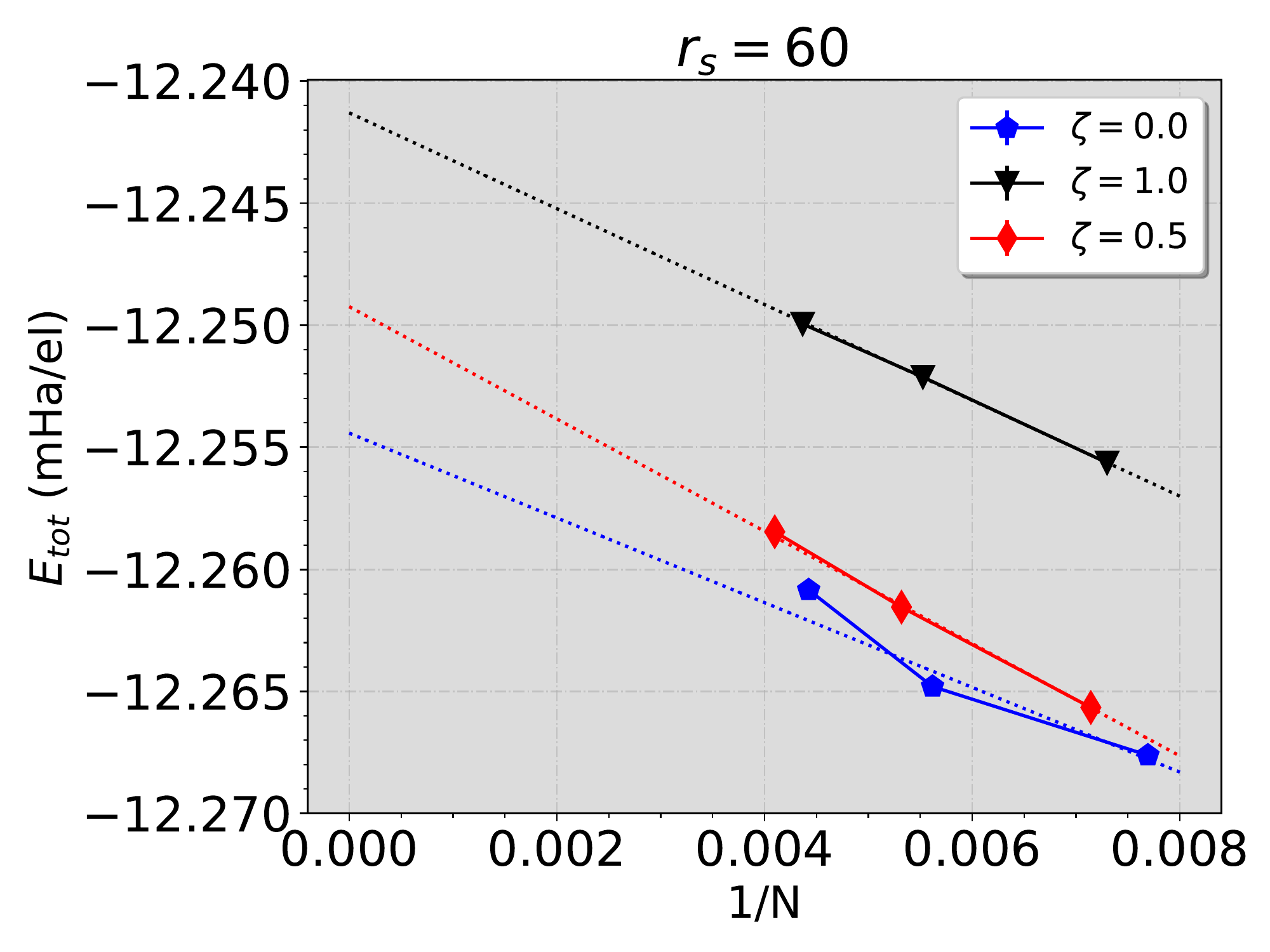}&
    \includegraphics[clip,scale=0.21]{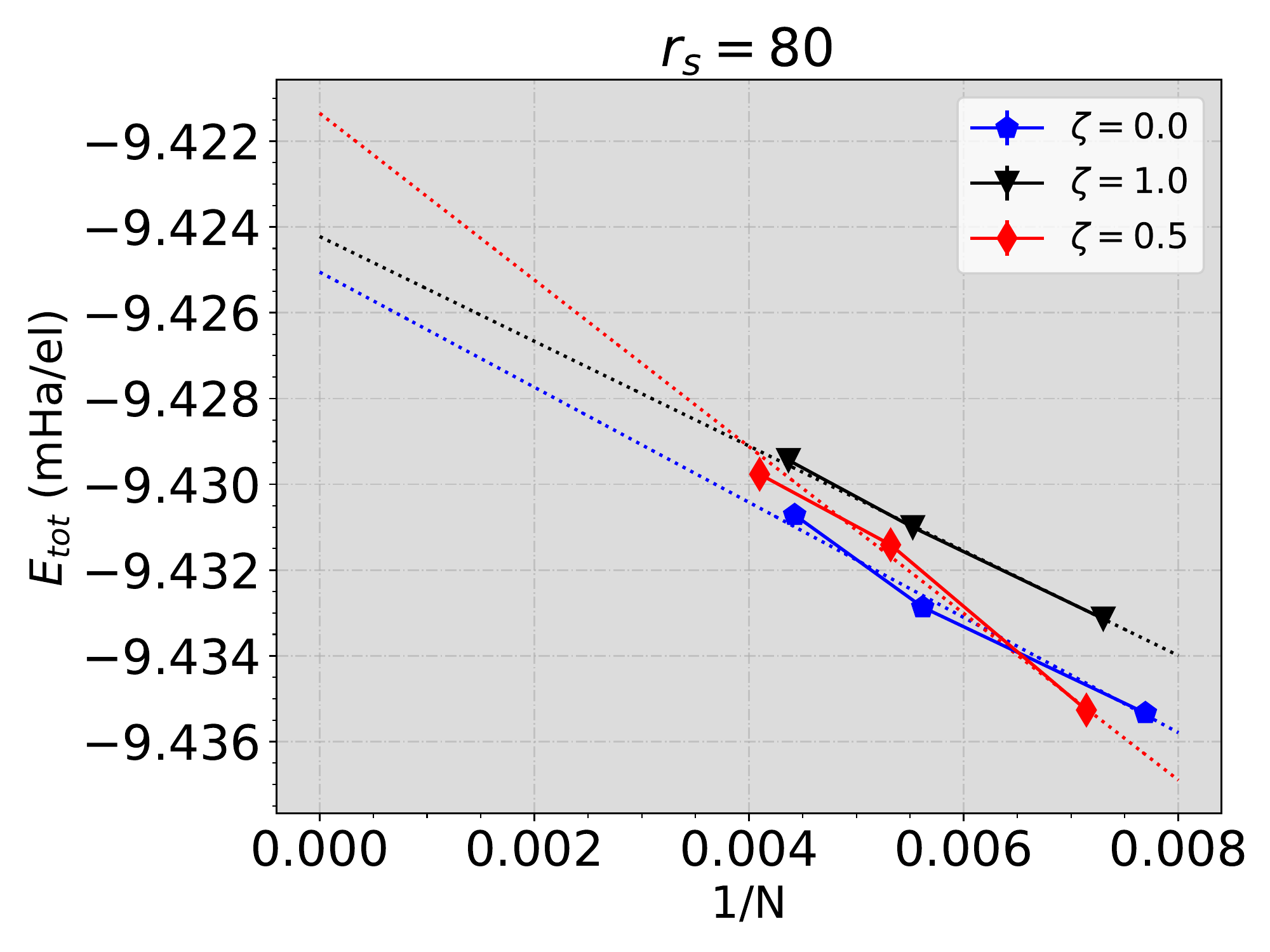}\\
    \includegraphics[clip,scale=0.21]{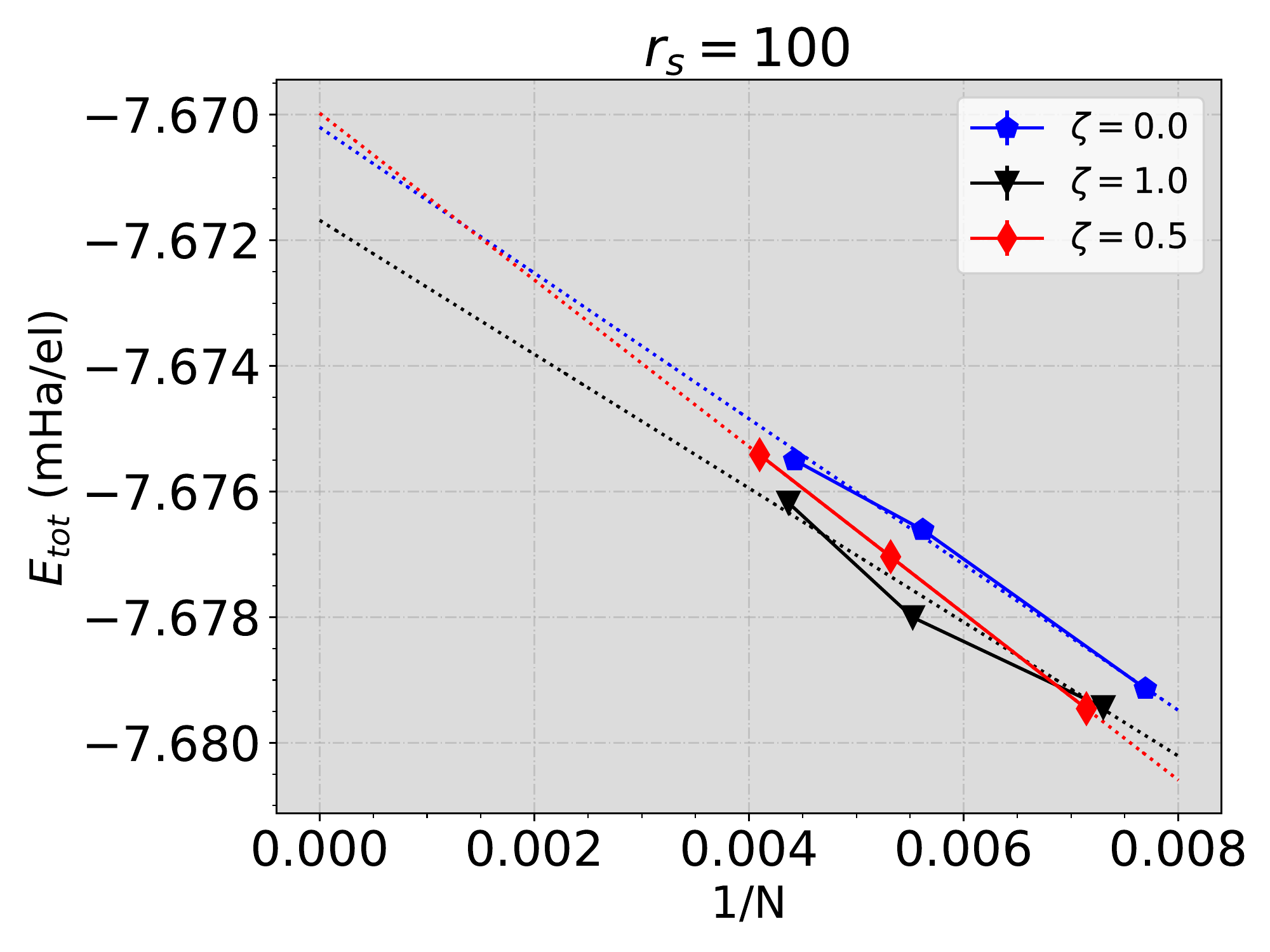}\\
    \end{tabular}
    \caption{\label{DMCNB} TA SJB-DMC energies against system size $N$
      for different spin polarizations $\zeta$. The dotted lines
      represent extrapolations to infinite system size.  The DMC
      energies were not extrapolated to zero time step.}
\end{figure}
The TA SJB-DMC energies of the fluid phase for various density and
polarization are plotted in Fig.\ \ref{DMCNB} against system size. The
SJB-DMC energies extrapolated to infinite system are presented in
Table \ref{EzetaB}.
\begin{table}[!htbp]
\centering
\caption{\label{EzetaB} TA SJB-DMC energies of the Fermi fluid
  extrapolated to the thermodynamic limit from different system sizes
  ($130 < N < 274$).  Energies are in mHa/el. The numbers in
  parentheses indicate statistical errors. The DMC energies were not
  extrapolated to zero time step.}
 \begin{tabular}{lccc}
 \hline\hline

 \diagbox{$r_\text{s}$}{$\zeta$} & 0.0 & 0.50 & 1.0   \\

 \hline

  30  & $-22.617(8)$ & $-22.5862(5)$  & $-22.4804(6)$ \\

  40  & $-17.612(4)$ & $-17.597(2)$   & $-17.555(2)$ \\

  60  & $-12.254(3)$ & $-12.2492(4)$  & $-12.2413(1)$ \\

  80  & $-9.4250(9)$ & $-9.421(1)$    & $-9.4242(2)$ \\

  100 & $-7.6702(4)$ & $-7.669976(7)$ & $-7.6717(9)$ \\

 \hline\hline
 \end{tabular}
\end{table}

\subsection{Finite-size effects in crystal phases}

In our broken-symmetry model of a Wigner crystal there is an
additional FS error due to the center-of-mass kinetic energy.

At low density, individual electrons occupy Gaussian orbitals
$\phi_{{\bf R}_\text{p}}({\bf r})=\exp(-C|{\bf r}-{\bf
  R}_\text{p}|^2)$, where ${\bf R}_\text{p}$ is a bcc primitive-cell
lattice vector, and $C$ is a Gaussian exponent.  Let ${\bf s}$ be the
offset to the center-of-mass position.  Assuming a rigid displacement
of the lattice by ${\bf s}$, the center-of-mass wave function is
$\Phi({\bf s})=\exp(-NCs^2)$.  The resulting center-of-mass kinetic
energy is
\begin{equation} T_\text{CM} = -\frac{1}{2N}\frac{\int \Phi^* \nabla_{\bf s}^2
\Phi \, d^3{\bf s}}{\int |\Phi|^2 \, d^3{\bf s}} =
  \frac{3C}{2}.  \end{equation} Hence the center of mass kinetic
energy per particle falls off as $3C/(2N)$.  There is therefore an
additional FS correction, on top of those discussed in
Sec.\ \ref{sec:fs_fluid}, to be applied to the energy per particle of
a Wigner crystal:
\begin{equation} \Delta T_\text{CM} = -\frac{T_\text{CM}}{N} = -\frac{3C}{2N}
\approx -\frac{0.33}{2 N r_\text{s}^{3/2}}, \end{equation} where in
the last step we have inserted the approximate expression for the
Gaussian exponent $C$ in a bcc Wigner crystal obtained by minimizing
the DMC energy in a large simulation cell \cite{Neil04},
Eq.\ (\ref{eq:C_formula}).  The center-of-mass kinetic energy
correction partially offsets the leading-order FS correction
of Eq.\ (\ref{eq:fs_corrs}).

Where the crystal orbitals are highly localized within the supercell,
twist averaging cannot have much effect on the energy per particle.
If the simulation-cell Bloch vector ${\bf k}_\text{s}$ is nonzero then
the crystal orbitals are
\begin{equation} \phi_{{\bf R}_\text{p}}({\bf r}) = \sum_{{\bf R}_\text{s}}
\exp\left(-C\left|{\bf r}-{\bf R}_\text{p}-{\bf
  R}_\text{s}\right|^2\right) \exp\left(i{\bf k}_\text{s} \cdot {\bf
  R}_\text{s}\right), \label{eq:crystal_bloch_orbs} \end{equation}
where ${\bf R}_\text{p}$ is a primitive-cell lattice point within the
supercell (which indexes the orbital) and ${\bf R}_\text{s}$ is a
simulation-cell lattice point. This is the usual prescription for
creating Bloch orbitals from localized functions; one can easily check
that $\phi_{{\bf R}_\text{p}}({\bf r}+{\bf R}_\text{s}')=\exp(i{\bf
  k}_\text{s} \cdot {\bf R}_\text{s}') \phi_{{\bf R}_\text{p}}({\bf
  r})$.

For a large simulation cell at low density, at most one of the
Gaussian functions $\exp\left(-C\left|{\bf r}-{\bf R}_\text{p}-{\bf
  R}_\text{s}\right|^2\right)$ in Eq.\ (\ref{eq:crystal_bloch_orbs})
is non-negligible at any given point ${\bf r}$ in the simulation cell.
So the $\exp(i{\bf k}_\text{s} \cdot {\bf R}_\text{s})$ factor just
contributes an unobservable phase to each orbital within the
simulation cell.  Hence we do not twist average our crystal energies.

\begin{figure}[htbp!]
    \centering
    \begin{tabular}{c c}
    \includegraphics[clip,scale=0.21]{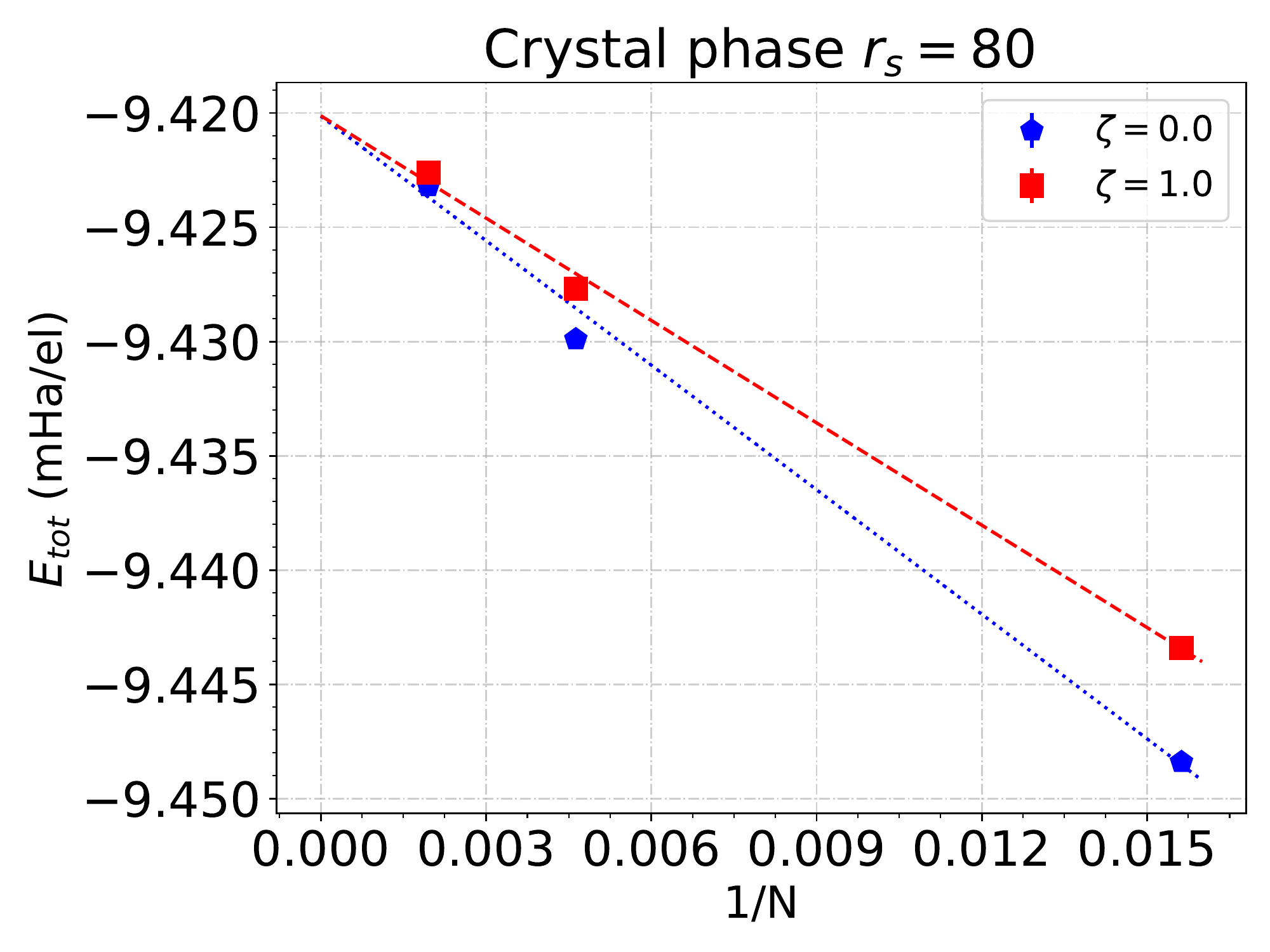}&
    \includegraphics[clip,scale=0.21]{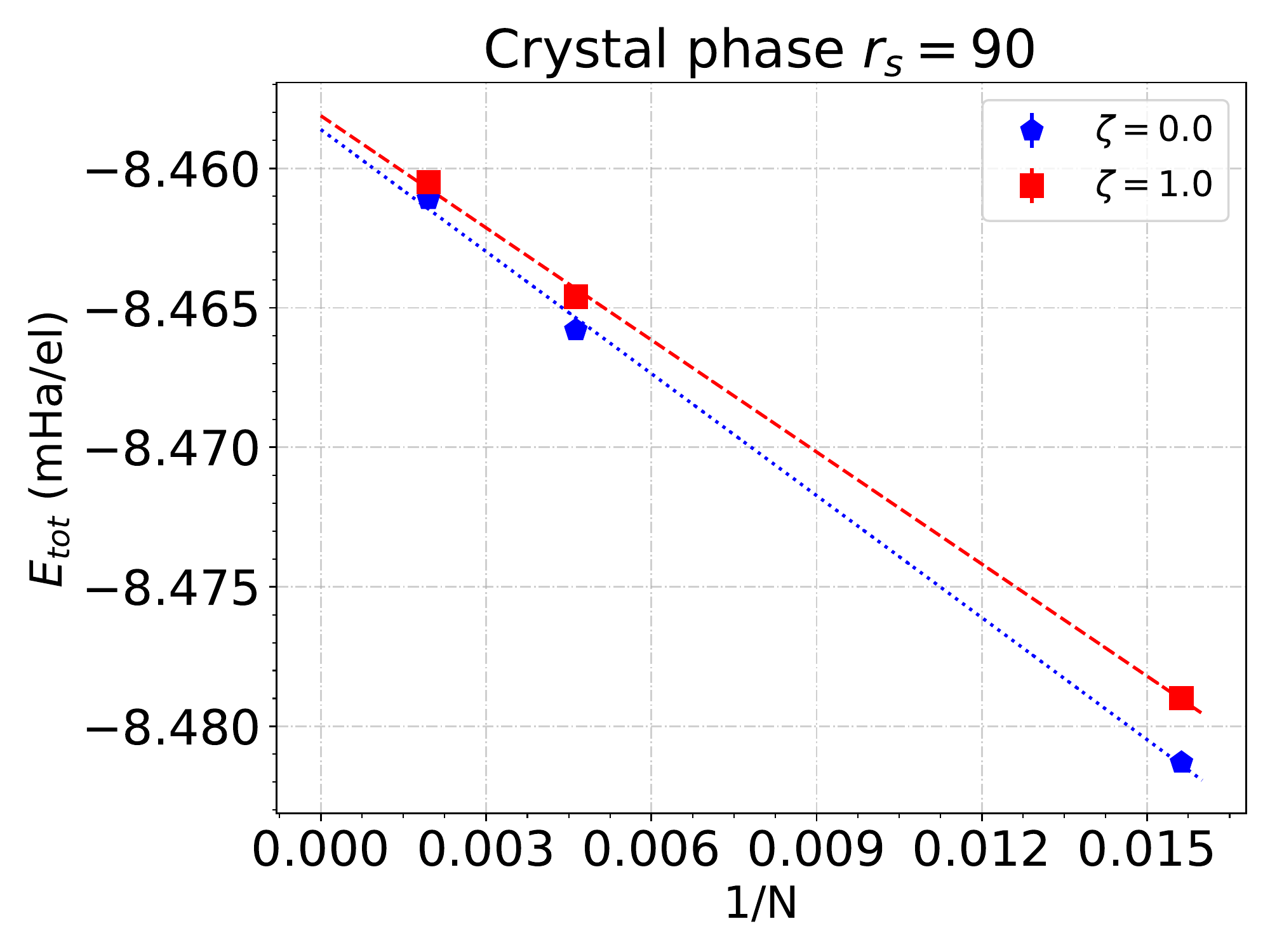}\\
    \includegraphics[clip,scale=0.21]{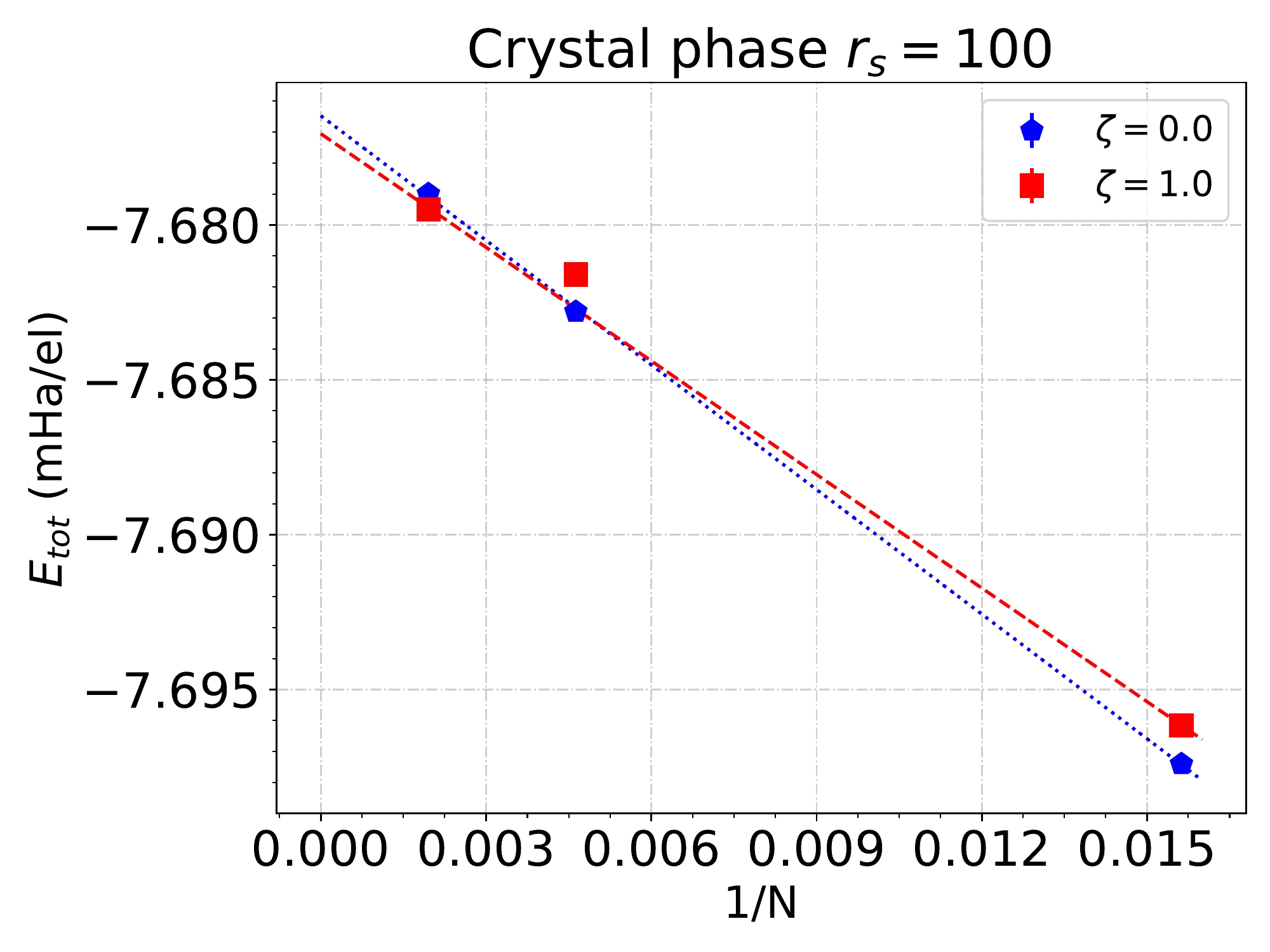}&
    \includegraphics[clip,scale=0.21]{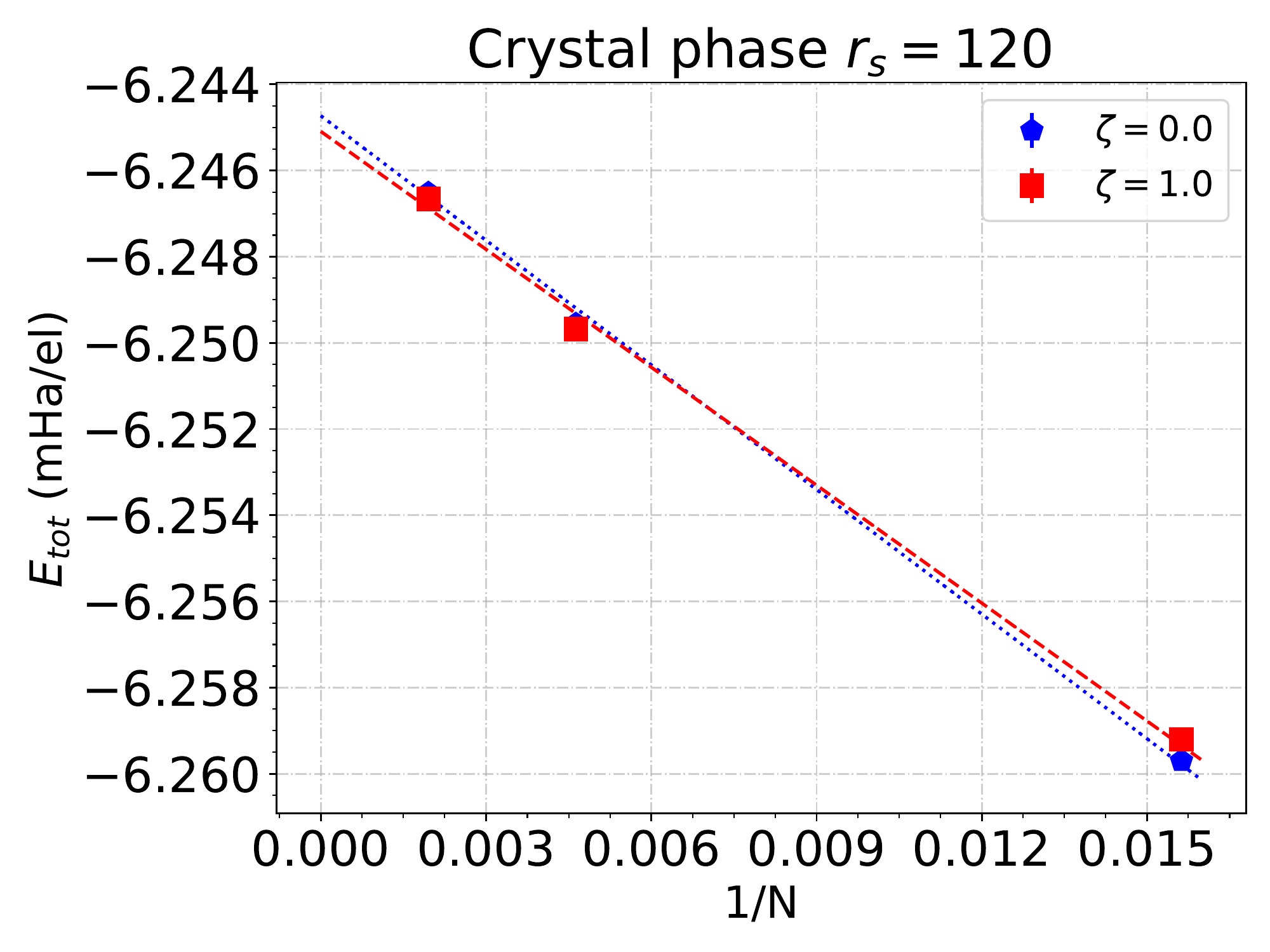}\\
    \end{tabular}
    \caption{\label{fs_wigner} SJ-DMC energy against the inverse of
      system size $N$ for the crystal phase. The dotted line shows the
      extrapolation to infinite system size. $\zeta=0$ and 1 are the
      spin polarizations of the system (antiferromagnetic and
      ferromagnetic crystals, respectively). The DMC energies have
      been extrapolated to zero time step.}
\end{figure}

The SJ-DMC energies of ferromagnetic Wigner crystals at different
densities $r_\text{s}=80$, 90, 100, and 125 as functions of system
size are plotted in Fig.\ \ref{fs_wigner}.  The energies extrapolated
to infinite system size are listed in Table
\ref{DMC_Crystal_therm_lim}.

\begin{table}[!htbp]
\centering
\caption{\label{DMC_Crystal_therm_lim} SJ-DMC energy of the crystal
  phase extrapolated to zero time step and the thermodynamic limit of
  infinite system size.  The numbers in parentheses indicate
  statistical and fitting errors.}
 \begin{tabular}{lcc}
 \hline\hline

& \multicolumn{2}{c}{Energy (mHa/el.)} \\

\raisebox{2ex}[0pt]{$r_\text{s}$} & Antiferromagnetic & Ferromagnetic \\

 \hline

  80   & $-9.4200(2)$  & $-9.41999(9)$ \\

  90   & $-8.4585(1)$  & $-8.4581(1)$ \\

  100  & $-7.6769(1)$  & $-7.6774(1)$ \\

  125  & $-6.24472(6)$ & $-6.24508(6)$ \\

 \hline\hline
 \end{tabular}
\end{table}

\begin{figure}[htbp!]
    \centering
    \includegraphics[clip,scale=0.5]{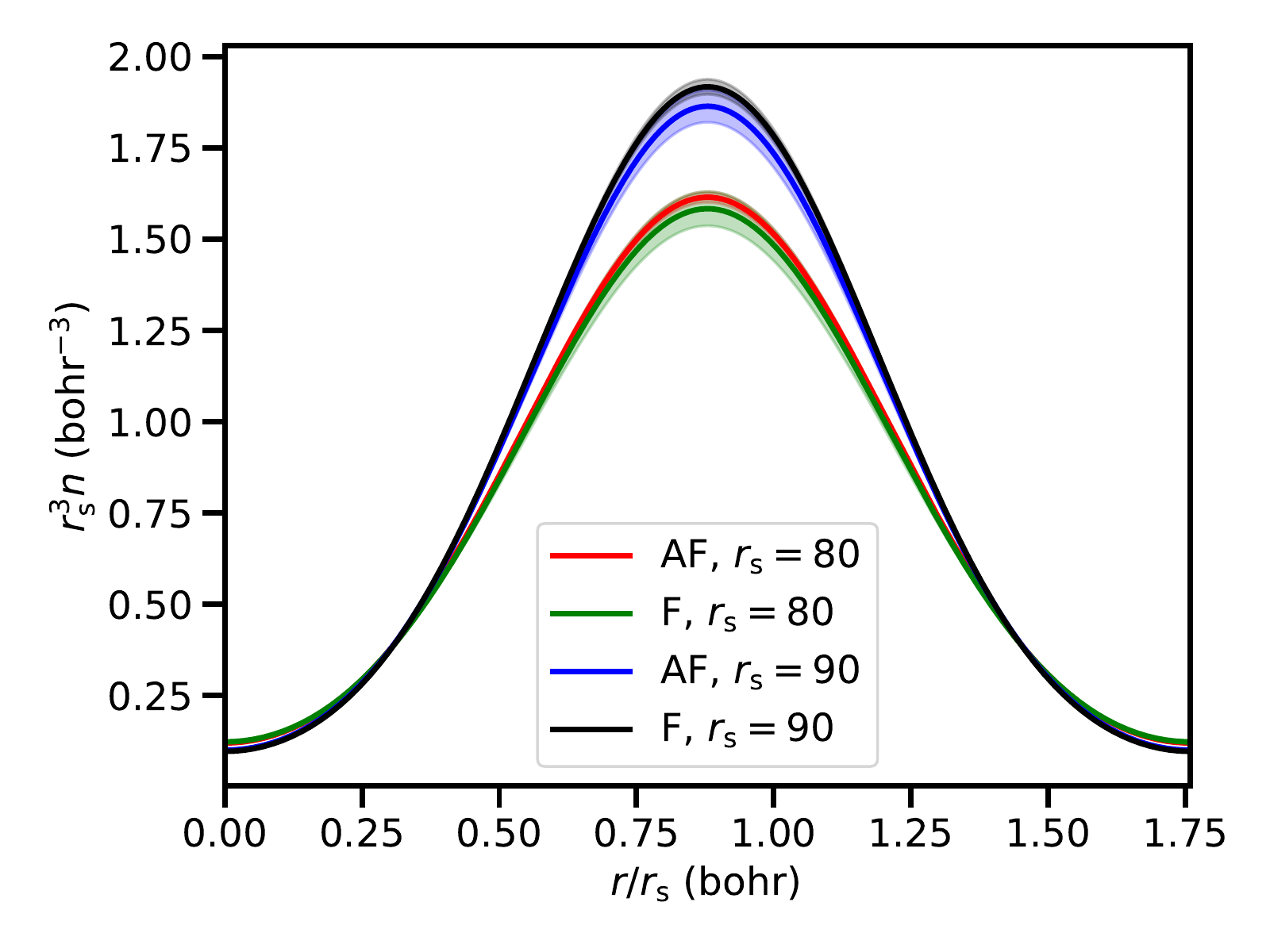}
    \caption{\label{fig:wigner_cden} Extrapolated estimates of the
      electronic charge densities of antiferromagnetic (AF) and
      ferromagnetic (F) bcc Wigner crystals plotted along a straight
      line in the $[100]$ direction.  Results are shown for two
      density parameters in the vicinity of the crystallization
      density.  The charge densities have been obtained by
      extrapolated estimation (twice the DMC charge density minus the
      VMC charge density), which largely removes errors that are
      linear in the error in the trial wave function.  Furthermore,
      the charge densities have been extrapolated to infinite system
      size, assuming the FS error goes as $N^{-1}$.  The
      shaded regions indicate one standard error about the mean.}
\end{figure}

The electronic charge densities of antiferromagnetic and ferromagnetic
crystals are plotted in Fig.\ \ref{fig:wigner_cden}.  The charge
densities were extrapolated to infinite system size by fitting
$\rho_\infty({\bf r})+b({\bf r})N^{-1}$ to the charge density data
$\rho_N({\bf r})$ at each point ${\bf r}$.  Here, $\rho_\infty({\bf
  r})$ and $b({\bf r})$ were fitting parameters at each point
$\mathbf{r}$.  $O(N^{-1})$ FS errors in the charge density arise due
to the center-of-mass kinetic energy, which leads to a tendency for
the orbitals to delocalize in a finite cell.

It is clear that the charge density is nonuniform at the
crystallization density, providing some numerical evidence that the
phase transition is first order.  The differences between
ferromagnetic and antiferromagnetic crystals are too small to resolve,
but it is clear that increasing $r_\text{s}$ has the effect of making
the charge density relatively localized, as expected.

\subsection{DMC time-step bias}\label{sec:dmc_dt}

\begin{figure}[htbp!]
    \centering
    \begin{tabular}{c c }
    \includegraphics[clip,scale=0.21]{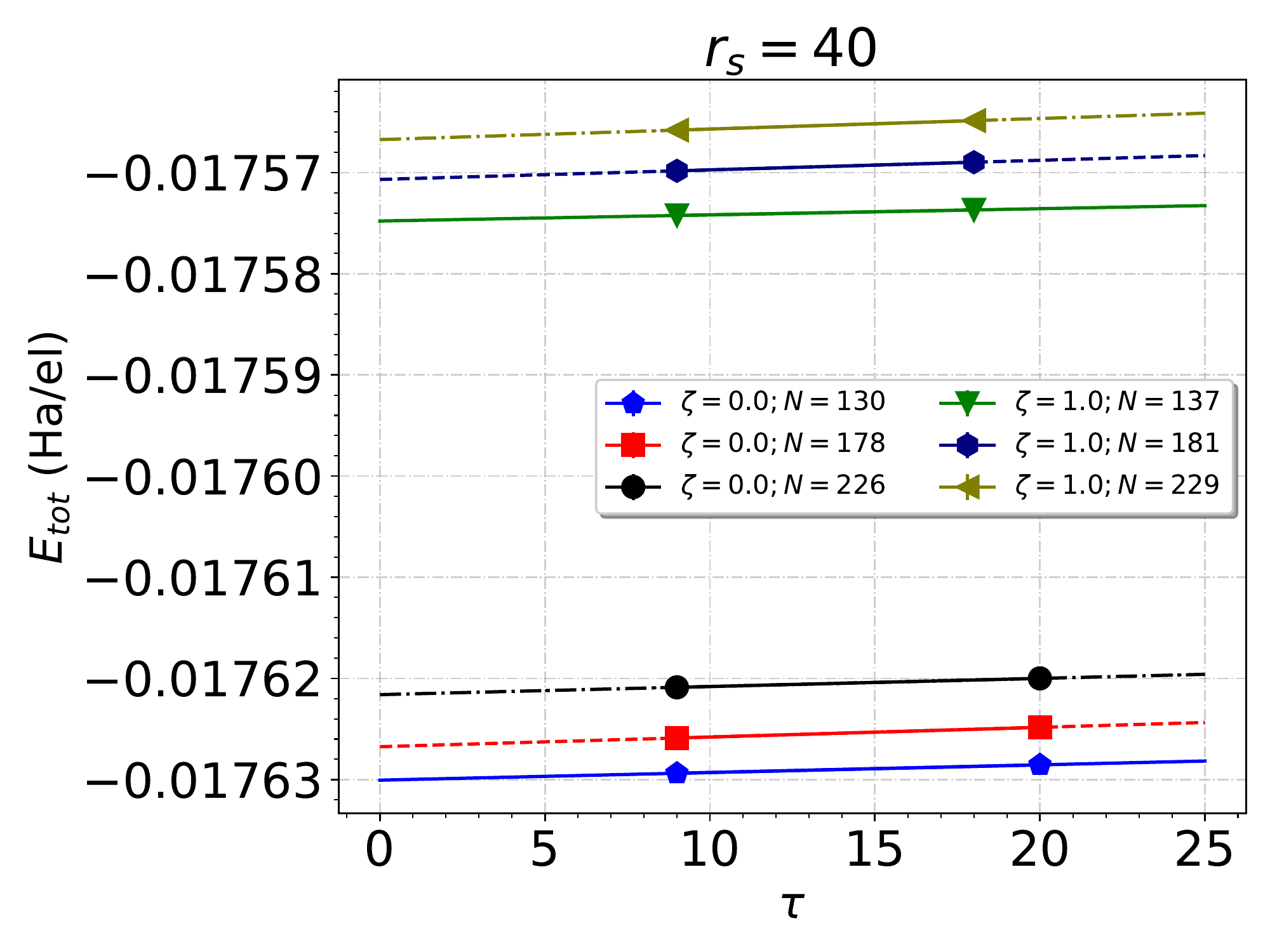}&
    \includegraphics[clip,scale=0.21]{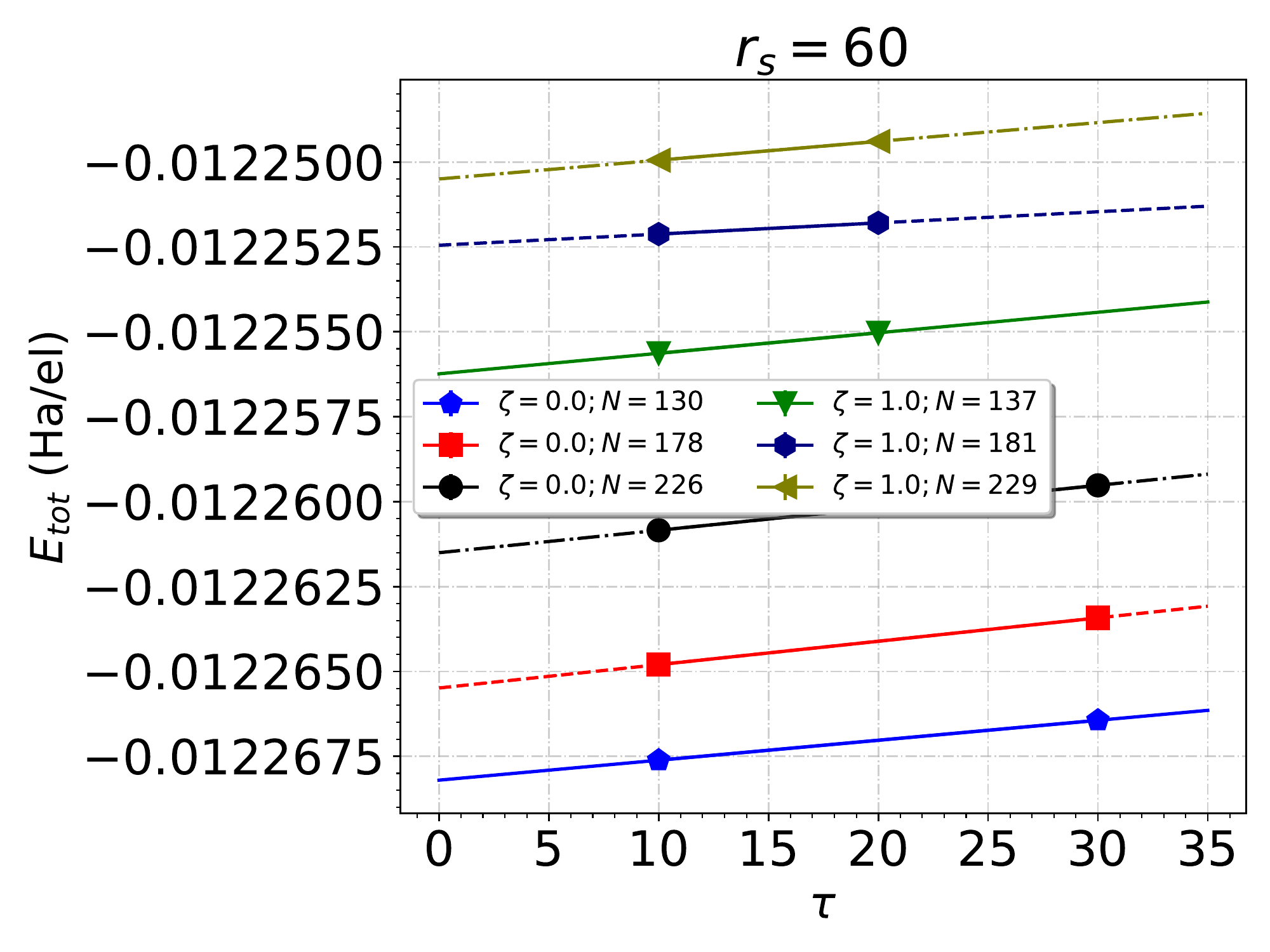}\\
    \includegraphics[clip,scale=0.21]{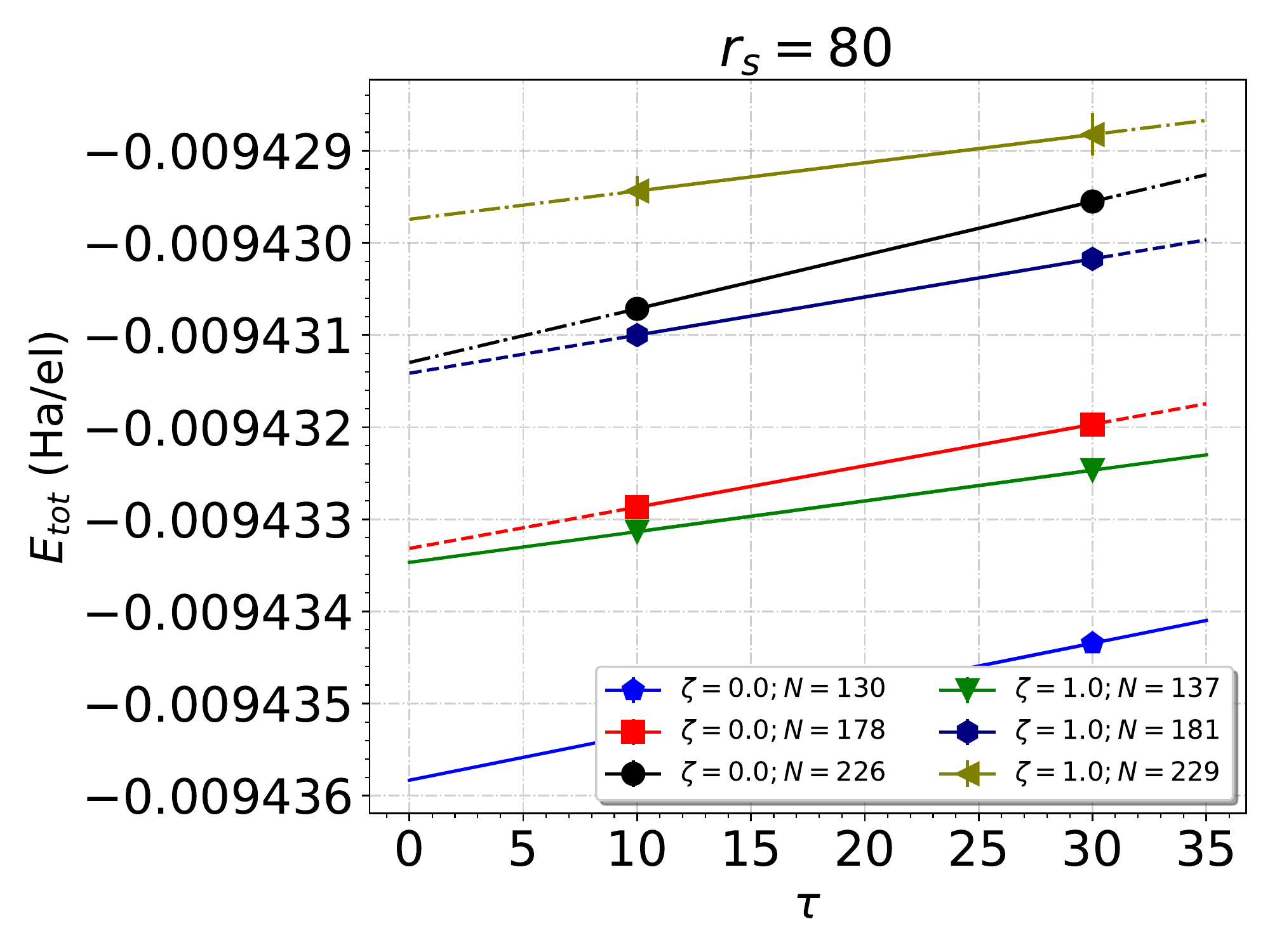}&
    \includegraphics[clip,scale=0.21]{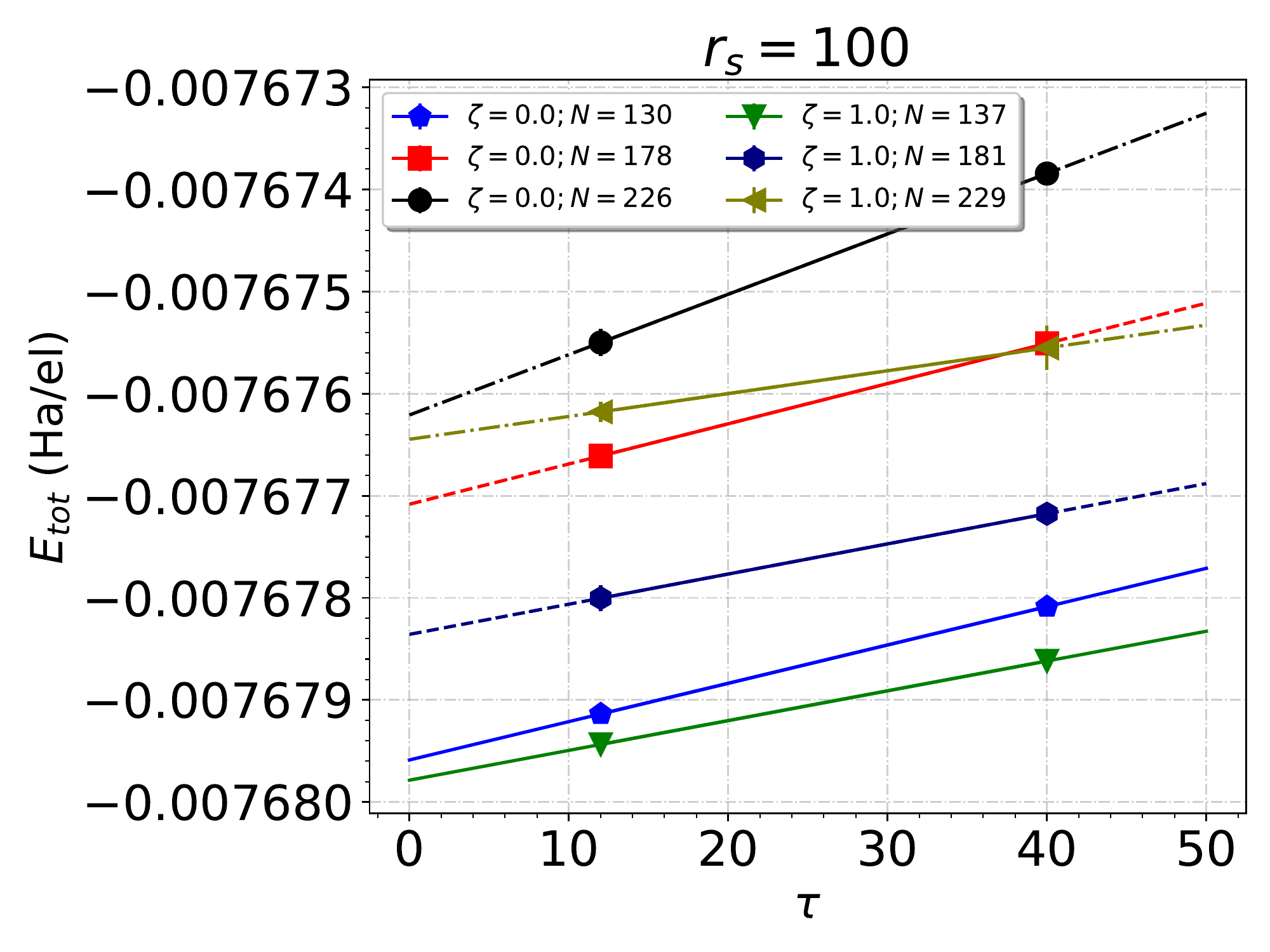}\\
    \end{tabular}
    \caption{\label{DMCvsTau} TA SJB-DMC energy against time step
      $\tau$ for the paramagnetic ($\zeta=0$) and ferromagnetic
      ($\zeta=1$) fluid phases at different system sizes $N$.}
\end{figure}

The variation of the DMC energy with time step is investigated in this
section.
Figure \ref{DMCvsTau} shows TA SJB-DMC energies against time step. The
population is varied in inverse proportion to the time step.  For all
the studied system sizes, densities, and spin polarizations the bias
at finite time step is always positive.
Our final results were linearly extrapolated to zero time step (and
hence infinite population) in every case.

\begin{figure}[htbp!]
    \centering
    \begin{tabular}{c c}
    \includegraphics[clip,scale=0.21]{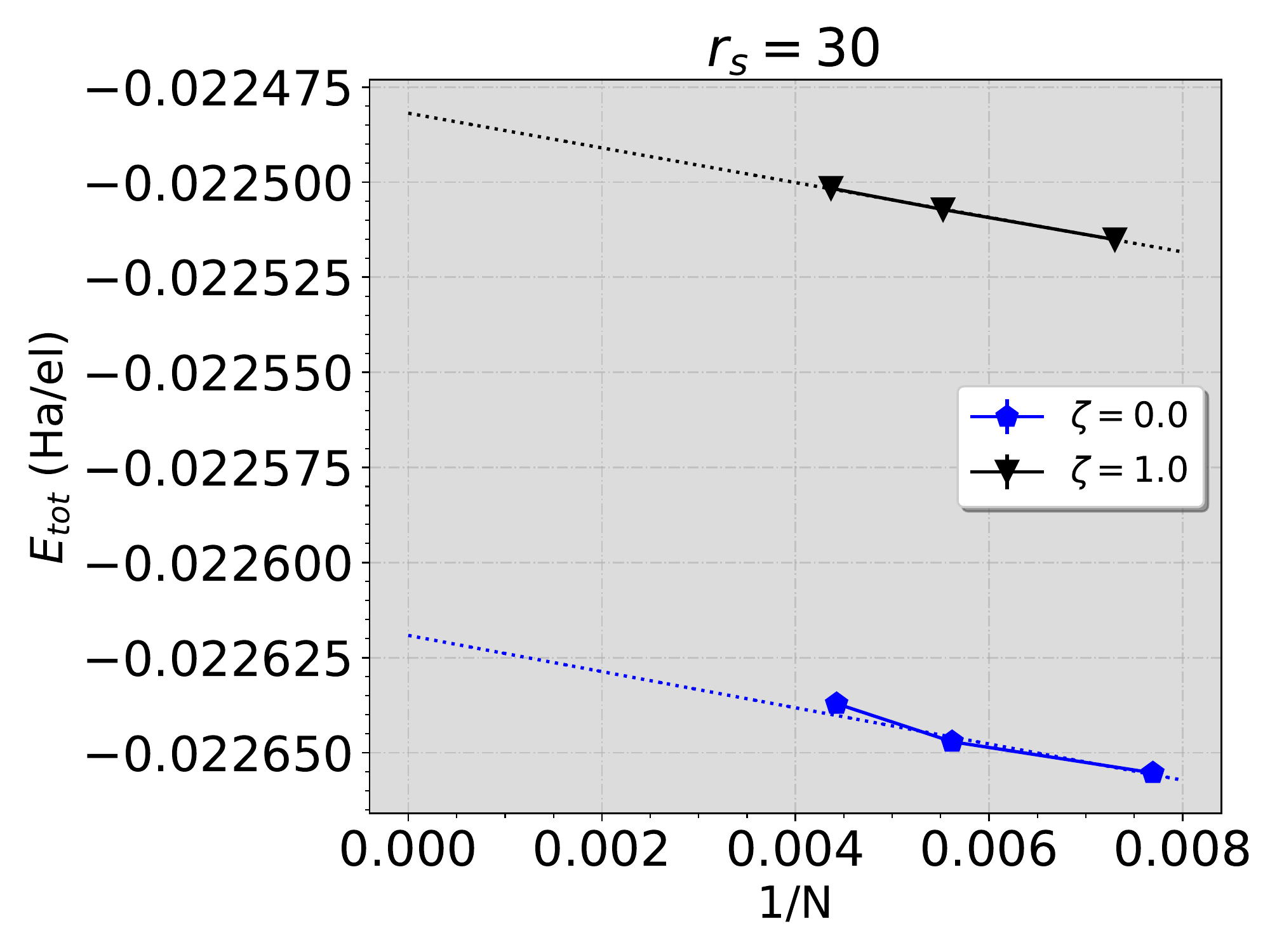}&
    \includegraphics[clip,scale=0.21]{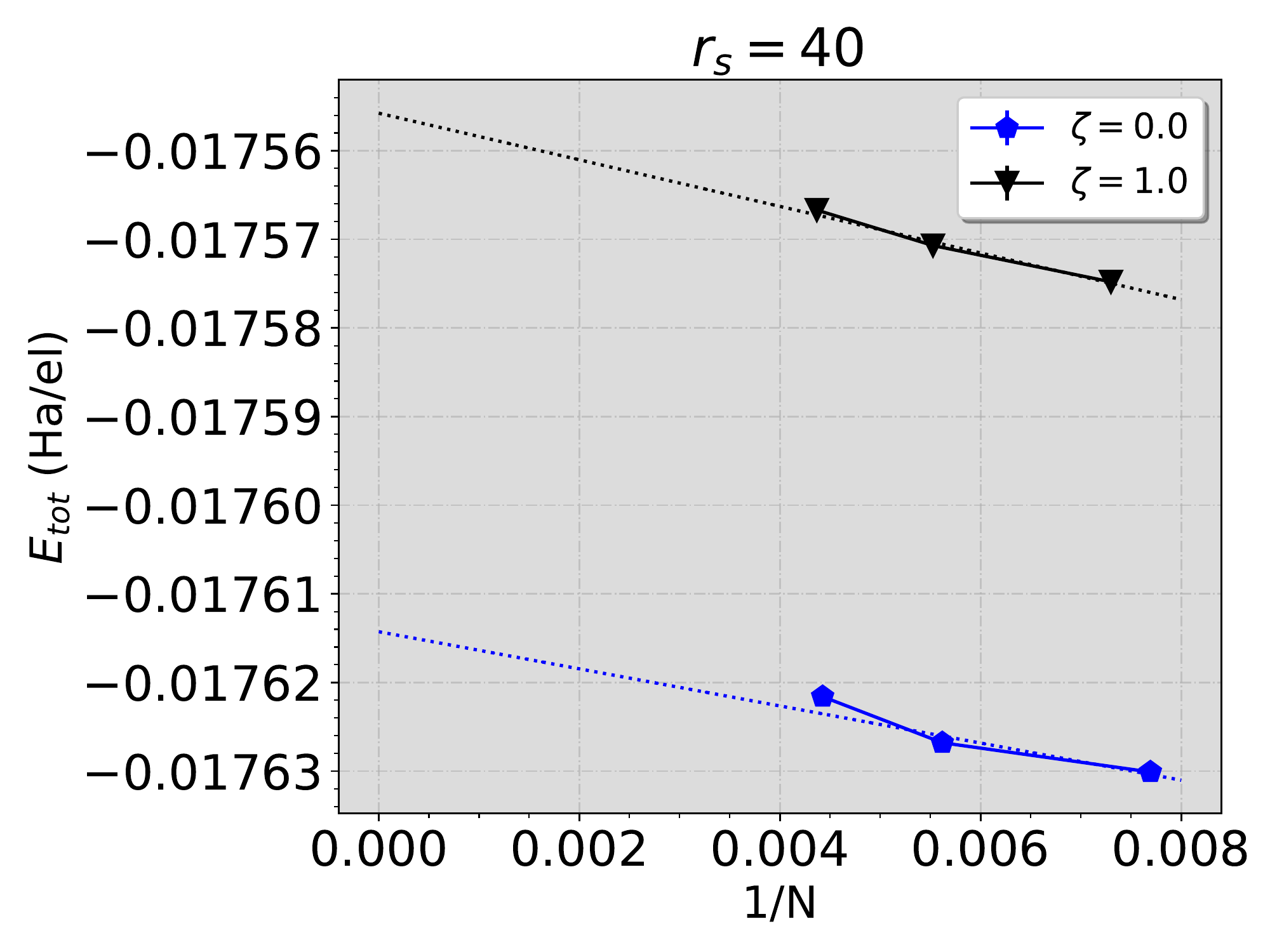}\\
    \includegraphics[clip,scale=0.21]{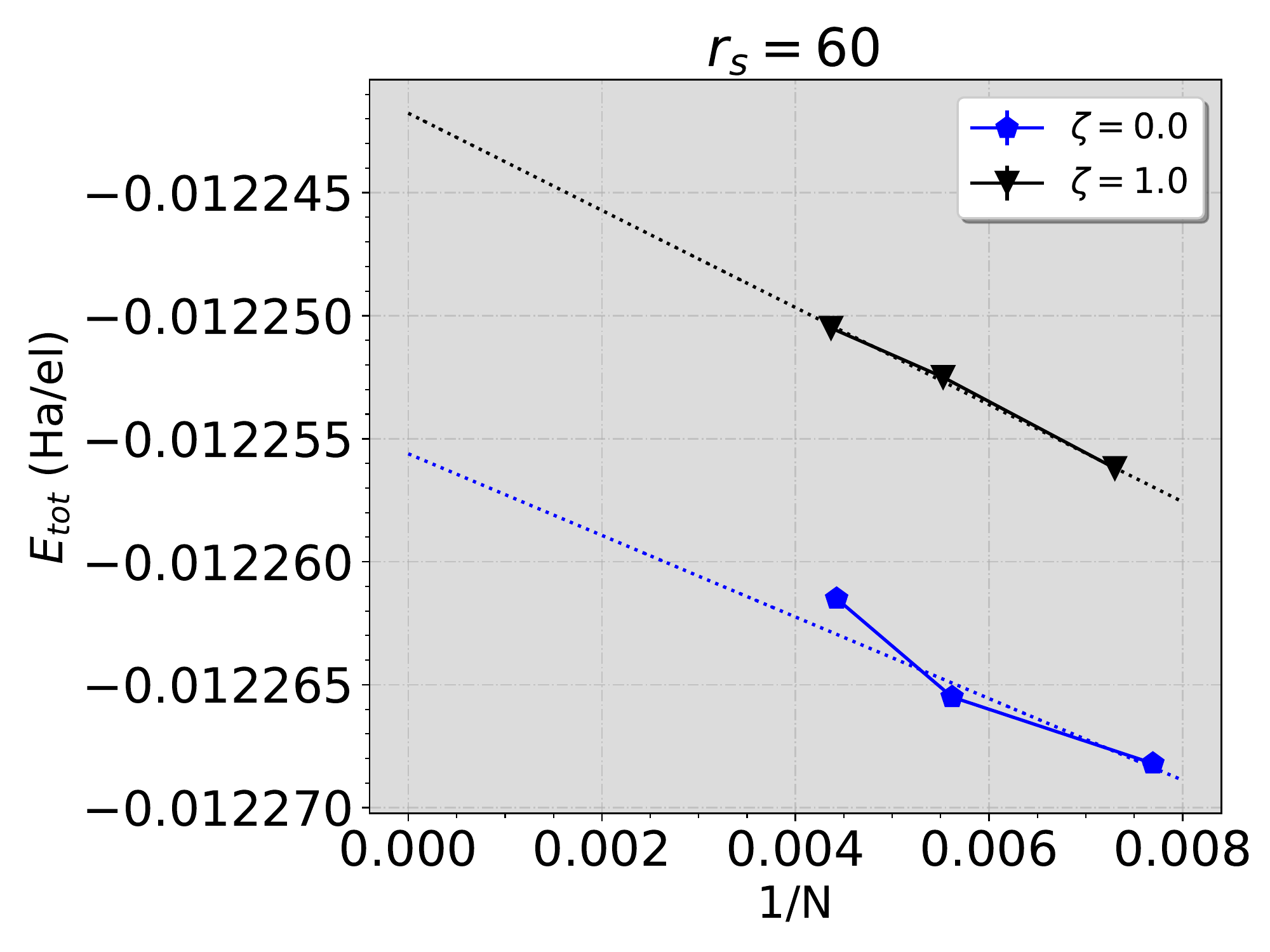}& \includegraphics[clip,scale=0.21]{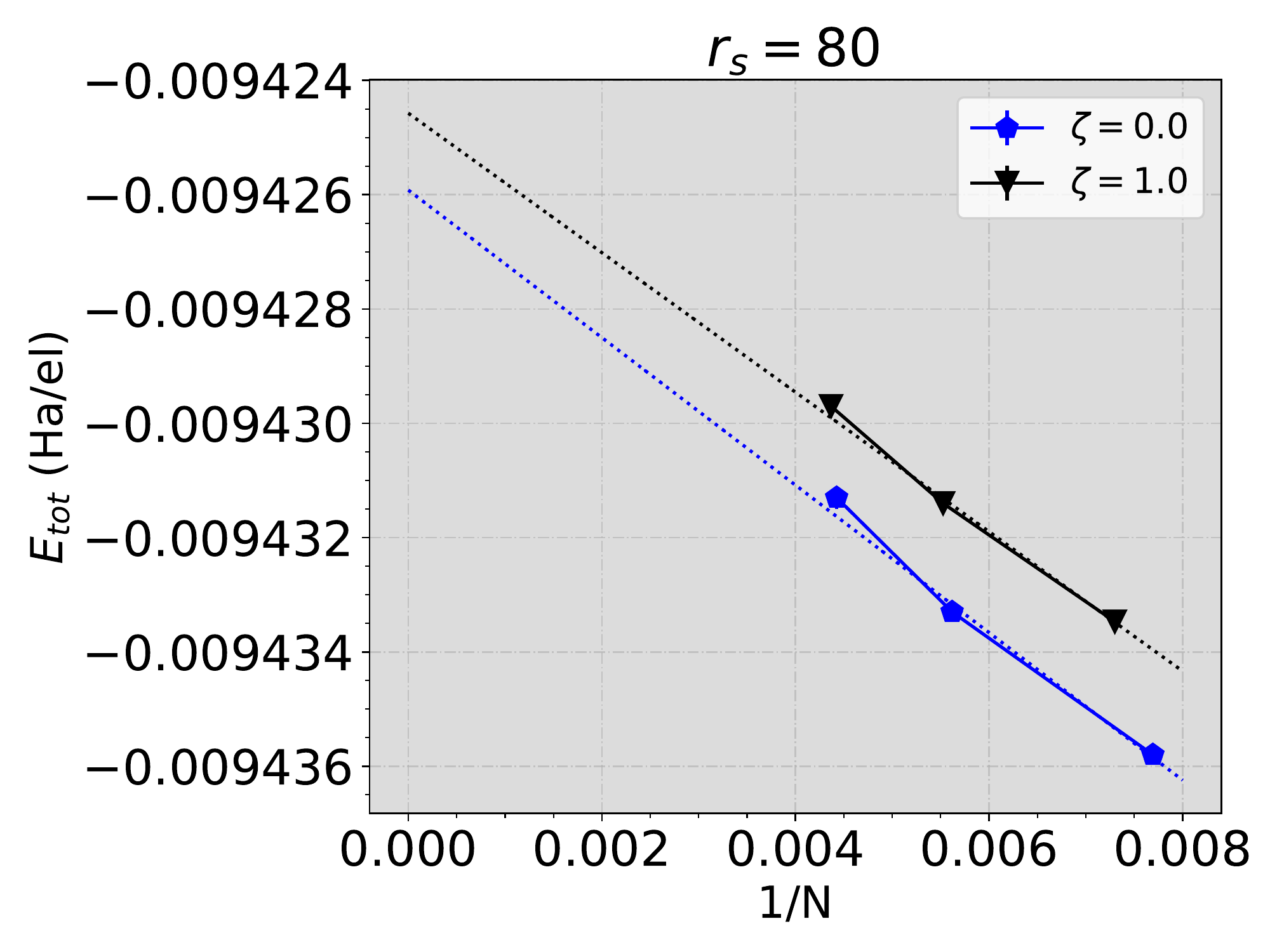}\\
    \includegraphics[clip,scale=0.21]{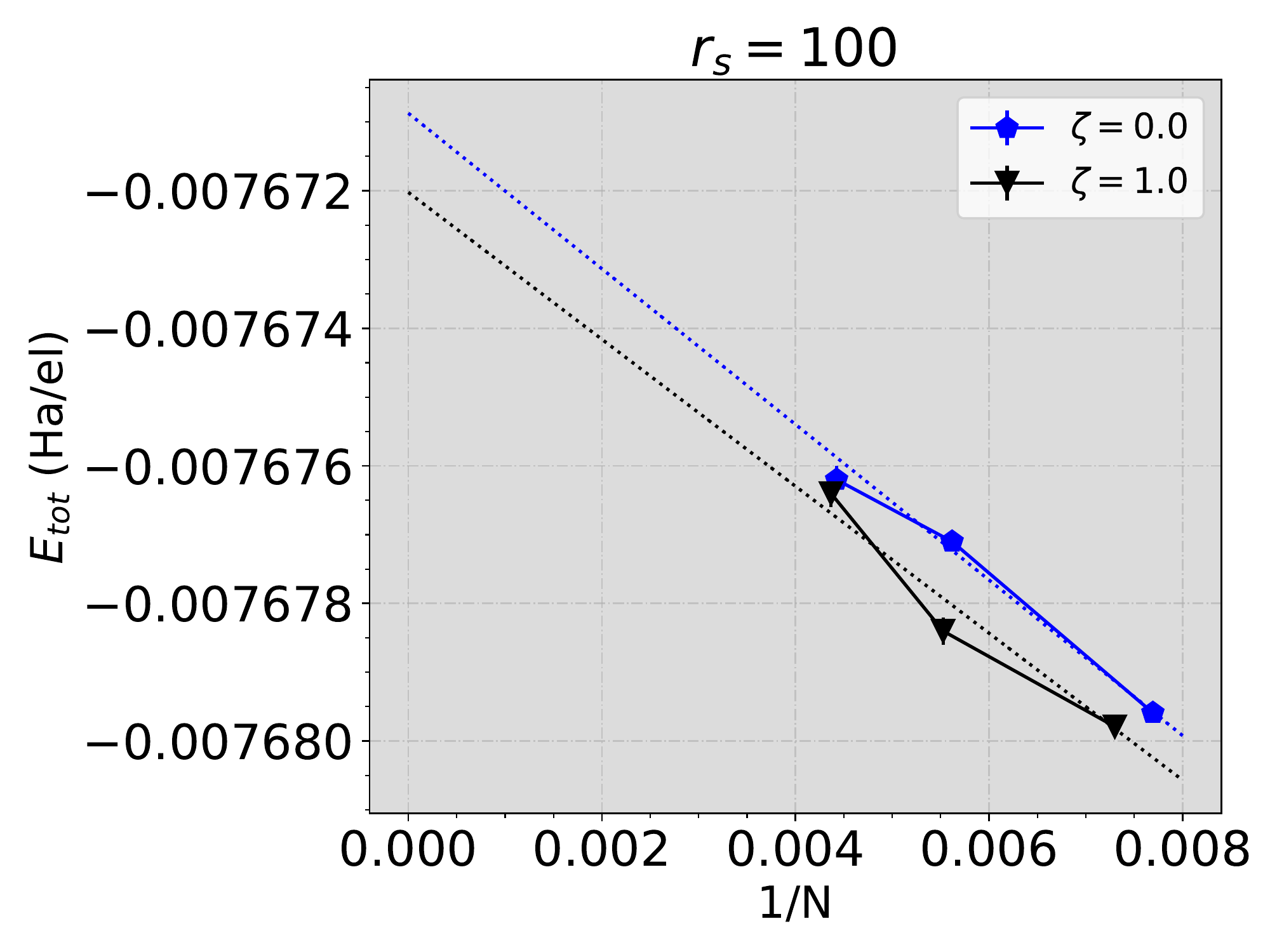}\\
    \end{tabular}
    \caption{\label{DMCNBTau} TA SJB-DMC energies against system size
      $N$ for different spin polarizations $\zeta$. The DMC energies
      are extrapolated to zero time step. The dotted line represents
      the extrapolation to infinite system size.}
\end{figure}

Figure \ref{DMCNBTau} shows SJB-DMC energies of paramagnetic and
ferromagnetic fluid phases, extrapolated to zero time step, at three
different system sizes.
\begin{table}[!htbp]
\centering
\caption{\label{EzetaBTau} TA SJB-DMC energies of the fluid phases of
  the 3D-HEG extrapolated to the thermodynamic limit from different
  system sizes ($130 < N < 274$).  Energies are in mHa/el. The numbers
  in parentheses indicate statistical errors. The DMC energies were
  extrapolated to zero time step.}
 \begin{tabular}{lccc}
 \hline\hline

 \diagbox{$r_\text{s}$}{$\zeta$} & 0.0  & 1.0   \\

 \hline

  30  & $-22.6191(7)$ &  $-22.4819(7)$ \\

  40  & $-17.6143(3)$ &  $-17.5558(7)$ \\

  60  & $-12.2556(3)$ &  $-12.2418(5)$ \\

  80  & $-9.4259(4)$ &  $-9.4246(3)$ \\

  100 & $-7.6709(3)$ &  $-7.6720(4)$ \\

 \hline\hline
 \end{tabular}
\end{table}
TA SJB-DMC energies extrapolated to infinite system size and zero time
step are listed in Table \ref{EzetaBTau}.

\begin{figure}[htbp!]
    \centering
    \begin{tabular}{c c }
    \includegraphics[clip,scale=0.16]{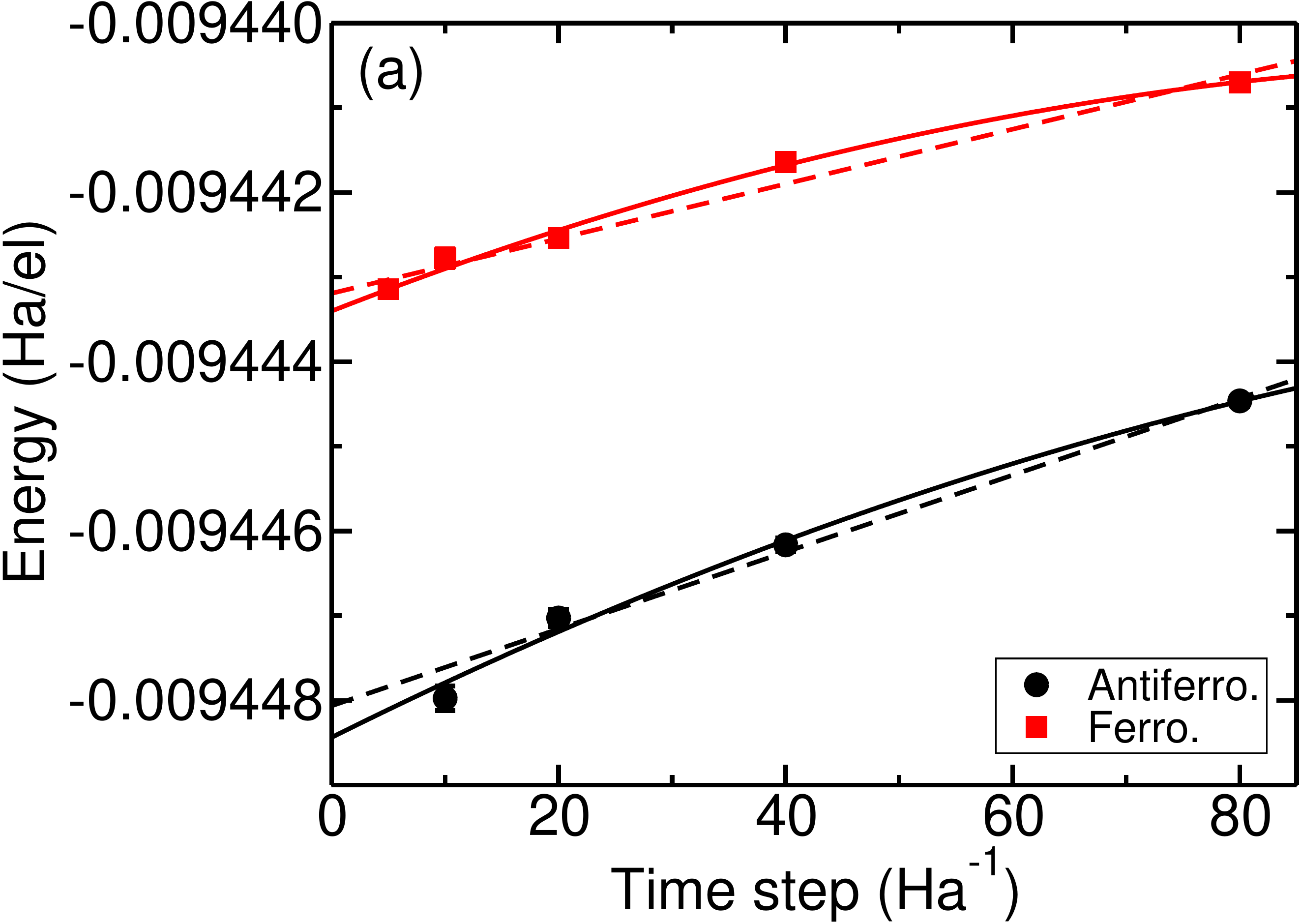}&
    \includegraphics[clip,scale=0.16]{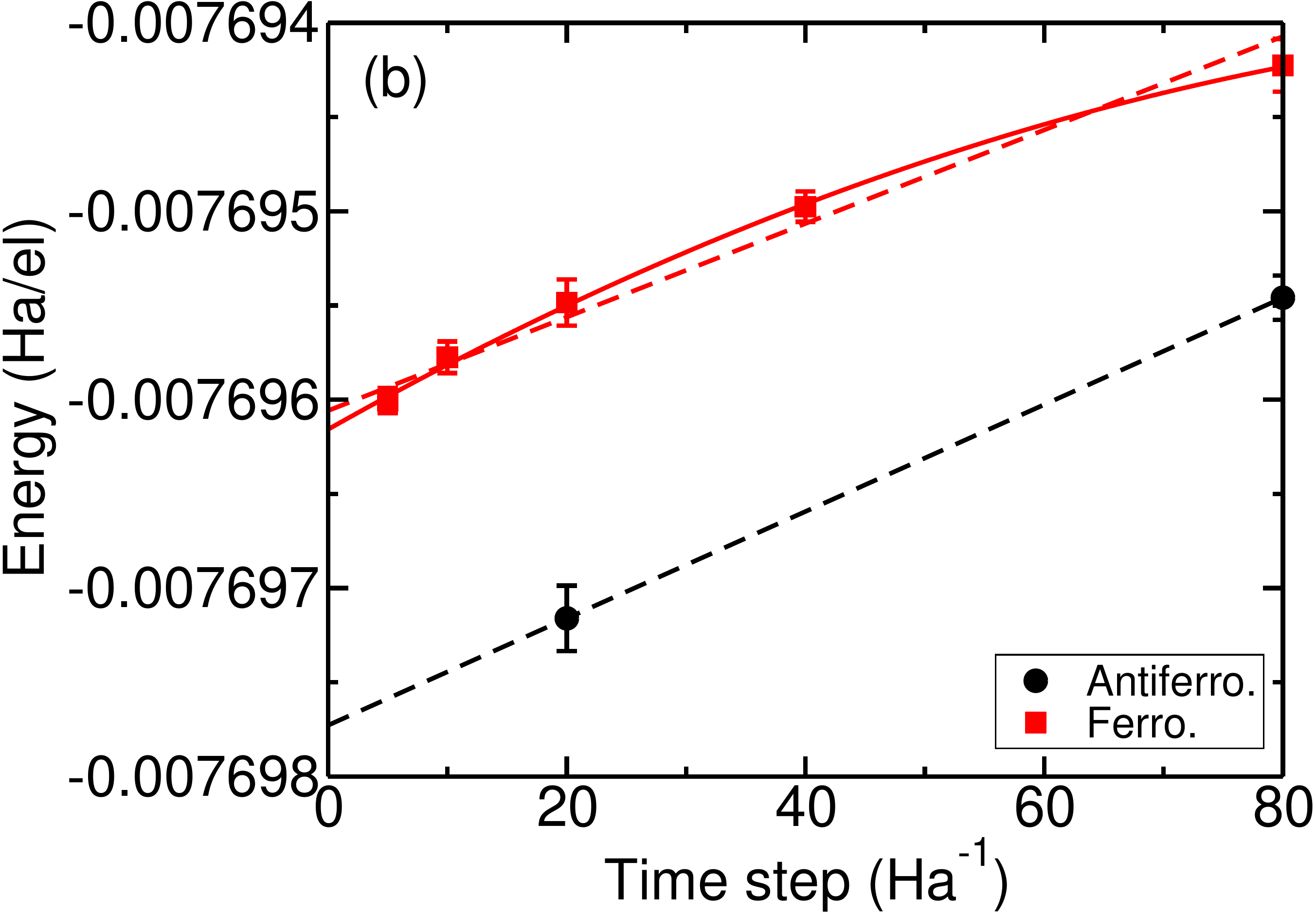}\\
    \end{tabular}
    \caption{SJ-DMC energy against time step for bcc Wigner crystals
      at (a) $r_\text{s}=80$ and (b) $r_\text{s}=100$.  Results are
      shown for both antiferromagnetic and ferromagnetic crystals. The
      system size is $N=64$ electrons. The target population is varied
      in inverse proportion to the time
      step.  \label{fig:wigner_dt_bias}}
\end{figure}

The SJ-DMC energy of bcc Wigner crystals is plotted against time step
in Fig.\ \ref{fig:wigner_dt_bias}.  Time steps in the range 10--80
Ha$^{-1}$ were used in our calculations.  At $r_\text{s}=80$ the data
are better fitted by a quadratic function of time step than a linear
function of time step; however, the quadratic fit is no better than
the linear fit at $r_\text{s}=100$.  However, even at $r_\text{s}=80$,
the difference between the results of linear and quadratic time-step
extrapolation is not statistically significant.

\section{Results and discussion}

\subsection{SJ-DMC magnetic phase diagram for the fluid phases}

\begin{figure}[htbp!]
    \centering
    \begin{tabular}{c c}
    \includegraphics[clip,scale=0.21]{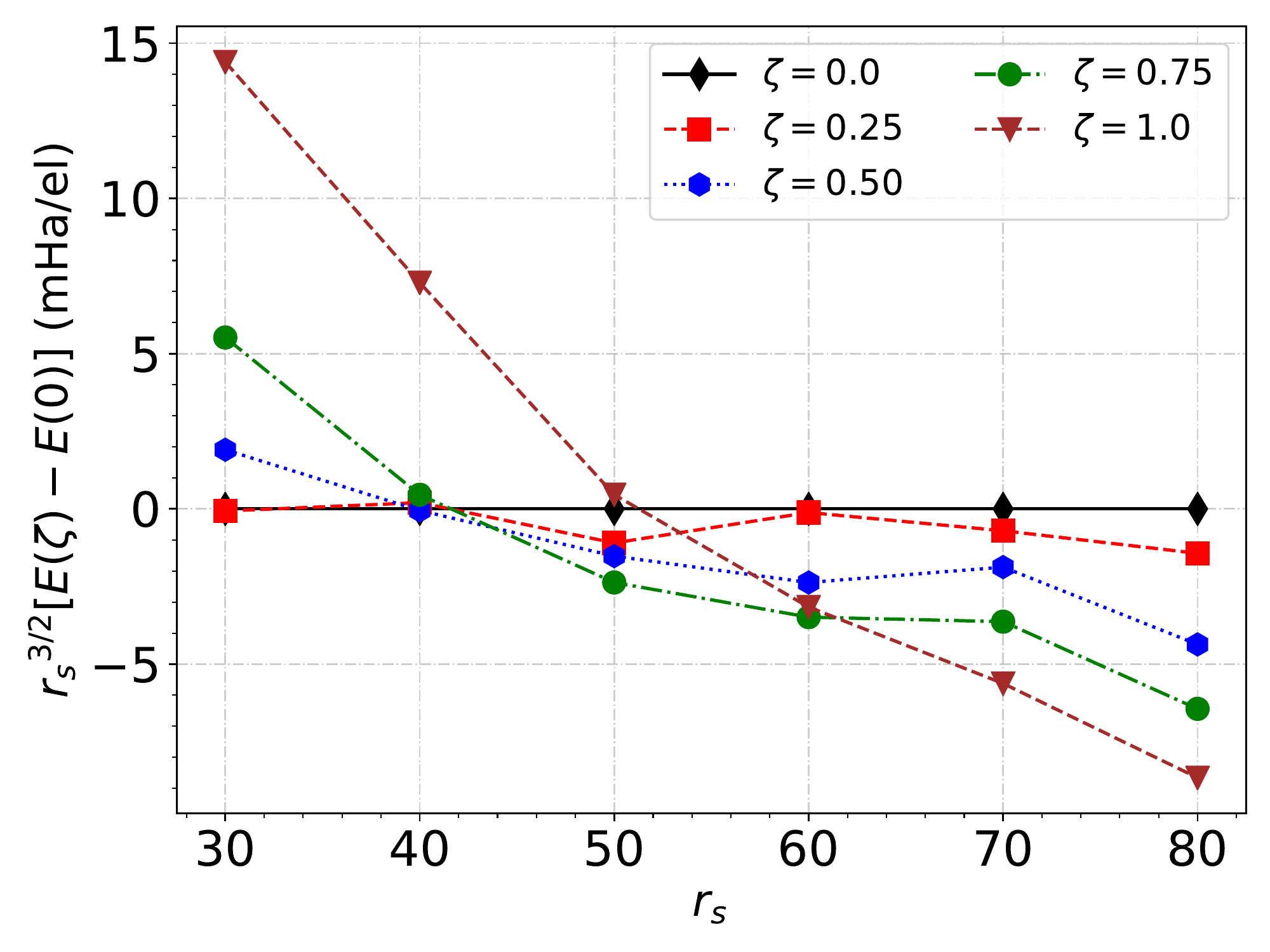}&
    \includegraphics[clip,scale=0.21]{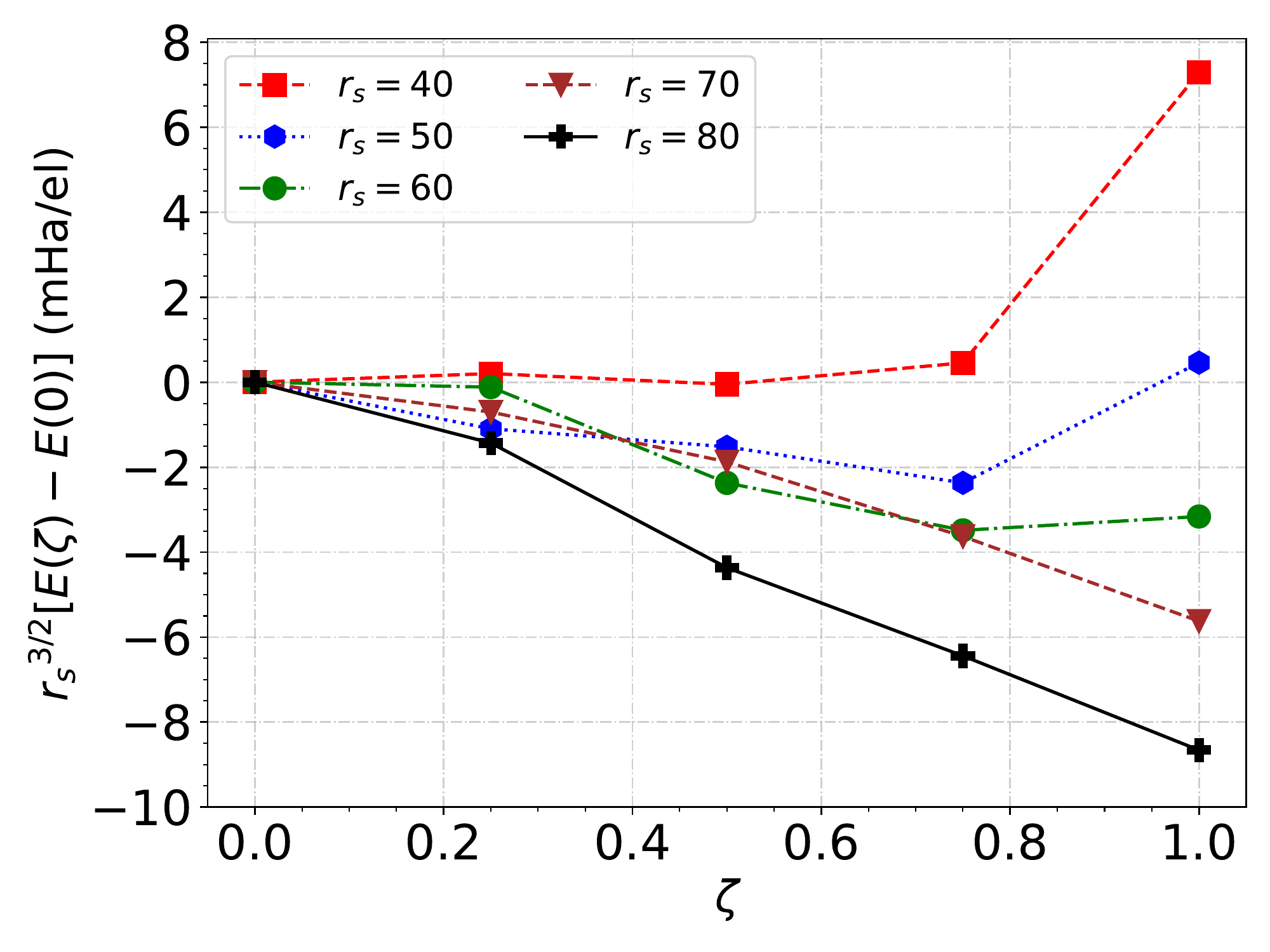}\\
    \end{tabular}
\caption{\label{polarization} (Left panel) SJ-DMC spin polarization
  energy of the 3D-HEG multiplied by $r_\text{s}^{3/2}$ against
  density parameter $r_\text{s}$. (Right panel) SJ-DMC spin
  polarization energy of the 3D-HEG multiplied by ${r_\text{s}}^{3/2}$
  against spin polarization $\zeta$ at various densities.}
\end{figure}

The SJ-DMC phase diagram (Fig.\ \ref{polarization}, left panel) shows
that the spin polarized state $\zeta=0.5$ has lower energy than the
paramagnetic phase at $r_\text{s}\geqslant40$. The fluid with
$\zeta=0.75$ becomes more stable than the fluid with $\zeta=0.5$ at
$r_\text{s} \approx 46$, and the 3D-HEG system adopts a fully
polarized state $\zeta=1$ at $r_\text{s} \approx 62$.

\begin{figure}[htbp!]
    \centering
    \begin{tabular}{c c}
         \includegraphics[clip,scale=0.17]{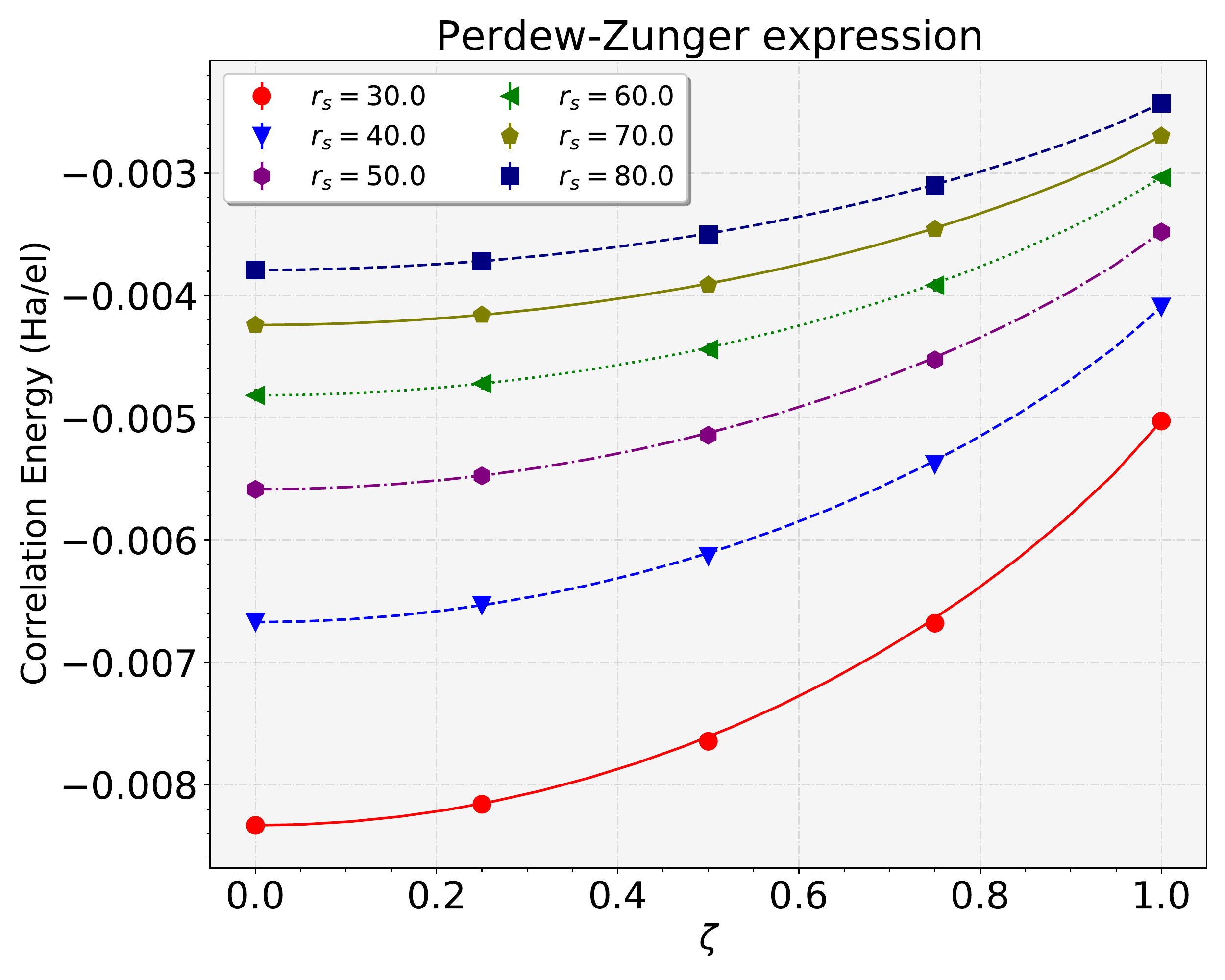}&
         \includegraphics[clip,scale=0.17]{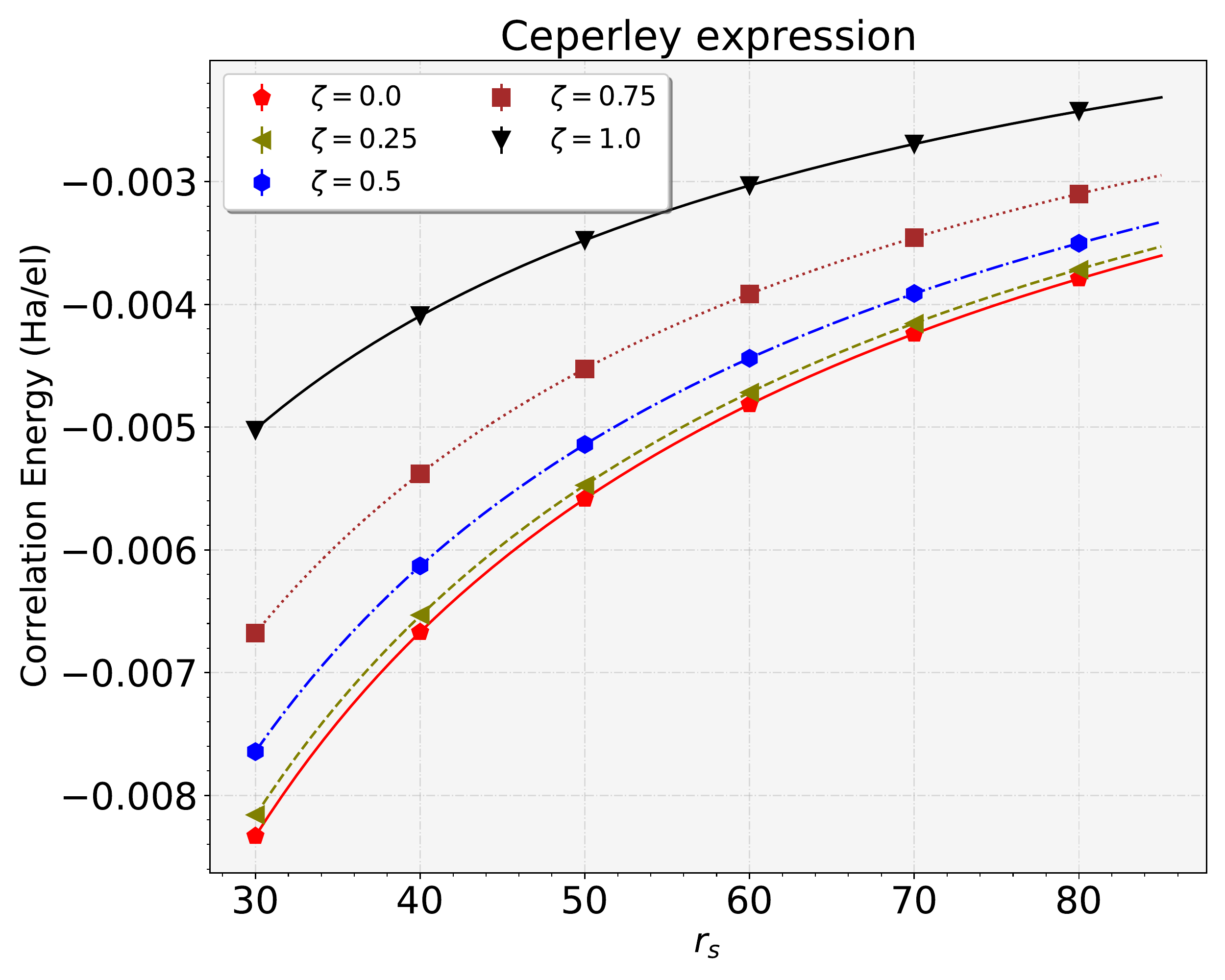}\\
         \\
    \end{tabular}
\caption{\label{E_Corr} Correlation energies obtained in SJ-DMC
  calculations for the fluid phase (symbols), together with the fit
  to Eq.\ (1) of the main text (lines). Error bars on the QMC data are
  shown, but are smaller than the symbols. The DMC energies were not
  extrapolated to zero time step.}
\end{figure}

\begin{table}[!htbp]
\centering
\caption{\label{FitParamSJ} Parameters in Eq.\ (2) of the main text,
  obtained by fitting to SJ-DMC data that have been TA and
  extrapolated to infinite system size, but not to zero time step.}
 \begin{tabular}{lcccc}
 \hline \hline

 $\zeta$ & $\gamma$ & $\beta_1$ & $\beta_2$ & $\chi^2$ \\

 \hline

 0.0  & $-0.122(5)$  & $0.923(45)$  & $0.287(10)$ &  $1.25$ \\

 0.25 & $-0.104(7)$  & $0.758(67)$  & $0.252(15)$ & $10.573$  \\

 0.5  & $-0.110(5)$  & $0.937(51)$  & $0.275(10)$ &  $3.137$ \\

 0.75 & $-0.091(14)$ & $0.938(189)$ & $0.250(36)$ &  $0.412$ \\

 1.0  & $-0.052(5)$  & $0.808(97)$  & $0.166(13)$ &  $6.58$ \\

\hline\hline
\end{tabular}
\end{table}

The correlation energy of a Fermi fluid is defined as the difference
between the Hartree-Fock energy per electron and the exact
ground-state energy per electron, where the latter is approximated by
our DMC results. The procedure developed by von Barth and Hedin \cite{Barth_1972} and
Perdew and Zunger \cite{Perdew_1981} to interpolate between $\zeta = 0$ and
$\zeta = 1$ was applied to our SJ-DMC correlation energies
(Fig.\ \ref{E_Corr}, top panel).
The well-known Perdew-Zunger expression for
the correlation energy per electron of the 3D-HEG is \cite{Perdew_1981}
\begin{equation}
 E_\text{c}(r_\text{s}, \zeta) = E_\text{c}^\text{para}(r_\text{s}) +
 \left[E_\text{c}^\text{ferro}(r_\text{s}) -
   E_\text{c}^\text{para}(r_\text{s})\right]f(\zeta),
   \label{E_corr}
 \end{equation}
which has the same $\zeta$-dependence as the exchange energy
\cite{Barth_1972},
$f(\zeta)=\frac{{(1+\zeta)}^{4/3}+{(1-\zeta)}^{4/3}-2 }{
  2\left(2^{1/3}-1\right)}$.
This interpolation is exact for the
exchange part of the energy and is likely to be reasonable for the
correlation energy too. The interpolation scheme is very successful at
low densities, as illustrated in Fig.\ \ref{E_Corr}.

Following Ceperley \cite{Ceperley78} we fit
\begin{equation} E_\text{c}^\text{para,ferro}(r_\text{s}) =
\frac{\gamma^\text{para,ferro}}{1 +
  \beta_1^\text{para,ferro}\sqrt{r_\text{s}} +
  \beta_2^\text{para,ferro}r_\text{s}}
\label{Ceperley}
\end{equation}
to our Fermi fluid correlation energies.

The fitting parameters $\gamma$, $\beta_1$, and
$\beta_2$ for different polarizations are listed in Table
\ref{FitParamSJ}.

\subsection{SJB-DMC magnetic phase diagram for the fluid phases}
\begin{figure}[htbp!]
    \centering
    \begin{tabular}{c c}
    \includegraphics[clip,scale=0.23]{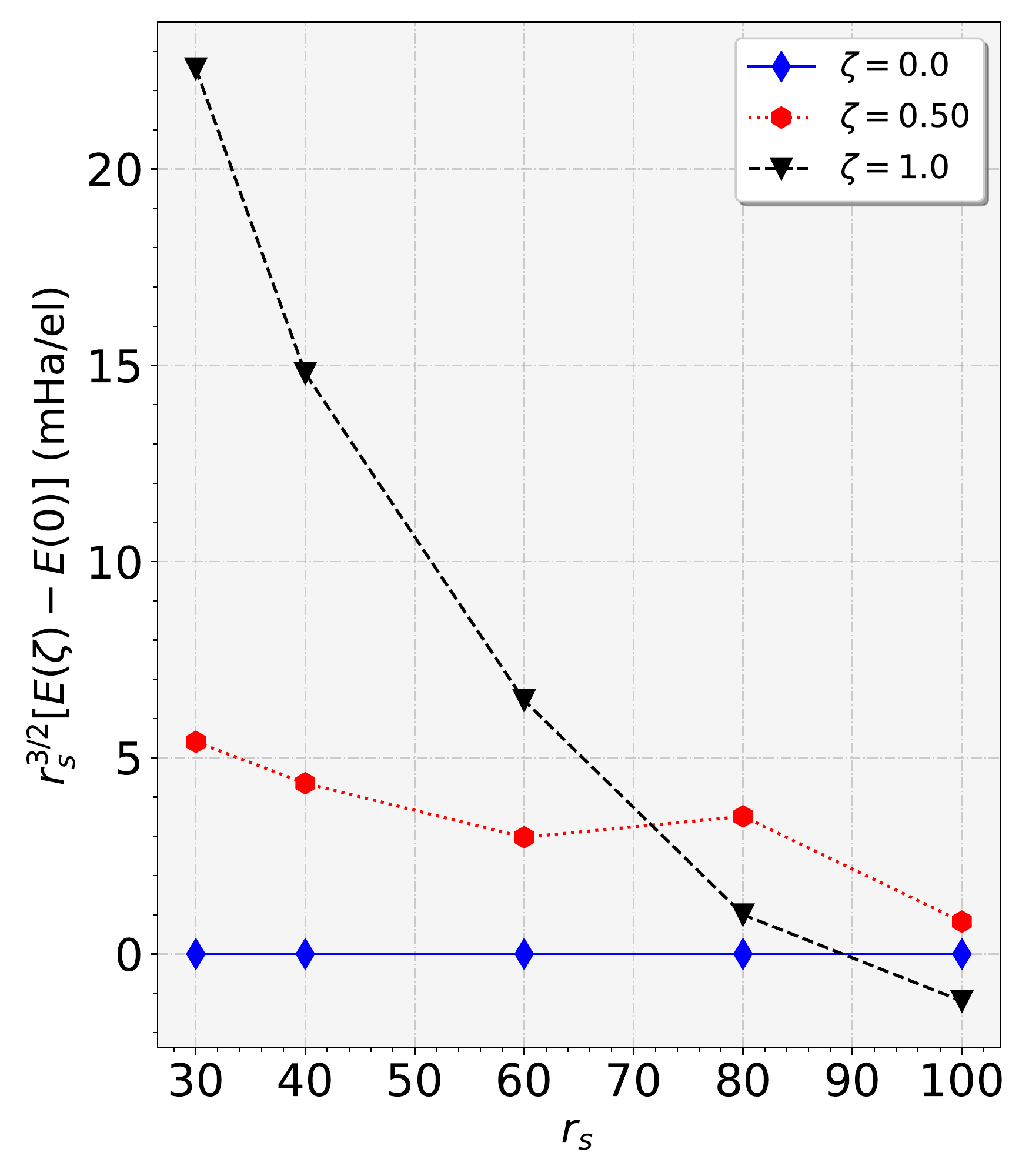}&
    \includegraphics[clip,scale=0.23]{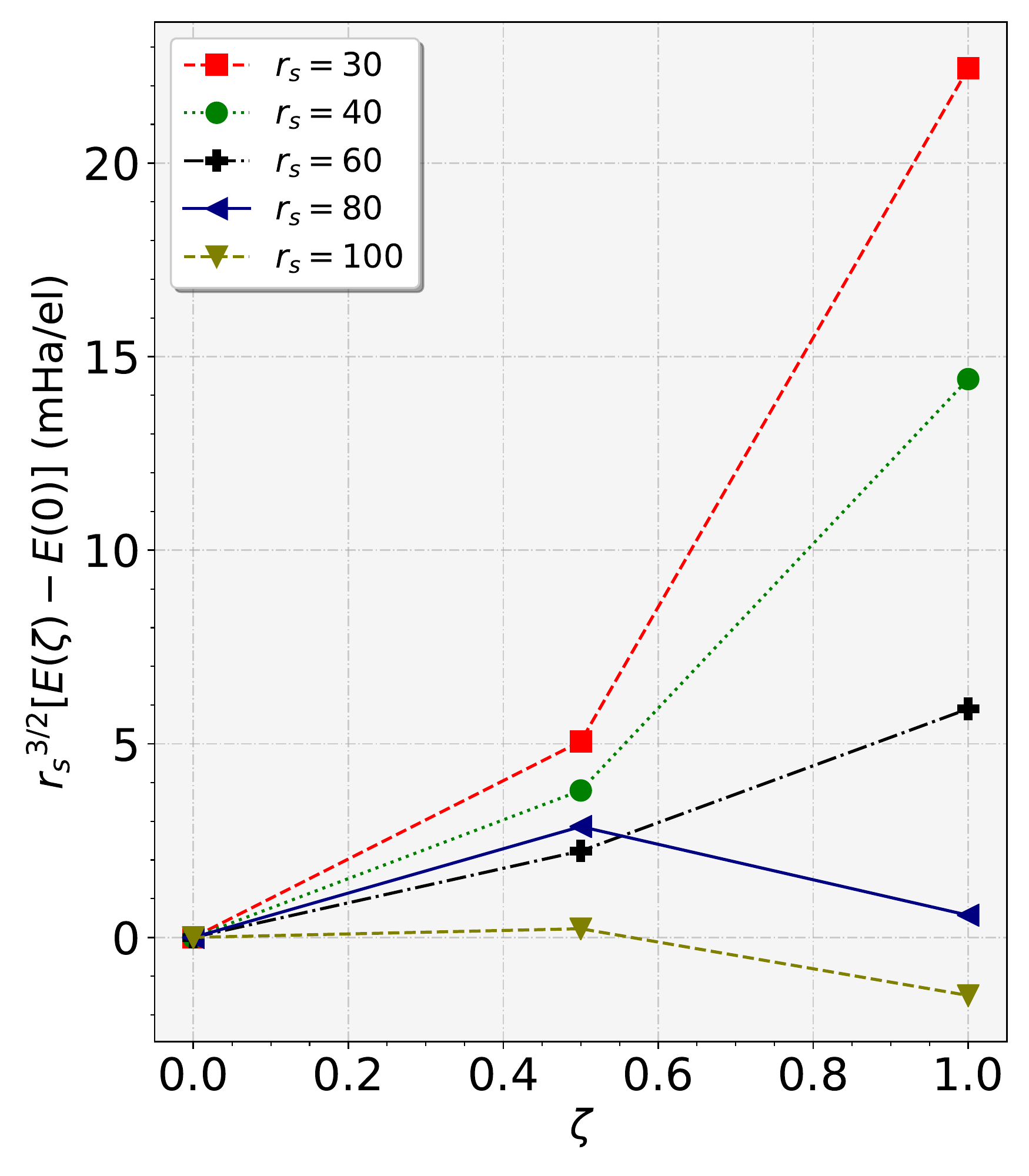}\\
    \end{tabular}
 \caption{\label{polarizationB} (Left panel) SJB-DMC relative energies
   of fluid phases of the 3D-HEG are plotted against $r_\text{s}$ for
   different spin polarizations $\zeta$ in the thermodynamic limit of
   infinite system size. (Right panel) SJB spin polarization energy of
   the 3D-HEG multiplied by ${r_\text{s}}^{3/2}$ at various
   densities. The DMC energies were not extrapolated to zero time
   step.}
\end{figure}

According to our SJB-DMC results for the fluid phases, the
paramagnetic fluid phase ($\zeta=0$) is stable for the entire density
range $r_\text{s} < 85.5(2)$ (Fig.\ \ref{polarizationB}, left
panel). There is no room for stability of the partially polarized
$\zeta = 0.5$ fluid phase.

\begin{figure}[htbp!]
\begin{tabular}{c}
    \includegraphics[clip,scale=0.3]{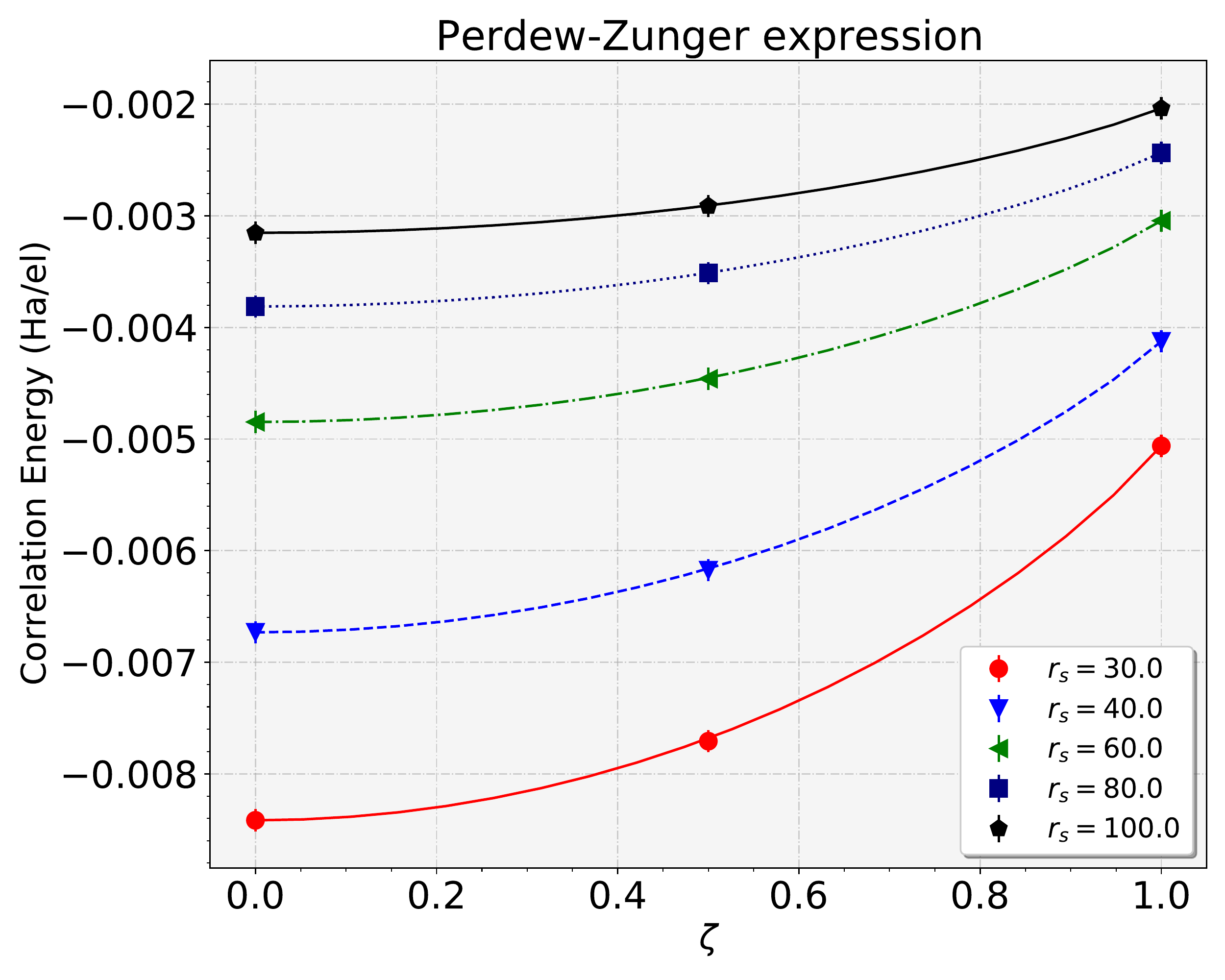} \\
    \includegraphics[clip,scale=0.3]{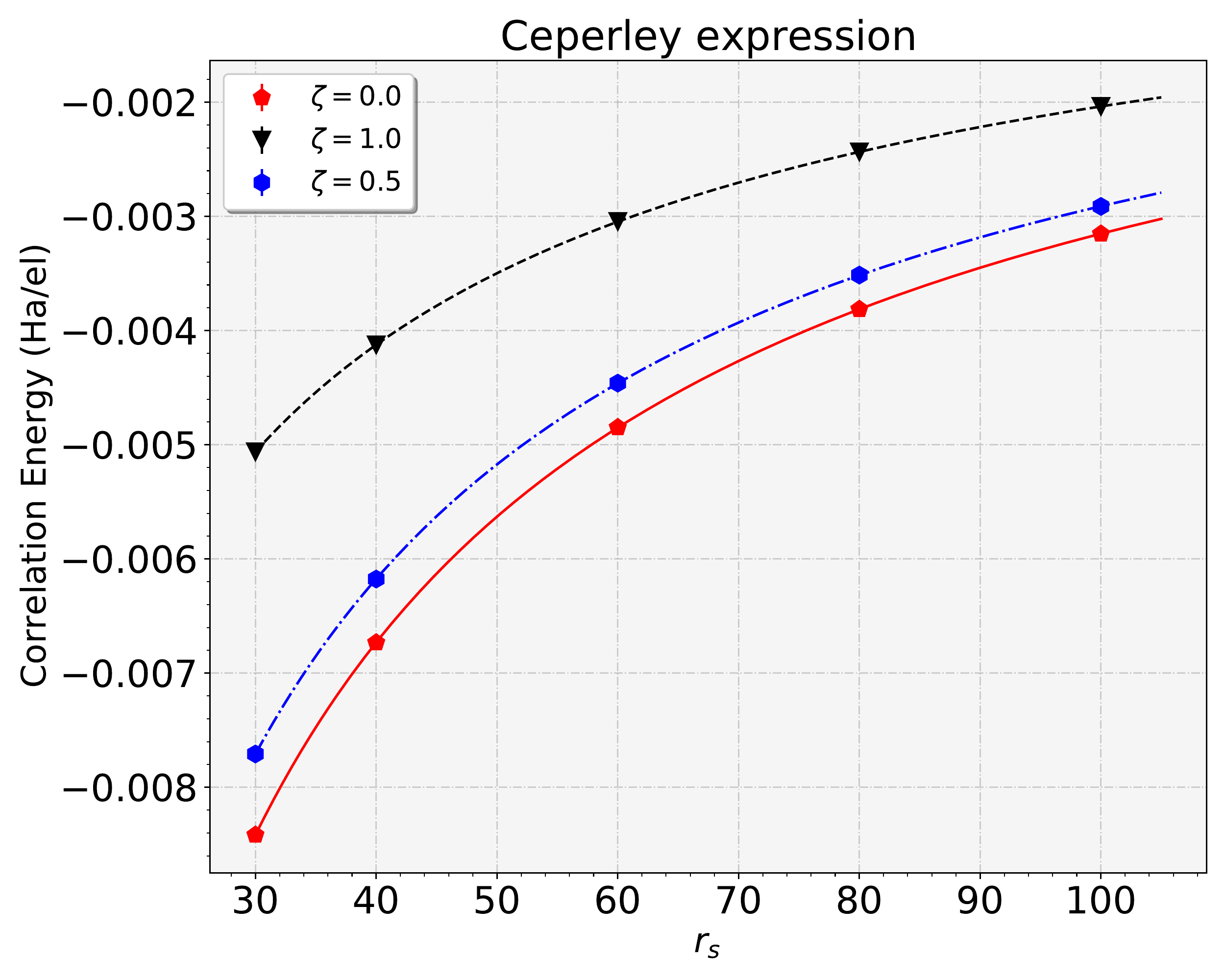}
\end{tabular}
\caption{\label{SJBCorr} Correlation energies obtained in SJB-DMC
  calculations for the fluid phase (symbols), together with the fit of
  Eq.\ (\ref{E_corr}) (lines).  Error bars on the QMC data are shown,
  but are smaller than the size of the symbols.}
\end{figure}

Figure \ref{SJBCorr} shows the SJB-DMC correlation energy of the fluid
phase as a function of spin polarization at different densities.
BF correlations lower the energy of the paramagnetic
fluid more than the ferromagnetic fluid.  SJ-DMC predicts that a fully
spin-polarized fluid phase becomes stable at $r_\text{s}=63(1)$,
whereas our SJB-DMC results do not show a statistically significant
region for the ferromagnetic fluid phase.  Same-spin electrons are
kept apart by the antisymmetry of the many-body wave function, while
opposite-spin electrons are only separated by correlation effects, so
that an accurate treatment of correlations lowers the energies of
paramagnetic phases more than ferromagnetic phases. Hence any future
improvements in QMC trial wave functions are expected further to
stabilize the paramagnetic fluid relative to spin-polarized fluids.
In a Wigner crystal electrons are kept apart by the localization of
orbitals on lattice sites, so there is relatively little scope for BF
correlations to lower the DMC energy significantly.  Indeed, the DMC
energy data in the Supplemental Material confirm that the effects of
BF on Wigner crystal DMC energies are small and do not significantly
alter the phase diagram \cite{Suppl}.  Our final energies are obtained
using SJB-DMC for Fermi fluid phases and SJ-DMC for Wigner crystal
phases.

\begin{table}[!htbp]
\centering
\caption{\label{FitParam} Parameters in the interpolation formula of
  Eq.\ (\ref{Ceperley}) for the correlation energy of a Fermi fluid,
  obtained by fitting to our SJB-DMC energy data at different spin
  polarizations $\zeta$.  The reduced $\chi^2$ values for the fits are
  also shown.}
 \begin{tabular}{lcccc}
 \hline \hline

 $\zeta$ & $\gamma$ (Ha/el) & $\beta_1$ & $\beta_2$ & $\chi^2$ \\

 \hline

 $0$ (para.) & $-0.13(1)$  & $1.0(1)$ & $0.32(3)$ & $0.92$ \\

 $0.5$ & $-0.15(1)$  & $1.3(1)$ & $0.36(2)$ & $10.96$ \\

 $1$ (ferro.) & $-0.062(6)$ & $0.97(8)$ & $0.19(1)$ & $5.51$  \\

\hline\hline
\end{tabular}
\end{table}

\subsection{Phase diagram}

\begin{figure*}[htbp!]
    \includegraphics[clip,scale=0.5]{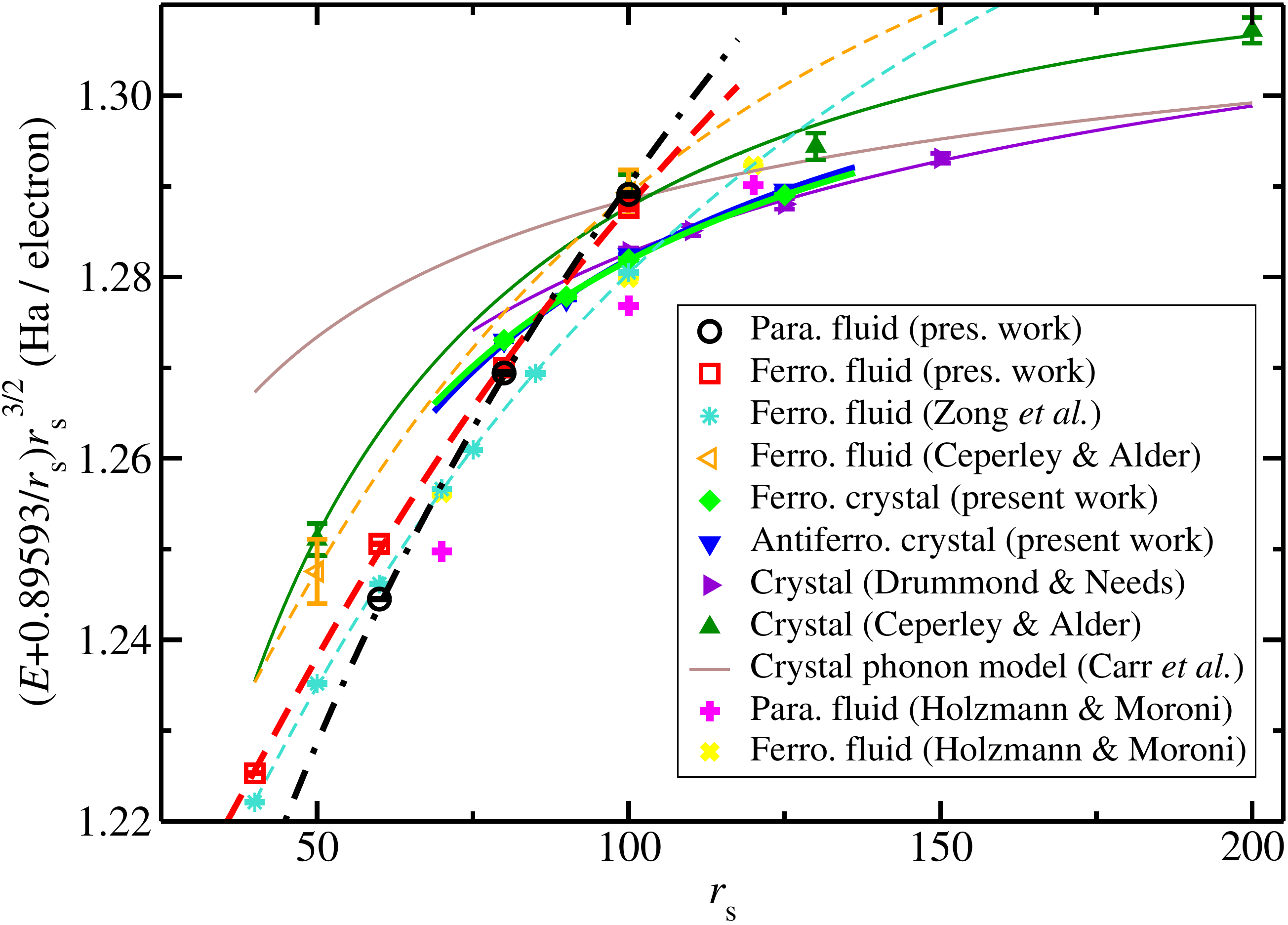}
\caption{\label{phasediagram} Energies per particle of the Fermi
  fluid and Wigner crystal phases at low density. The Madelung energy
  of the bcc lattice has been subtracted off and the resulting
  energies have been rescaled by $r_\text{s}^{3/2}$ to highlight the
  differences between phases.}
\end{figure*}

The DMC energies of different phases of the 3D-HEG, extrapolated to
the thermodynamic limit, are plotted against $r_\text{s}$ in
Fig.\ \ref{phasediagram}.  Our Wigner crystal energies are in good
agreement with the results reported in
Ref.\ \onlinecite{Neil04}. However, our Fermi fluid energies are
substantially higher than those of Ref.\ \onlinecite{Zong}, leading to
a significant revision of the crystallization density, which is now
predicted to occur at $r_\text{s}=86.6(7)$.  We investigate possible
reasons for the disagreement with Ref.\ \onlinecite{Zong} in the
next section, finding that the treatment of finite-size
effects is the most likely source of disagreement.

The ferromagnetic fluid becomes more stable than the paramagnetic fluid in
the immediate vicinity of the crystallization density; hence we do not
predict a region of stability for itinerant ferromagnetism in the
3D-HEG\@. The absence of a region of stability for the ferromagnetic
fluid has also recently been predicted by Holzmann and Moroni, who
performed DMC calculations for the fluid phases of the 3D-HEG in a
66-electron simple cubic cell and applied finite-size corrections to
their data \cite{Holzmann_2020}.
The curves fitted to our DMC energy data for ferromagnetic and
antiferromagnetic bcc crystals cross at $r_\text{s}=93(3)$, which is
just inside the region of stability for the Wigner crystal.  However,
in the region of stability, the differences between our ferromagnetic
and antiferromagnetic crystal DMC energies are statistically
insignificant.

Path integral Monte Carlo calculations of the exchange
coupling constants of bcc Wigner crystals \cite{Candido_2004} show
that at $r_\text{s}=100$ the energy difference between ferromagnetic
and antiferromagnetic configurations is only $1.2\times 10^{-8}$ Ha/el
(an order of magnitude smaller than our DMC error bars), and
demonstrate that the 3D Wigner crystal is antiferromagnetic
\cite{Thouless_1965}.  Given that the energy difference between
antiferromagnetic and ferromagnetic crystals is significant at high
density and exponentially small at low density, fitted energy-density
curves are liable to cross spuriously.

\subsection{Investigation of disagreement with F.\ H.\ Zong, C.\ Lin, and
D.\ M.\ Ceperley, Phys.\ Rev.\ E {\bf 66}, 036703 (2002)
(Ref.\ \onlinecite{Zong})}

As can be seen in Fig.\ \ref{phasediagram}, our SJB-DMC energies in
the thermodynamic limit are higher than those of
Ref.\ \onlinecite{Zong}.  Here we try to identify the cause of the
disagreement.

According to Ref.\ \onlinecite{Zong}, the energy of a paramagnetic
($\zeta=0$) 3D-HEG at $r_\text{s}=40$ computed using TA SJB-DMC
simulations and extrapolated to the thermodynamic limit from $N=54$
and $N=108$ is $-0.0176187(3)$ Ha/el.  To try to reproduce this result
we used SJB wave functions for the paramagnetic ($\zeta=0$) 3D-HEG at
$r_\text{s}=40$ and the same system sizes as
Ref.\ \onlinecite{Zong}. The DMC time step was 10 Ha$^{-1}$.  We used
1200 walkers for $N=54$ and 2400 walkers for $N=108$.  The numbers of
twists for $N=54$ and $N=108$ were 700 and 110, respectively.  Our DMC
energy at the thermodynamic limit, which is obtained by extrapolation
of the TA DMC energies in Table \ref{N54N108}, is $-0.0176038(1)$
Ha/el. This is 14.9(3) $\mu$Ha/el.\ higher than the result obtained in
Ref.\ \onlinecite{Zong}.

\begin{table}[!htbp]
\centering
\caption{\label{N54N108} Energies and energy variances
  (${\sigma_\text{v}}^2$) of the paramagnetic ($\zeta=0$) 3D-HEG at
  $r_\text{s}=40$. The TA DMC energies are calculated using an SJB
  wave function in which the Jastrow factor and backflow function were
  optimized at ${\bf k}_\text{s}={\bf 0}$. Energies are in Ha/el.\ and
  variances are in Ha$^2$.}
 \begin{tabular}{lccc}
 \hline\hline

  $N$ &  $E_{\rm VMC}$ &  ${\sigma_\text{v}}^2$ &  $E_{\rm DMC}$ \\

  \hline

  $54$  & $-0.0175860(9)$ & $0.0000553(6)$ & $-0.01766933(6)$ \\

  $108$ & $-0.0175770(5)$ & $0.0000898(8)$ & $-0.01763656(7)$ \\

\hline\hline
 \end{tabular}
\end{table}

\begin{figure}[!htbp]
\centering
 \includegraphics[clip,scale=0.75]{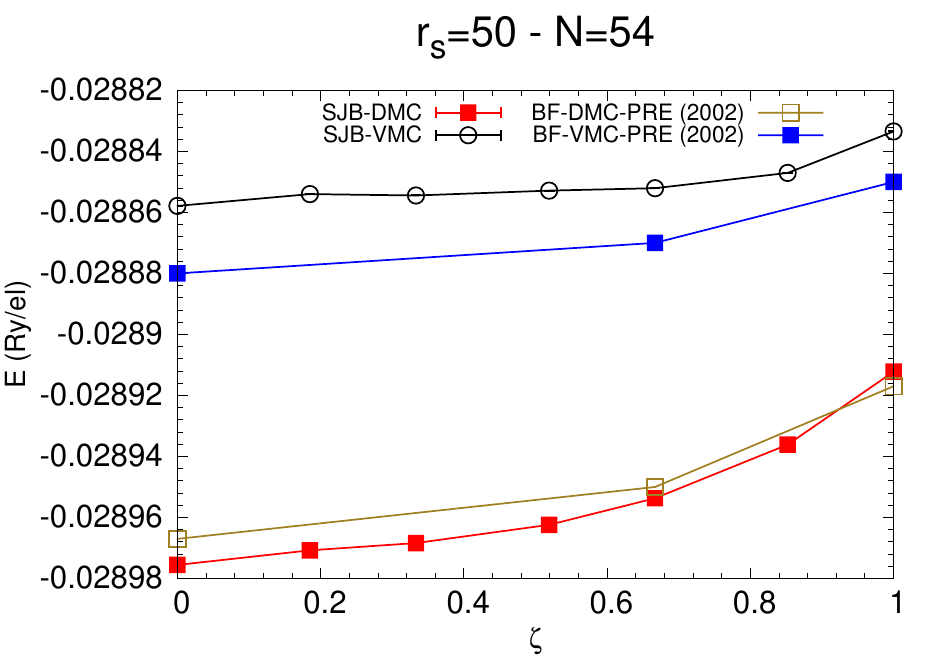}
 \caption{\label{N54_Fig1_PRE} Twist-averaged SJB-VMC and SJB-DMC
   energies per electron $E$ against spin polarization $\zeta$ at
   $r_\text{s}=50$ for $N=54$ electrons. Our results were obtained
   using a face-centered cubic simulation cell.  The results reported
   in Ref.\ \onlinecite{Zong} [``PRE (2002)''] may have been obtained
   in a simple cubic simulation cell, although the difference between
   the ground-state energies in face-centered and simple cubic cells
   is small, as shown in Table \ref{FCC_SC}.}
\end{figure}

To investigate further, we have studied $54$-electron 3D-HEGs at
$r_\text{s}=50$ with the same set of spin polarizations as
Ref.\ \onlinecite{Zong}. We used our SJB wave function with several
hundred twists to reach a precision of $10^{-7}$
Ha/electron. Figure~\ref{N54_Fig1_PRE} shows our VMC and DMC energies
compared with data extracted from Fig.\ 1 of
Ref.\ \onlinecite{Zong}. Because they used a three-body term in their
Jastrow factor, their VMC energies are $\sim 0.02$ mRy/electron lower
than our VMC energies. However, the three-body Jastrow term does not
directly affect the nodal surface of the wave function and indeed our
DMC energies agree well with those of Ref.\ \onlinecite{Zong}. We
performed two test calculations in which a polynomial three-body term
was included in our Jastrow factor. The resulting SJB-VMC energies for
the paramagnetic and ferromagnetic fluid phases are $-0.028913(1)$ and
$-0.028906(1)$ Ry/electron, respectively, which are lower than the
corresponding VMC energies of Ref.\ \onlinecite{Zong} (see
Fig.\ \ref{N54_Fig1_PRE}).  However, including the three-body term did
not change our DMC energies significantly.  Indeed,
Fig.\ \ref{N54_Fig1_PRE} shows that the source of the discrepancy
between our final SJB-DMC results and those of Ref.\ \onlinecite{Zong}
is neither the form of our trial wave function nor the optimization
scheme, because the SJB-DMC results agree at $N=54$.

We calculated the energy of the 3D-HEG at $r_\text{s}=50$ in the
infinite system size limit using TA and extrapolation from data at
$N=54$ and $N=108$; these are the system sizes used in
Ref.\ \onlinecite{Zong}. According to Ref.\ \onlinecite{Zong}, the TA
SJB-DMC energies of paramagnetic fluids at $r_\text{s}=50$ at a system
size of $N=54$ and at the thermodynamic limit are $-0.028967(1)$ and
$-0.0288990(6)$ Ry/electron, respectively. Hence they find the
absolute difference between the SJB-DMC energy at infinite system size
and in a 54-electron cell to be just 0.068(1) mRy/electron.

Our SJB-DMC energies of the paramagnetic fluid at $r_\text{s}=50$ at a
system size $N=54$ and at the thermodynamic limit are $-0.0289755(1)$
and $-0.0288756(4)$ Ry/electron, respectively, and the absolute
difference between them is 0.1000(4) mRy/electron, much larger than
predicted in Ref.\ \onlinecite{Zong}.

According to Ref.\ \onlinecite{Chiesa_2006}, the leading order
correction for systematic long-range FS errors in the energy is
$\Delta_N = \omega_\text{p}/(2N) = \sqrt{3/r_\text{s}^3}/(2N)$ in
atomic units, where $\omega_\text{p} = \sqrt{3/r_\text{s}^3}$ is the
plasma frequency. Using $N=54$ and $r_\text{s}=50$ gives $\Delta_N =
0.09072$ mRy/electron, which is relatively close to our estimate of
the difference between the energy per particle at infinite system size
and in a 54-electron cell.

\begin{figure}[!htbp]
\centering
 \includegraphics[clip,scale=0.75]{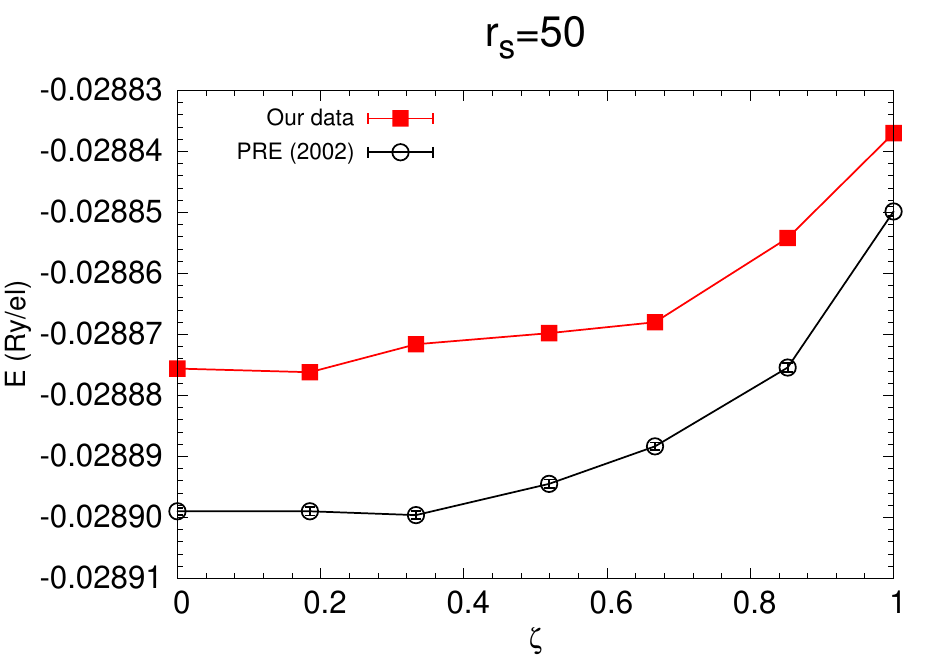}
 \caption{\label{DMC_vs_zeta} Energy per electron as a function of
   spin polarization $\zeta$ at $r_\text{s}=50$. The energies are
   obtained using extrapolation from TA energies at $N=54$ and $N=108$
   to infinite system size. }
\end{figure}
Figure~\ref{DMC_vs_zeta} compares our DMC energies with those of
Ref.\ \onlinecite{Zong} for 3D-HEGs at $r_\text{s}=50$ in the
thermodynamic limit. The differences between our DMC energies and
those of Ref.\ \onlinecite{Zong} at $\zeta=0.0$, 0.185, 0.333, 0.519,
0.667, 0.852, and 1.0 are 0.023(7), 0.023(7), 0.0280(7), 0.0247(6),
0.0203(6), 0.0212(9), and 0.013(1) mRy/electron, respectively.

We investigated other factors that could affect our DMC energies:
\begin{itemize}
 \item We reoptimized the backflow function and Jastrow factor at
   different twist vectors ${\bf k}_\text{s}$ for the paramagnetic
   fluid. The SJB-DMC energies at infinite system size, which are
   obtained by extrapolation from TA data at $N=54$ and $N=108$, show
   that the energy change due to optimizing the backflow function at
   different twists is small (see Table~\ref{BF_reopt}).
\begin{table*}[!htpb]
\centering
\caption{\label{BF_reopt} SJB-DMC energy in Ha/electron of a
  paramagnetic fluid at $r_\text{s}=50$ in the thermodynamic limit of
  infinite system size ($N \rightarrow \infty$) using three different
  simulation-cell Bloch vectors ${\bf k}_\text{s}$ for the
  optimization of the backflow function and Jastrow factor.  The
  SJB-DMC energies were obtained by extrapolation from TA SJB-DMC
  energies obtained at system sizes of $N=54$ and $N=108$.}
 \begin{tabular}{l|ccc|ccc}
 \hline \hline

\multicolumn{1}{c}{} & \multicolumn{6}{c}{SJB-DMC energy (Ha/electron)} \\

  & & $\zeta=0$ & & &$\zeta=1$ &  \\

  \hline

 ${\bf k}_\text{s}$  & $N=54$ & $N=108$ & $N \rightarrow \infty$ & $N=54$ & $N=108$ & $N \rightarrow \infty$\\

 \hline

  $\Gamma$ & $-0.0144877(1)$ & $-0.0144628(1)$ & $-0.0144378(1)$ & $-0.0144561(2)$ & $-0.0144373(3)$&  $-0.0144185(3)$ \\

  $L$      & $-0.0144870(1)$ & $-0.0144627(1)$ & $-0.0144383(1)$ & $-0.0144553(2)$ & $-0.0144369(3)$ & $-0.0144185(3)$ \\

  $X$      & $-0.0144866(1)$ & $-0.0144622(1)$ & $-0.0144378(1)$ & $-0.0144555(2)$ & $-0.0144366(3)$ & $-0.0144178(3)$ \\

  \hline \hline
 \end{tabular}
\end{table*}

 \item All the results of this section were obtained with a fixed DMC
   time step $\tau=10$ a.u., which is appropriately small at density
   parameter $r_\text{s}=50$. As shown in
   Sec.\ \ref{sec:dmc_dt}, the resulting time step errors are
   much smaller than the difference between our energy data and those
   of Ref.\ \onlinecite{Zong}.

 \item The shape of the simulation cell affects the DMC energy at
   finite system size. We used face centered cubic (fcc) simulation
   cells for our fluid calculations. It is not clear to us what shape
   of simulation cell was used in Ref.\ \onlinecite{Zong}.  $N=54$ is
   a magic number of electrons for paramagnetic 3D-HEGs in both simple
   cubic (sc) and fcc cells subject to periodic boundary
   conditions. We compare TA SJB-DMC energies for the paramagnetic
   fluid phase in fcc and sc cells at $r_\text{s}=50$ in
   Table~\ref{FCC_SC}. The difference between these DMC energies in
   the thermodynamic limit is negligible, as expected.
\begin{table}[!htbp]
\centering
\caption{\label{FCC_SC} TA SJB-DMC energy of the paramagnetic fluid
  phase at $r_\text{s}=50$ using fcc and sc simulation cells. Energies
  are in Ha/electron.}
 \begin{tabular}{lccc}
 \hline \hline
 Simulation cell  & $N=54$ & $N=108$  & $N \rightarrow$ $\infty$ \\
 \hline
  fcc & $-0.0144878(1)$ & $-0.0144628(1)$ & $-0.0144378(1)$ \\
  sc  & $-0.0144832(1)$ & $-0.0144592(1)$ & $-0.0144352(1)$ \\
  \hline \hline
 \end{tabular}
\end{table}

\end{itemize}

Since our SJB-DMC results agree with Ref.\ \onlinecite{Zong} at $N=54$
but not in the thermodynamic limit, and since Ref.\ \onlinecite{Zong}
states that finite-size effects are small in contradiction with the
analytic theory of finite-size effects \cite{Chiesa_2006}, we conclude
that FS extrapolation is the most likely cause of the disagreement
between our results and Ref.\ \onlinecite{Zong}.

\subsection{Comparison with M.\ Holzmann and S.\ Moroni,
  Phys.\ Rev.\ Lett.\ \textbf{124}, 206404 (2020)
  (Ref.\ \onlinecite{Holzmann_2020}) }

Reference \onlinecite{Holzmann_2020} disagrees with
Ref.\ \onlinecite{Zong} and agrees with our finding that the
ferromagnetic fluid has no region of stability.  Nevertheless, there
remains a quantitative disagreement over the crystallization density.
Reference \onlinecite{Holzmann_2020} uses computationally expensive
recursive backflow wave functions in SJB-DMC calculations at a fixed,
relatively small system size ($N=66$, in an sc cell).  Furthermore,
they extrapolate their SJB-DMC energy data to zero VMC energy
variance.  In general such an extrapolation is error-prone, possibly
introducing nonvariational errors (consider, for example, the effects
of switching between optimizing wave functions by variance and energy
minimization); however, in Ref.\ \onlinecite{Holzmann_2020} great care
has been taken to ensure the extrapolation is as reliable as possible.
On the other hand, rather than extrapolating energy data to infinite
system size, Ref.\ \onlinecite{Holzmann_2020} relies on analytic FS
correction formulas.

In Table \ref{table:para_sc_N66_rs100} we compare our TA SJB-DMC
energies for the paramagnetic 66-electron Fermi fluid in a sc cell at
$r_\text{s}=100$ with the results reported in
Ref.\ \onlinecite{Holzmann_2020}.  Our (nonrecursive) SJB trial wave
function gives a lower TA DMC energy than their nonrecursive SJB wave
function (``BF0'' in Table \ref{table:para_sc_N66_rs100}).  However,
the recursion of backflow transformations followed by extrapolation to
zero VMC variance results in lower TA SJB-DMC energies than ours.
This difference between the fluid energies in our work and
Ref.\ \onlinecite{Holzmann_2020} is sufficient to explain about half
the difference between the predicted crystallization densities
[$r_\text{s}=86.6(7)$ and 113(2), respectively].  The rest of the
difference can be ascribed to the fact that we extrapolate to infinite
system size from larger simulation cells.

\begin{table}
\centering
\caption{TA SJB-DMC energies for a 66-electron paramagnetic Fermi
  fluid of density parameter $r_\text{s}=100$ in a sc cell. The
  results without citation were obtained in the present work.
  ``BF$n$'' denotes an SJB wave function obtained using $n$ recursive
  backflow transformations.  \label{table:para_sc_N66_rs100}}
\begin{tabular}{lc}
\hline \hline

Trial wave function & SJB-DMC energy (mRy/el.) \\

\hline

BF0 \cite{Holzmann_2020} & $-15.37588(9)$ \\

No 3-body Jastrow term & $-15.37721(11)$ \\

With 3-body Jastrow term & $-15.37810(8)$ \\

BF1 \cite{Holzmann_2020} & $-15.38345(7)$ \\

BF4 \cite{Holzmann_2020} & $-15.38683(4)$ \\

Extrap.\ to zero var.\ \cite{Holzmann_2020} & $-15.38914(17)$ \\

\hline \hline
\end{tabular}
\end{table}

\section{Conclusion}

In conclusion, we have revisited the phase diagram of the 3D-HEG using
state-of-the-art QMC methods.  The energies of Wigner crystals are
similar to previous QMC calculations.  However, we find ferromagnetic
fluid energies that are significantly higher than previous
calculations, leading to a higher crystallization density, at
$r_\text{s}=86.6(7)$.  We find no statistically significant region of
stability for itinerant ferromagnetism.  The zero-temperature phase
diagram of the 3D-HEG is therefore found to be qualitatively similar
to that of the two-dimensional homogeneous electron gas
\cite{Drummond_2009}.

\acknowledgments{S.\ Azadi acknowledges PRACE for awarding us access to the
High-Performance Computing Center Stuttgart, Germany, through the
project 2020235573.  Our Wigner crystal calculations were performed
using Lancaster University's High End Computing cluster.  This work
used the ARCHER2 UK National Supercomputing Service
(\url{https://www.archer2.ac.uk}). We acknowledge useful conversations
with Matthew Foulkes, Gino Cassella, and David Ceperley.}

\end{document}